\documentclass[11pt]{article}
\usepackage[utf8]{inputenc}
\usepackage[a4paper,bindingoffset=0.2in,
left=0.5in,right=0.7in,top=1in,bottom=1.2in]{geometry}
\usepackage{jheppub}
\usepackage{amsmath,amssymb,amsfonts,graphicx,slashed,color,amsthm,mathtools,upgreek, enumerate, tensor, scalerel, subfig}
\usepackage[colorlinks=true, linkcolor=blue, citecolor=blue, urlcolor=blue]{hyperref}
\usepackage{arydshln}
\usepackage[dvipsnames]{xcolor}
\usepackage[normalem]{ulem}
\usepackage{bbold}

\usepackage{cancel}

\newcommand{\beq}{\begin{equation}}
\newcommand{\eeq}{\end{equation}}
\newcommand{\beqn}{\begin{eqnarray}}
\newcommand{\eeqn}{\end{eqnarray}}

\renewcommand{\d}{\mathrm{d}}
\newcommand{\eps}{\epsilon}

\newcommand{\cH}{\mathcal{H}}

\newcommand{\bra}[1]{\langle #1 \vert}
\newcommand{\ket}[1]{\vert #1 \rangle}

\newenvironment{malign}{\begin{equation}\begin{aligned}}{\end{aligned}\end{equation}\noindent\ignorespacesafterend}

\newcommand{\Tr}{\text{Tr}}

\newcommand{\jt}{\tilde{J}}

\newcommand{\qt}{\tilde{q}}

\def\le{\left}
\def\ri{\right}
\newcommand\ov{\over}
\newcommand\p{\ensuremath{\partial}}
\DeclarePairedDelimiter{\abs}{\lvert}{\rvert}
\newcommand{\es}[2] {\begin{equation} \label{#1} \begin{split} #2 \end{split} \end{equation}}

\title{Entanglement membrane in the Brownian SYK chain}
\author[a]{M\'ark Mezei} 
\author[b,c]{and Harshit Rajgadia}
 \affiliation[a]{Mathematical Institute, University of Oxford, Woodstock Road, Oxford, OX2 6GG, United Kingdom} 
 \affiliation[b]{Department of Theoretical Physics, Tata Institute for Fundamental Research, 1 Homi Bhabha Road, Mumbai, Maharashtra 400005, India}
 \affiliation[c]{Kavli Institute for Theoretical Physics, University of California, Santa Barbara, CA 93106}
\date{}
\abstract{
    There is mounting evidence that  entanglement dynamics in chaotic many-body quantum systems in the limit of large subsystems and long times is described by an entanglement membrane effective theory. In this paper, we derive the membrane description in a solvable chaotic large-$N$ model, the Brownian SYK chain. This model has a collective field description in terms of fermion bilinears connecting different folds of the multifold Schwinger-Keldysh path integral used to compute R\'enyi entropies. The entanglement membrane is a traveling wave solution of the saddle point equations governing these collective fields. The entanglement membrane is characterised by a velocity $v$ and a membrane tension ${\cal E}(v)$ that we calculate. We find that the membrane has finite width for $v<v_B$ (the butterfly velocity), however for $v > v_B$, the membrane splits into two wave fronts, each moving with the butterfly velocity. Our results provide a new viewpoint on the entanglement membrane and uncover new connections between quantum information dynamics and scrambling.  }

\begin{document}
\maketitle

\section{Introduction}

Quantum chaotic dynamics \cite{Srednicki_1994, Srednicki1999Approach, DAlessio2016AdvPhys} is in the focus of many current developments in physics, ranging from quantum simulations \cite{Li2017PRX_OTOC_NMR,Kaufman2016Science_ThermalizationEntanglement} to the black hole information puzzle \cite{Sekino_2008, Hayden:2007cs, Harlow_2016}. It is important to uncover what aspects of the dynamics are universal and how different chaotic phenomena across time scales are related. 

Random matrix universality \cite{Wigner1955AnnMath, BohigasGiannoniSchmit1984PRL, AltlandZirnbauer1997PRB,Cotler:2016fpe} is a central concept in this quest, and hydrodynamic transport of conserved densities \cite{Kovtun2012Lectures,Lucas2015HydroTransport} is another. In recent years much insight was gained  by quantifying scrambling through operator growth. In a wide range of chaotic models  \cite{Roberts_2015,RobertsStanford_2015, RobertsSwingle2016LRButterfly, vonKeyserlingk2018OperatorHydrodynamics} the support of simple operators grows ballistically with a characteristic velocity $v_B$, the velocity of the butterfly effect.\footnote{$v_B$ is closely related, but distinct from the Lieb-Robinson velocity \cite{LiebRobinson1972FiniteVelocity,Roberts:2016wdl}.} In large-$N$ systems a measure of operator size, the out-of-time order correlator (OTOC) has an effective description in terms of universal scramblon modes \cite{Maldacena:2016upp,BlakeLeeLiu2018QHydro,Gu_2022,Stanford:2021bhl,Choi_2023,StanfordVardhanYao2024ScramblonLoops} responsible both for the OTOC's exponential decay in time of and its structure in space.

There is a large body of evidence that quantum information dynamics also exhibits universality across a wide range of models \cite{Liu_2014,Nahum2017PRX, Zhou:2018myl,Mezei_2017}.  We can track the dynamics of  quantum information 
through the R\'enyi entropies of geometric subregions, a one parameter generalisation of the von Neumann entropy. 
In chaotic systems we expect that there is a local thermalisation time scale $t_\text{loc}$ and a corresponding distance scale $\ell \sim v_B  \,t_\text{loc}$. For subregions much larger than $\ell$ and at times much larger than $t_\text{loc}$ we expect that a simplified effective description of entropy dynamics should apply. This description is proposed to be the entanglement membrane effective theory \cite{JonayHuseNahum2018}. 
 By now there is a variety of setups where the membrane description has been derived: in random quantum circuits \cite{Nahum2017PRX,Zhou:2018myl}, in holographic field theories \cite{Mezei2018PRD} (so far only in the von Neumann case), in Floquet systems~\cite{Zhou:2019pob}, and in Brownian spin chains \cite{vardhan_moudgalya2024}. Our work adds the large-$N$ Brownian SYK chain\footnote{Ref.~\cite{Jian:2021hve} discusses a special case of entanglement membranes (which they refer as spacetime domain walls) in the context of measurement-induced phase transition in the Brownian SYK chain.} \cite{Saad:2018bqo, Sunderhauf:2019djv, Jian:2021hve, Stanford:2021bhl, Balasubramanian:2023xdp} to this list and provides a new viewpoint on the entanglement membrane. In the rest of the introduction we assume basic familiarity with the entanglement membrane. We provide a brief review of it in section~\ref{sec:membrane_review}, which can be read as a self-contained introduction.

 To compute the $n$-th R\'enyi entropy we consider $n$ replicas of the original system, and to describe real time dynamics we use Schwinger-Keldysh techniques, which lead to another doubling of degrees of freedom. An SYK chain has $N$ fermion flavours on each lattice site interacting through a  $q$-local interaction.
 The collective fields of the Brownian SYK chain are equal time fermion two-point functions between the $2n$ copies averaged over the $N$ flavours. The saddle point equations form a multi-field reaction–diffusion system that generalises 
 the Fisher-Kolmogorov-Petrovsky-Piskunov (FKPP) equation \cite{Fisher, kolmogorov1937, BramsonLectures} used to model population growth and wave propagation.
 We find that  the entanglement membrane is a new saddle point, a localised traveling wave solution of this equation. 
 
 By changing the subregion we change boundary conditions on the saddle point equations, which changes the velocity $v$ of the traveling wave.\footnote{Unlike the FKPP equation, our multi-field reaction–diffusion system does not have a lower bound of $v$ for traveling waves.} This velocity becomes the velocity of the entanglement membrane, and by evaluating the on-shell action of the traveling wave we obtain the membrane tension ${\cal E}^{(n)}(v)$, which is an input (Wilson function) of the membrane effective theory. We can carry out this computation only by numerically solving a set of ordinary differential equations for general values of $q$-locality parameter, but we managed to obtain analytic results for large $q$.
 These are our main technical results: ${\cal E}^{(n)}(v)$ for $q=4$ and $n=2,3$ is plotted in figure~\ref{fig:membrane_tension}, while in equation~\eqref{E2v_largeq} we give an analytic expression for ${\cal E}^{(2)}(v)$ in the large $q$ limit, which we also plot in figure~\ref{fig:membrane_tension_largeq}. 
 
 We find that the traveling wave cannot travel faster than the butterfly velocity $v_B$, and its size diverges in the limit $v \rightarrow v_B$. If we try to force the wave to go faster, it splits into two fronts, each moving with $v_B$. This leads to a breakdown of localised membrane description. The two fronts are separated by a new domain of width $(v-v_B)T$, where $T$ is the length of the time evolution. 
 
 Since scramblons are shock wave solutions in the same collective variables, in the Brownian SYK chain we can uncover the interplay between the chaotic processes of entanglement and operator growth and explain how the butterfly velocity $v_B$ controls aspects of both. We also argue that for $v>v_B$ the aforementioned wide domain between two fronts hosts scramblon fluctuations that become large for $T>t_\text{scr}\sim \log N$.

 While we focus on the chain geometry, our results straightforwardly carry over to higher dimensional lattices:
the membrane tension is isotropic and ${\cal E}^{(n)}(v)$ is independent of the dimension and is given by our result. The entanglement membrane is particularly powerful in higher dimensions, where it computes the entropy of arbitrary shaped subregions. 

The dynamics of R\'enyi entropies in Brownian systems with continuous time evolution has been studied before \cite{Zhang:2023vpm,Swann:2023vpg,vardhan_moudgalya2024, parrikar2025}. In \cite{parrikar2025}, HR and collaborators studied the spread of information stored in $k$ qubits injected at some site of the Brownian SYK chain in the large $N$ limit. When $k \ll N$, the information spread can be computed perturbatively in the ratio $k/N$. It was found that at leading order in $k/N$, this is governed by the operator growth equation, which is the FKPP equation. As a consequence of this, the information spreads on the chain with the butterfly velocity $v_B$. Our work is a generalization of this setup, where, instead of a small qubit, we study the spread of information stored in a large subregion as the chain evolves in time. In this case, the information spreading is more nontrivial and it is governed by the membrane formula as discussed in section~\ref{sec:membrane_review}.

In~\cite{Swann:2023vpg} the $N=1$ Brownian SYK chain was analysed with rather different methods from ours. Their main focus was on the free theory that is quadratic in the fermions, where no membrane was found. When interactions (for four neighbouring sites) were added, they found a saddle point with a localised membrane, indicating the crossover from diffusive to ballistic spreading of information. They only analysed the $v=0$ case and their findings are in full agreement with ours.\footnote{We checked that upon converting conventions the value ${\cal E}^{(2)}(0)$ matches between their paper and ours.} {\bf Note added:} Upon completion of the draft of this paper we learned about ongoing work, where Swann and Nahum analyse the $v\neq 0$ case~\cite{Swann_draft}. They have independently discovered the splitting of the membrane into two fronts for $v>v_B$.

In \cite{vardhan_moudgalya2024}, the authors map the calculation of the $n$-th R\'enyi entropy of a subsystem in a Brownian spin chain to a transition amplitude under a Euclidean evolution by an effective Hamiltonian on $2n$ copies of the system. Their work does not rely on large $N$. The membrane picture emerges at late times as a consequence of some low-lying excitations of the Hamiltonian.  These low-lying excitations correspond to plane wave excitations of locally dressed domain walls that separate two different ground states of the Hamiltonian. The traveling wave solution found in our work precisely corresponds to a domain wall that propagates between two different fixed points of the large $N$ equations of motion of the Brownian SYK chain.

While the Brownian SYK chain is chaotic, it only captures the infinite temperature dynamics, in contrast to the SYK chain \cite{Gu_2017}, where the dynamics is temperature dependent. The dynamics of the R\'enyi entropy on the SYK chain were studied in \cite{Gu_lucas_2017}, see \cite{zhang2022syk} for a review. The authors computed the entanglement growth of a two-sided subregion in the thermofield double state at low temperature and weak intersite coupling. They found that the R\'enyi entropy grows linearly at early times: this is indicative of a membrane effective description. However, there are open questions about the saturation behaviour of the entropies. It would be interesting to generalise our findings to this setup.

 The outline of the paper is as follows: In section~\ref{sec:membrane_review} we review the entanglement membrane effective theory and an operator entanglement setup that makes the computation of the membrane tension particularly convenient. In section~\ref{sec:Renyi2}, we set up the Schwinger-Keldysh path integral to compute R\'enyi entropies. We focus on the second R\'enyi entropy to avoid complicated bookkeeping, and treat higher R\'enyi entropies in appendices~\ref{app:vgvb},~\ref{app:Renyi3}, and~\ref{app:highern}. The main result of this section is the determination of boundary conditions obeyed by the SYK collective fields. In section~\ref{sec:BSYK} we review the collective field path integral in Brownian SYK on the multifold Schwinger-Keldysh contour, we obtain the saddle point equations that are a multi-field reaction–diffusion system in the continuum limit, and using symmetries we reduce the number of independent fields. In section~\ref{sec:solitons} we argue that in the long time limit the saddle point has an emergent translation symmetry in spacetime only broken by boundary layers near the initial and final boundaries. Away from these layers we find that the saddle point is a traveling wave. For $v>v_B$ there are two fronts traveling with $v_B$ separated by a domain that hosts scramblons. For $v<v_B$ instead we find one localised traveling wave that we identify with the entanglement membrane. In section~\ref{sec:action} we evaluate the on-shell action. This computation is nontrivial because the collective field action contains a nonlocal in time Pfaffian term, which however can be treated in the adiabatic approximation of a few-fermion quantum mechanics.
We end the main text with a discussion in section~\ref{sec:discussion} and relegate several technical details and higher R\'enyi index computations to the appendices.

\section{A brief review of the membrane effective theory}\label{sec:membrane_review}
\begin{figure}[h]
    \centering
\includegraphics[width=0.8\linewidth]{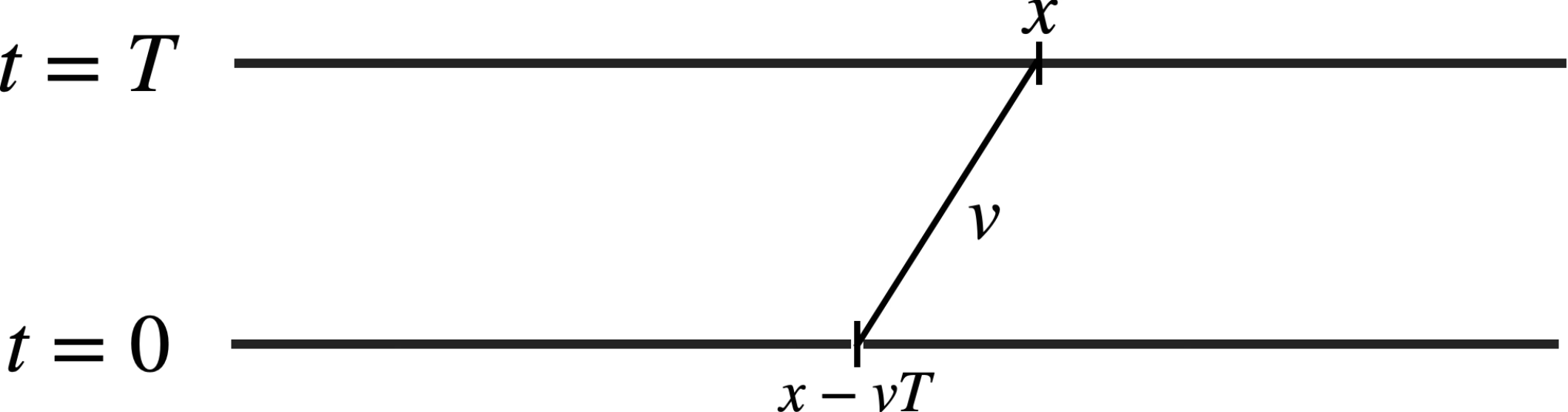}
    \caption{The $n$-th R\'enyi entropy of sites to the left of site $x$ at $t = T$ is related to a membrane of velocity $v$ that intersects the chain at $t = 0$ at site $x- vT$. The velocity of the membrane is determined by the minimization in equation~\eqref{eq:entanglement_growth_conjecture}. }
\label{fig:membrane_quench}
\end{figure}
 Consider a quantum system  on a one-dimensional chain with Hilbert space $\cH$. We are interested in the entanglement growth of some subregion in the time evolved state $\ket{\omega(T)} = U(T) \ket{\omega}$ for some $\ket{\omega} \in \cH$. For simplicity, let us consider the subregion containing all sites to the left of site $x$ and denote the $n$-th R\'enyi entropy of the subregion at time $T$ by $S^{(n)}(x,T)$. According to the membrane conjecture \cite{JonayHuseNahum2018}, $S^{(n)}(x,T)$ is given by the following formula (see figure~\ref{fig:membrane_quench}):
 \begin{malign} \label{eq:entanglement_growth_conjecture}
     S^{(n)}(x,T) = \min_{v} \left( s_{\text{eq}} \,\mathcal{E}^{(n)}(v)\, T +  S^{(n)}(x - vT,0) \right)\,.
 \end{malign}
The function $\mathcal{E}^{(n)}(v)$ is called the membrane tension and $s_{\text{eq}}$ is the local entropy density in thermal equilibrium. 

The membrane conjecture imposes some nontrivial constraints on the membrane tension. We can derive some simple constraints from the condition that for a state in thermal equilibrium, the R\'enyi entropy must follow the Page curve~\cite{Page:1993df}:
\begin{malign}
    \lim_{T \rightarrow \infty} S^{(n)}(x,T)  = \begin{cases}
        s_{\text{eq}} (x + L), \quad -L < x < 0\,, \\ s_{\text{eq}} (L  -x), \quad 0 < x < L\; .
    \end{cases}  
\end{malign}
In this case, the membrane formula suggests, for $x < 0$ that
\begin{malign}
    S^{(n)}(x,T)  = s_{\text{eq}} \min_v \left(\mathcal{E}^{(n)}(v) \, T + (x + L - v T)  \right)\,.
\end{malign}
The above function is minimised when $\mathcal{E'}^{(n)}(v_0) = 1$. The condition that the R\'enyi entropy should follow a Page curve implies that at $v = v_0$:
\begin{malign}
    \mathcal{E}^{(n)} (v_0) = v_0\,.
\end{malign}
Ref.~\cite{JonayHuseNahum2018} conjectured that $v_0 = v_B$ (the butterfly velocity). Therefore, the membrane tension must satisfy
\begin{malign}  \mathcal{E}^{(n)}(v_B) = v_B, \quad \mathcal{E'}^{(n)}(v_B) = 1\,.
\end{malign}together with the convexity condition:
\begin{malign}\label{eq:tension_membrane} 
\mathcal{E''}^{(n)}(v) \geq 0\,.
\end{malign}

\begin{figure}[h]
    \centering
    \includegraphics[height=4cm]{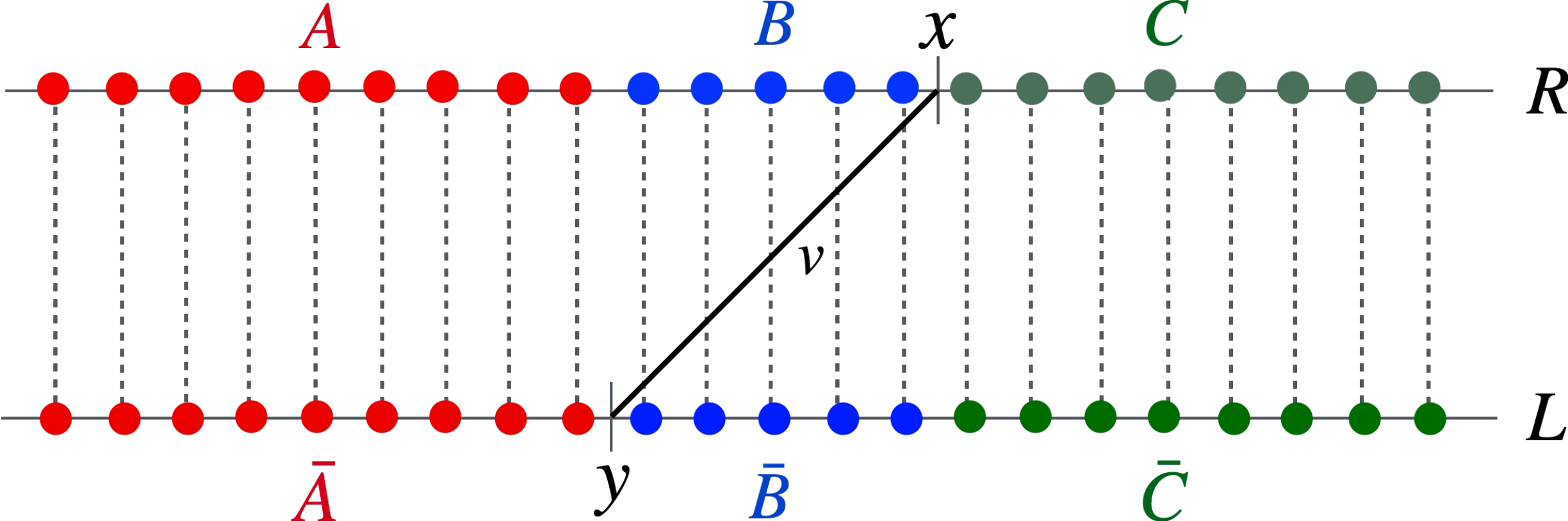}
    \caption{ {A maximally entangled state in $\cH_R \otimes \cH_L$. The dashed lines illustrate maximal entanglement between sites in $R$ and $L$ at $T = 0$. $S^{(n)}(x,y,T)$ is the entanglement entropy of the region $A \cup B \cup \bar A$. (Note that the bar stands for the subsystem in the $L$ chain and not for complement.) We illustrate the entanglement membrane by a black line of slope $v$ connecting $x$ and $y$.} }
\label{fig:setup}
\end{figure}
 Ref.~\cite{JonayHuseNahum2018} also conjectured that $\mathcal{E}^{(n)}(v)$ (for an infinite temperature state) is given by the ``operator R\'enyi entropy" of the unitary operator $U(T)$ defined as follows: Consider a quantum system consisting of two infinite chains $R \cup L$ with Hilbert spaces $\cH_R\otimes\cH_L$ such that $\cH_R \cong \cH_L \cong \cH$. Define a  state $\ket{\Psi}$ in which every site in chain $R$ is maximally entangled with a site in chain $L$ as shown in figure~\ref{fig:setup}. Now, consider the time evolution of chain $R$ by the unitary operator $U(T) \otimes I$
\begin{equation}\label{eq:maxstate}
    \ket{\Psi(T)} = U(T) \otimes I \ket{\Psi}\,.
\end{equation}
The operator R\'enyi entropy $S^{(n)}(x,y,T)$ is the R\'enyi entropy of the subregion containing all sites to the left of $x$ on chain $R$ and all sites to the left of $y$ on chain $L$ in the state $\ket{\Psi(T)}$ such that \begin{equation}\label{eq:membrane_velocity}
    v = \frac{x -y}{T}\,.
\end{equation}
The operator entanglement defines an ``effective" membrane tension $\mathcal{E}^{(n)}_\text{eff}$ which is related to $\mathcal {E}^{(n)}(v)$ as \begin{malign}  \label{eq:membrane_conjecture}
 \,\mathcal{E}^{(n)}_{\text{eff}}(v) = \frac{S^{(n)}(x,y,T)}{s_{\text{eq}} T} = \begin{cases}
     \mathcal{E}^{(n)}(v) \quad &v \leq v_B, \\ 
     v \quad  & v > v_B \,.
 \end{cases}
\end{malign}
For an infinite temperature state $s_{\text{eq}} = \log d$, where $d$ is the local Hilbert space dimension. 

In the rest of the paper we will use the above relation to calculate the R\'enyi membrane tensions for the range $v < v_B$ in the Brownian SYK chain. We will also see that a localised membrane exists for $v<v_B$, while for $v>v_B$,  $\mathcal{E}_\text{eff}^{(n)}(v)$ is associated to a macroscopic region in spacetime that is bounded by two wave fronts. The two wave fronts may be thought of as a membrane split into two halves.

\section{Second R\'enyi entropy} \label{sec:Renyi2}

To compute the $n$-th R\'enyi entropy using the path integral method, we need to consider $2n$ copies of the system: the reduced density matrix is computed from a bra and a ket that requires two copies, and raising it to the $n$-th power requires $n$ copies.
To illustrate our methods in the simplest setting, we will focus on the second R\'enyi entropy in the main text, and relegate the treatment of higher $n$ to Appendices~\ref{app:Renyi3},~\ref{app:highern}, and~\ref{app:tension_vgvb}. While we only focus on the Brownian SYK chain in this paper, in this section we explain the general setup of real time R\'enyi entropy computations in fermionic systems and we do not need to specify the Hamiltonian. Hence we postpone specialising to the  Brownian SYK Hamiltonian in section~\ref{sec:BSYK}.

\subsection{Schwinger-Keldysh contour}

As discussed in the previous section, the membrane tension is given by the formula~\eqref{eq:membrane_conjecture}. 
 Therefore, we are interested in computing $S^{{(2)}}(x,y,T)$. To calculate $S^{(2)}(x,y,T)$ it is convenient to divide  chain $R$ (respectively  $L$) into three subregions $A, B, C$ (respectively  $\bar A, \bar B, \bar C$) as shown in figure~\ref{fig:setup}. Then $S^{(2)}(x,y,T)$ corresponds to the second R\'enyi entropy of the subregion $A \cup B \cup \bar A$. To calculate the second R\'enyi entropy of $A \cup B \cup \bar A$ in the state $\ket{\Psi(T)}$, we need $\Tr\,\rho^2(T)_{A \cup B \cup \bar A}$. 

\begin{figure}[ht]  
    \centering
    \includegraphics[width =0.99\textwidth]{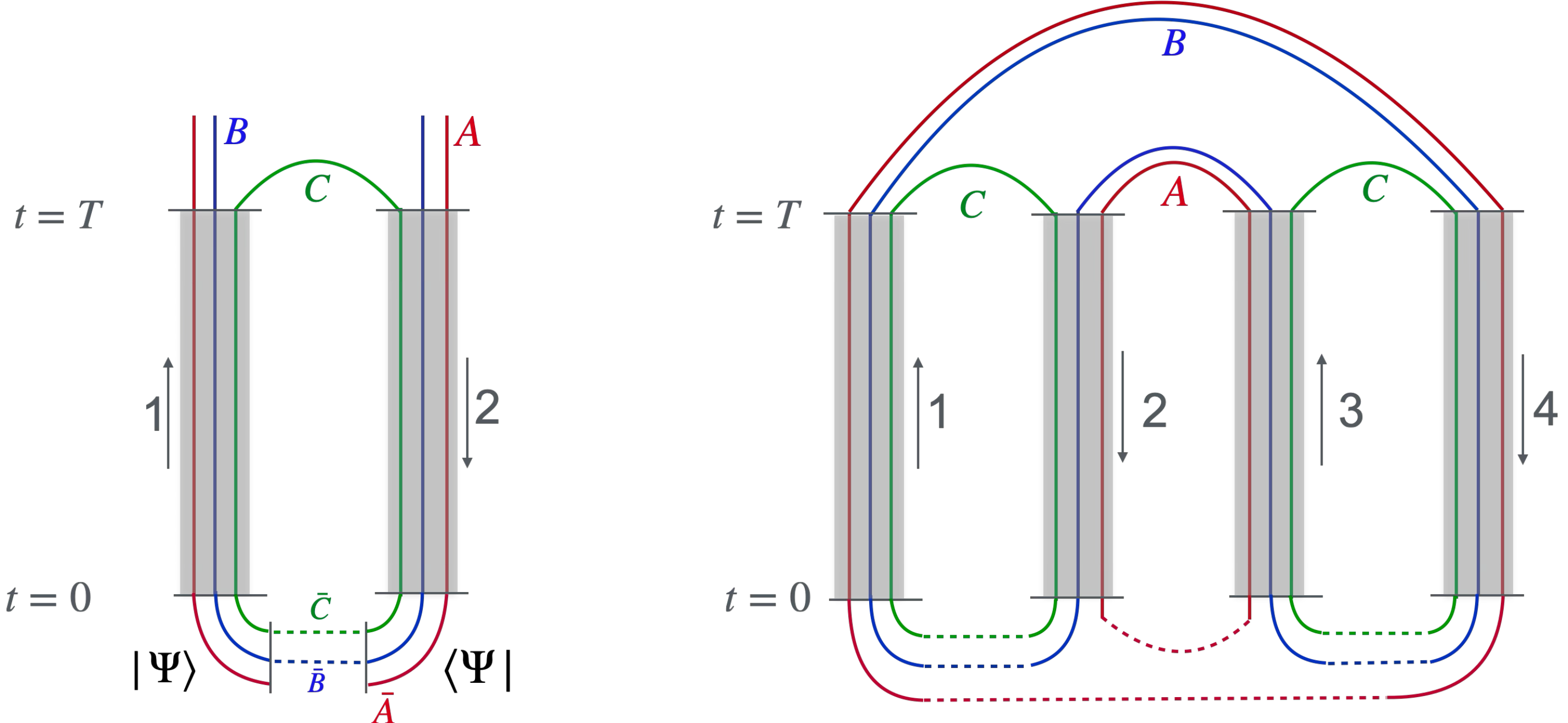}
    \caption{{ \textbf{Left:} Path integral contour for preparation of the state $\rho(T)_{A \cup B \cup \bar A}$. The solid arcs at the bottom of the contour illustrate maximally entangled states between the subregions of the chains $R$ and $L$. The grey boxes illustrate the joint unitary evolution of fermions in chain $R$. The open legs on the left and right denote the ket and bra states of the density matrix $\rho(T)_{A \cup B \cup \bar A}$ respectively. \textbf{Right:} We consider two copies of $\rho(T)_{A \cup B \cup \bar A}$ and cyclically contract all the open legs in $A \cup B \cup \bar A  $. This defines the Schwinger-Keldysh contour for $\Tr\,\rho^2(T)_{A \cup B \cup \bar A}$.}}
  \label{fig:density_matrices}
\end{figure}

The contour representation of $\Tr(\rho^2_{A \cup B \cup \bar A})$ is shown in figure~\ref{fig:density_matrices}.
Let us begin with the contour representation of the maximally entangled state $\ket{\Psi}\bra{\Psi}$. $\ket{\Psi} \bra{\Psi}$ can be prepared by cutting the fermionic path integral contour corresponding to $\Tr(e^{-0 \times H})$ into two arcs. This is represented by the solid arcs below the $t = 0$ line in figure~\ref{fig:density_matrices}. The ends of the arcs attached to the $t =0$ line denote the state of chain $R$ while the other ends of the arcs attached to the vertical line below $t = 0$ denote the state of chain $L$. 
To construct the state $\ket{\Psi(T)} \bra{\Psi(T)}$ we evolve $\ket{\Psi} \bra{\Psi}$ by the unitary operator $U(T) \otimes I$. The solid vertical lines between $t = 0$ and $t = T$ illustrate the evolution of the state on chain $R$. The reduced density matrix on $A \cup B \cup \bar A$ is obtained by identifying the open ends of the legs in $C \cup \bar B \cup  \bar C$.

The Schwinger-Keldysh contour of $\Tr(\rho^2(T)_{A \cup B \cup \bar A})$ is then obtained by taking two copies of $\rho(T)_{A \cup B \cup \bar A}$ and cyclically identifying the open legs as shown in figure~\ref{fig:density_matrices}. Notice that by construction, all contours are closed. Therefore, all fermions obey antiperiodic boundary conditions. 

Labeling a fermion on site $u$ that runs along a vertical leg $k$ by $\psi^k_u$, the path integral can be written as
\begin{equation} \label{eq:path_integral}
 \Tr (\rho_{A \cup B \cup \bar A}^2) =  \frac{1}{(\sqrt{2})^{2NL}} \int  \prod_{u } \prod_{j=1}^{4} \mathcal{D} \psi_u^{j}  \,  \exp \left[i \sum_{k = 1}^{2} \int_0^T \d t \left(  \mathcal{L}(t,\psi_u^{2k-1}(t)) - \mathcal{L}(t,{\psi}_u^{2k}(t) \right) \right]\,,
\end{equation}
The factor of $1/\sqrt{2}^{2NL}$ is an overall normalization that arises from the path-integral representation of the maximally entangled state.\footnote{Every Majorana fermion along a closed contour contributes an overall factor of $\sqrt{2}$.} The Lagrangian $\mathcal{L}$ is given by: \begin{align}
    \mathcal{L}(t,\psi(t))= i \sum_{u} \frac{\psi_u (t)\partial_t \psi_u(t)}{2} - H(t,\psi(t))\,.
\end{align}

Here, $H(t)$ is the time-dependent Hamiltonian of the Brownian SYK chain.  
The antiperiodic boundary condition is ensured if we use the following boundary conditions at $t = 0,T$.\footnote{The choice of boundary condition at $t = 0$ and $t = T$ is not unique.  Any other choice of boundary conditions that ensures antiperiodic boundary conditions for all fermions works equally well.}
\begin{malign} \label{BCinit}
u \in A:& \quad \psi_u^1(0) =  \psi_u^4(0), \quad \psi_u^2(0) = - \psi_u^3(0) \\ 
u \in B \cup C:& \quad  \psi_u^1(0) =  \psi_u^2(0), \quad    \psi_u^3(0) =  \psi_u^4(0)\,,
\end{malign}
and 
\begin{malign}  \label{BCfinal}
u \in A \cup B:& \quad \psi_u^1(T) =  -\psi_u^4(T), \quad \psi_u^2(T) =  \psi_u^3(T) \\ 
u \in  C:& \quad  \psi_u^1(T) =  -\psi_u^2(T), \quad    \psi_u^3(T) = - \psi_u^4(T)\,.
\end{malign}
The Schwinger-Keldysh contour described above has forward time evolution on legs $1,3$ and backward time evolution on legs $2,4$. By a change of variables $\psi_u^{2k} \rightarrow -i \psi_u^{2k}$ \cite{Stanford:2021bhl}, we can re-interpret the backward evolution on legs $2, 4$ as forward evolution. The resulting path integral is
\begin{equation}\label{eq:path_integral_modified}
 \Tr (\rho_{A \cup B\cup \bar A}^2) =  \frac{1}{(\sqrt{2})^{2NL}} \int  \prod_{u } \prod_{j=1}^{4} \mathcal{D} \psi_u^{j}  \,  \exp \left(i \int_0^T \d t \, \mathcal{L}(t)  \right)\,,
\end{equation}
where
\begin{malign} \label{eq:lagrangian_modified}
    \mathcal{L}(t) = i \sum_{k = 1}^{4} \frac{\psi_u^k(t) \partial_t \psi_u^k(t)}{2} - \sum_{k = 1}^2 \left(  H(t, \psi_u^{2k-1}(t)) - H(t, -i\psi_u^{2k}(t)) \right)\,.
\end{malign}

\subsection{Boundary Conditions}

The path integral defined in~\eqref{eq:path_integral_modified} computes a transition amplitude between an initial and a final state under a time-dependent Hamiltonian shown in equation~\eqref{eq:lagrangian_modified}. The initial and final states lie in the Hilbert space $(\mathcal{H}_R \otimes \mathcal{H}^*_R)^{\otimes 2}$ and they are fixed by the antiperiodic boundary conditions of the fermionic path integral.  

\subsubsection{Boundary conditions at $t = 0$}

Let us denote the state at $t = 0$ as $\ket{\Gamma_i}$. From~\eqref{BCinit} and the field redefinition for the even legs we read off that $\ket{\Gamma_i}$ satisfies
\begin{malign}  \label{eq:initial_state}
   u \in A:& \quad  \psi_u^1(0) \ket{\Gamma_i} = -i \psi_u^{4}(0) \ket{\Gamma_i},\quad  \psi_u^2(0) \ket{\Gamma_i} = -i \psi_u^{3}(0) \ket{\Gamma_i} \\
     u \in B \cup C:& \quad  \psi_u^1(0) \ket{\Gamma_i} = -i \psi_u^{2}(0) \ket{\Gamma_i},\quad  \psi_u^3(0) \ket{\Gamma_i} = -i \psi_u^{4}(0) \ket{\Gamma_i}\,.
\end{malign}

\subsubsection{Boundary conditions at $t = T$}
The boundary conditions at $t = T$ in~\eqref{BCfinal} similarly fixes the final state $\bra{\Gamma_f}$:
\begin{malign}  \label{eq:final_state}
    u \in A: \quad &\bra{\Gamma_f}  \psi_u^1(T)  =\bra{\Gamma_f}  i \psi_u^{4}(T),\quad  \bra{\Gamma_f} \psi_u^2(T)  = \bra{\Gamma_f}  i \psi_u^{3}(T)  \\
     u \in B \cup C: \quad  &\bra{\Gamma_f}  \psi_u^1(T)  = \bra{\Gamma_f} i \psi_u^{2}(T) ,\quad  \bra{\Gamma_f} \psi_u^3(T) = \bra{\Gamma_f} i \psi_u^{4}(T)\,.
\end{malign}
It is straightforward to check from the definition of $\ket{\Gamma_i}$ and $\bra{\Gamma_f}$ that all fermions satisfy the antiperiodic boundary conditions. 

\subsubsection{Boundary conditions for the dynamical variables}
In the next section, we will show that the effective dynamics of the Brownian SYK chain is described by the following bilinear operators of fermions:
\begin{equation}
    g_u^{i,j}(t) = \frac{1}{N} \sum_I \psi^i_{I,u}(t) \psi^j_{I,u}(t), \quad \text{for}  \quad 1 \leq i \neq j \leq 4\,,
\end{equation}
where the subscripts $I,u$ label $I$-th fermion flavour at site $u$. Such operators commute with the following flavour-wise parity symmetry~\cite{Stanford:2021bhl}:
\begin{malign}
        P_{I,u} = - 4\, \psi^1_{I,u} \psi^2_{I,u} \psi_{I,u}^3 \psi_{I,u}^4\,.
\end{malign}
Moreover, according to equations~\eqref{eq:initial_state} and~\eqref{eq:final_state}, the initial and final states satisfy
\begin{malign} 
    P_{I,u} \ket{\Gamma_i} = \ket{\Gamma_i},\, \quad \bra{\Gamma_f} P_{I,u} = \bra{\Gamma_f}\,.
\end{malign}
Therefore, the dynamics of the Brownian SYK chain is restricted to the $P_{I,u} = 1$ sector of the Hilbert space. The parity symmetry leads to a reduction of the number of independent  $g_u^{i,j}$. Let us illustrate how this comes about: we suppress indices that do not play a role and multiply $1=P$ with $g^{1,2}=\psi^1\psi^2$ to obtain:
\es{g12multi}{
g^{1,2}&=\psi^1\psi^2\,P=-4\psi^1\psi^2\psi^1\psi^2\psi^3\psi^4\\
&=4(\psi^1)^2(\psi^2)^2\psi^3\psi^4=\psi^3\psi^4=g^{3,4}\,,
}
where in the first line we plugged in the definitions and in the second line we anticommuted fermions and used the Majorana property $ (\psi)^2 = \frac12$.
Similarly, $g_u^{i,j}$ satisfy the following constraints:\footnote{For $n = 2$, flavour-wise parity symmetry automatically imposes replica symmetry. For $n > 2$, we must impose additional constraints on the saddle point. This is discussed in appendix~\ref{sec:replica_symmetry}.}
\begin{equation}\label{eq:replica_constraints}
    g^{1,2}_u = g_u^{3,4}, \quad g_u^{1,4} = g_u^{2,3}, \quad g^{1,3}_u = - g_u^{2,4}\,.
\end{equation}
Therefore, the effective dynamics is restricted to the following three variables
\begin{equation}  \label{eq:xyz}
 x_u  = -2i \, g_u^{1,2} , \quad   y_u =  2 g_u^{1,3}, \quad z_u=  -2i\, g_u^{1,4}\,.  
\end{equation}
From the definition of the initial and final states in~\eqref{eq:initial_state} and~\eqref{eq:final_state} we can find the boundary conditions for $x_u, y_u, z_u$. E.g.~let us consider $u\in A$ at $t=0$ and suppress indices: multiply the equation $0=(\psi^1+ i \psi^{4})\ket{\Gamma_i}$ by $2\psi^1$ to conclude
\es{zBC}{
0&=2\psi^1(\psi^1+ i \psi^{4})\ket{\Gamma_i}=\le(2(\psi^1)^2+ 2i \psi^1\psi^{4}\ri)\ket{\Gamma_i}\\
&=(1-z)\ket{\Gamma_i}\,,
}
where in going to the second line we used the Majorana property and the definition of $z$. We conclude that $z_u(0) =  1$.
The rest of the boundary conditions are found similarly and are shown below: 
\begin{enumerate}
    \item $t = 0$: \begin{malign}  \label{eq:boundary_conditions_xyz0}
      u \in A:\quad   z_u(0) &=  1, \quad x_u(0) = -y_u(0)\,, \\
       u \in B \cup C:\quad  x_u(0) &=  1, \quad z_u(0) = y_u(0)\,,
    \end{malign}
    \item $t = T$:  \begin{malign} \label{eq:boundary_conditions_xyzT}
        u \in A \cup B:\quad  z_u(T) &= 1,\quad x_u(T) = y_u(T)\,, \\ 
        u \in C:\quad x_u(T) &= 1,\quad z_u(T) = -y_u(T)\,.
    \end{malign}
\end{enumerate}
Note that the boundary conditions  satisfy 
\begin{equation}\label{eq:boundary_constraints}
    x_u^2 - y_u^2 +z_u^2 = 1, \quad \forall u
\end{equation}
everywhere. In the next section, we will show that the above relation is also preserved throughout the dynamics. Hence, the effective dynamics is described by two independent variables at every site.

The $x,y,z$ variables quantify the correlation between the different folds of the multifold Schwinger-Keldysh contour, e.g.~$x = 1$ corresponds to a maximally entangled state between contours 1 and 2; see~\cite{Stanford:2021bhl} for details and helpful cartoons. In the study of random quantum circuits~\cite{Zhou:2018myl,Zhou:2019pob} and Brownian models~\cite{vardhan_moudgalya2024} spin degrees of freedom valued in the permutation group emerge from averaging. We may think of $x,y,z$ as the fields obtained by coarse-graining these spins.

\section{R\'enyi entropy in the Brownian SYK chain}
\label{sec:BSYK}
The Brownian SYK chain consists of a one-dimensional lattice of nodes, where each node is comprised of $N$ Majorana fermions. These fermions interact with an onsite, all-to-all, $q$-body term with time-dependent random couplings, together with a nearest-neighbour, all-to-all, $\tilde{q}$-body hopping term, also with time-dependent, random couplings:  
\begin{malign}\label{eq:Hamiltonian_saddle}
   H(t) &= \sum_{u}\left(  i^{q/2} \sum_{I_1 < \dots < I_q} J^u_{I_1 \dots I_q}  (t)\psi_{I_1,u} \dots  \psi_{I_q,u} \right.  \\ & \qquad  \left.\qquad   +\,i^{\qt/2} \sum_{I_1 < \dots <  I_{\qt/2};J_1 < \dots< J_{\qt/2} }  J^{u,u+1}_{I_1  \dots  I_{\qt/2}; J_1  \dots  J_{\qt/2}} (t) \,\psi_{I_1,u}  \dots  \psi_{I_{\qt/2},u} \, \psi_{J_1,u+1}  \dots \psi_{J_{\qt/2},u+1}  \right)\,,
\end{malign}
where the label $u$ denotes the lattice site, while $I,J,\cdots$ are the fermion flavour indices; we have suppressed the contour indices in the above formula. The couplings in the Hamiltonian are chosen randomly and independently at each instance of time and for every lattice site from a Gaussian distribution with zero mean and the following variances:
\begin{malign}\label{eq:variance}
    \langle  J^{u}_{I_1 \dots I_q}(t) J^{u}_{I_1 \dots I_q}(t') \rangle &= \frac{J (q-1)!}{N^{q-1}}\delta(t-t') ,\\  \langle J^{u,u+1}_{I_1 \dots I_{\qt/2} J_1 \dots J_{\qt/2}}(t) J^{u,u+1}_{I_1 \dots I_{\qt/2} J_1 \dots J_{\qt/2}} (t') \rangle &= \frac{\jt \left(\frac{\qt}{2}!\right)^2}{\qt N^{\qt -1}} \delta(t-t')\,.
\end{malign}
In the following analysis, we will set $\qt =  2$. We expect similar conclusions to apply for other values of $\qt$.\footnote{In the Hamiltonian SYK context it is more common to choose $\qt=q$, so that the on-site and nearest-neighbour terms have the same scaling dimension and hence they are equally important in the infrared low-temperature regime. In the Brownian context we are effectively working at infinite temperature, hence such considerations do not apply.}

Since the couplings in the Brownian SYK theory are sampled from a random distribution, we can only hope to calculate the averaged moments of the density matrix in equation~\eqref{eq:path_integral_modified}.  Therefore, the quantity we must consider is $\mathbb{E}\left[\text{Tr}\,\rho^n_{\text{ref}\cup a\cup B}\right]$, %\begin{malign}

where $\mathbb{E}[\cdots]$ denotes ensemble average over the couplings. %In a genuine calculation of $\langle S_n\rangle $, we must average over $\log  \Tr (\rho^n)$, but in the following calculation, we will assume that $\langle \log \Tr (\rho^n) \rangle \approx \log \langle \Tr(\rho^n) \rangle$. 
In the standard SYK model, the averaging over couplings is done in a time-independent manner, and this results in an effective action which is nonlocal in time. The advantage in the Brownian SYK model is that the averaging happens independently at each instance of time, and as a result we get an effective action in the path integral that is local in time: 
\beq \label{eq:Renyi_path_integral}
\mathbb{E}\left[ \Tr \rho_{A\cup B \cup D}^n \right] =  \int \prod_{I,j,u,t}\mathcal{D}\psi^j_{I,u}(t) \; e^{-S},
\eeq
\beqn 
S&=& \int_0^T  \d t \sum_{u} \left(  \sum_{I=1}^N \sum_{j=1}^{2n} \frac{\psi^j_{I,u}(t) \partial_t \psi^j_{I,u}(t)}{2} 
+ H\right),
\eeqn
where we have now restored the contour index $j$, and
\begin{malign} \label{eq:pos_hamiltonian}
H &= \frac{J(q-1)!}{2 N^{q-1}}\left(i^{q/2} \sum_j s_j  \sum_{I_1 < \dots <  I_q} \psi_{I_1, u}^j \dots \psi_{I_q,u}^j\right)^2 + \frac{ \jt}{4 N} \sum_{I,J} \left(i \sum_j \psi_{I,u}^j \psi_{J,u+1}^{j}\right)^2    \\ &=
\frac{N J}{2 q}  \, \sum_{j,k=1}^{2n} s_j s_k \left(\sum_{I=1}^N \frac{ \psi^j_{I,u}(t) \psi^k _{I,u}(t)}{N} \right)^q  \\&\quad+\frac{N \jt}{4}  \sum_{j,k=1}^{2n}  \left( \sum_{I=1}^N \frac{\psi_{I,u}^j(t) \psi_{I,u}^k(t)  }{N} \right) \left( \sum_{J=1}^N  \frac{\psi_{J,u+1}^j (t) \psi_{J,u+1}^k(t)}{N} \right) + O(N^0)\,,
\end{malign}
and introduced the notation
\begin{equation}
    s_j = \begin{cases}
        1 & \text{if} \quad j \in \{1,3\}\,, \\
        -i^q & \text{if} \quad j \in \{2,4\}\,.    \end{cases}
\end{equation}
Since $H$ is the sum of squares of local Hermitian operators, it is positive. Therefore, the averaging over the couplings has turned the Lorentizan path integral into a Euclidean evolution by a positive Hamiltonian. This is a shared feature with  the study of random quantum circuits~\cite{Zhou:2018myl,Zhou:2019pob} and Brownian models~\cite{vardhan_moudgalya2024}.

Let us first focus on the Hamiltonian from~\eqref{eq:pos_hamiltonian}. Since the action is local in time and fermions enter the action in a  flavour-summed form, we can easily calculate the terms in the action where fermions on the same contour appear together using the operator identity $(\psi^j_{I,u})^2 = \frac{1}{2} $.  Such ``contour-diagonal'' terms give the following contribution to the action:
\es{eq:sdiag}{
  S_{\text{diag}} &=   \sum_j \int  \d t \left[ \frac{N J}{2 q} \left( 
\sum_{i=1}^N\frac{ \psi_{I,u}^j \psi_{I,u}^j}{N} \right)^q   + \frac{N \jt}{4}\left(\sum_{I=1}^N
 \frac{ \psi_{I,u}^j \psi_{I,u}^j}{N}  \right) \left( \sum_{J=1}^N\frac{ \psi_{J,u+1}^j \psi_{J,u+1}^j}{N}  \right)\right] \\&= \sum_{u} 2n \left(\frac{N JT}{2^{q+1} q }+ \frac{N \jt T}{16} \right) \,.}
The contribution to the action coming from fermion pairs from different contours is nontrivial. Together with the kinetic term, it is given by:
\begin{malign}\label{eq:srest}
    S_{\text{rest}} =  \int   &\d t \sum_{j} \sum_{I=1}^N \frac{\psi_{I,u}^j \partial_t \psi_{I,u}^j}{2} + \frac{N J}{2 q} \int \d t  \, \sum_{j \neq k} s_j s_k \left( \sum_{I=1}^N\frac{ \psi_{I,u}^j \psi_{I,u}^k}{N}  \right)^q\\& +\frac{N \jt}{4} \int \d t  \,\sum_{j\neq k}  \left( \sum_{I=1}^N\frac{ \psi_{I,u}^j \psi_{I,u}^k}{N} \right)\left(\sum_{J=1}^N\frac{ \psi_{J,u+1}^j \psi_{J,u+1}^k}{N}  \right)\,.
\end{malign}
The next step is to integrate out fermions from the path integral in~\eqref{eq:Renyi_path_integral}. This is done by introducing the $(g,\sigma)$ collective field variables as follows:
\begin{malign} 
    \int \prod_{I,j,u,t}\mathcal{D}\psi^j_{I,u}(t) \; e^{-S} = \int \prod_{I,j,u,t}\mathcal{D}\psi^j_{I,u}(t) \int \mathcal \prod _{i<j} \mathcal{D} g_u^{i,j} \mathcal{D} \sigma_u^{i,j}\; e^{-S - S_{g,\sigma}},  
\end{malign}
where 
\begin{equation}
    S_{g,\sigma} =  \sum_u\frac{N}{2}\sum_{i \neq j} \int \d t \, \sigma^{i,j}_u(t) \left(g^{i,j}_u(t) - \sum_{I = 1}^N \frac{\psi_{I,u}^i(t) \psi_{I,u}^j(t)}{N}\right)\,.
\end{equation}
$\sigma_u(t)$ plays the role of a Lagrange multiplier that imposes the constraint \footnote{Note that our definition of $g_u$ only contains off-diagonal elements. This is a convenient choice to make $g_u$ anti-symmetric.}
\begin{equation}
    g^{i,j}_u(t) = \sum_{I = 1}^N \frac{\psi_{I,u}^i(t) \psi_{I,u}^j(t)}{N}, \, \quad i \neq j. 
\end{equation}
This allows us to replace all the bilinears in   $S_{\text{rest}}$ by the corresponding $g$ variables. This turns the fermion path integral into a Gaussian, which we can perform exactly. The path integral over the $I$-th fermion at site $u$ gives
\begin{equation}
    \int\prod_j \mathcal{D}\psi_{u,I}^j \, e^{-\frac{1}{2} \int \d t \, \left(\sum_i \psi^i_{I,u} \partial_t \psi_{I,u}^i - \sum_{i,j} \sigma^{i,j}_u \psi^i_{I,u}\psi^j_{I,u} \right)} = e^{ \log \text{Pf} (\partial_t - \sigma_u) }\,.
\end{equation}
Integrating out all fermions, we get the following effective action in terms of $(g,\sigma)$ variables: 
\es{eq:g_sigma_action}{
   -S_{\text{eff}} =  &\sum_u N \log \text{Pf}(\partial_t - \sigma_u)  - \sum_{u}\frac{N}{2} \int \d t \, \sigma^{i,j}_{u} \, g^{i,j}_{u}    \\&- \frac{N J}{2 q}\sum_{u} \int \d t  \, \left( \sum_{j \neq k} s_i s_j  \,   \left(  g^{i,j}_{u}  \right)^q + \frac{2n }{2^{q}} \right) 
-\sum_{u}\frac{N \jt}{4} \int \d t  \left( \,\sum_{i\neq j}   \, g^{i,j}_{u } \;  g^{i,j}_{u+1} + \frac{2n }{4}\right)\,.}
In the large $N$ limit, the path integral can be evaluated in the saddle point approximation. The classical equations of motion for the $(g, \sigma)$ can be determined by varying the action with respect to $g,\sigma$. We get the following equations:\footnote{The first equation below is not what we get, if we simply vary the action. It requires nontrivial manipulations to bring the equation to this form, which are explained in~\cite{Stanford:2021bhl}.}
\begin{malign}\label{eq:g_sigma} 
    \partial_t g^{i,j}_{ u} &= \sum_k \left( \sigma^{i,k}_{u} g^{k,j}_{u} - g^{i,k}_{u} \sigma^{k,j}_{u} \right) = [\sigma_{u},g_{u}]^{i,j}, \\
    \sigma^{i,j}_{u} &= - J s_i s_j \,  (g^{i,j}_{u})^{q-1} -  \frac{\jt}{2}\, \left( g^{i,j}_{u-1} + g^{i,j}_{u+1} \right)\,.
\end{malign}

The first equation implies that $\Tr(g^{2m})$, with $m$ an integer, is conserved. Equivalently, the eigenvalues of $g_u$ do not evolve in time. In section~\ref{sec:action}, we will use this result to evaluate the saddle point action.

\subsection{Saddle point equations}

In equation~\eqref{eq:replica_constraints}, we reduced the number of independent $g_u^{i,j}$ due to the flavour-wise parity symmetry. This reduction applies even off-shell, and for consistency the saddle point equations should  preserve these equivalence relations under time evolution.
It is straightforward to check that the $(g,\sigma)$ equations in~\eqref{eq:g_sigma} indeed do so.  Therefore, we can consistently describe the dynamics using the $x_u,y_u,z_u$ variables defined in equation~\eqref{eq:xyz}. These variables satisfy the following equations of motion:
\begin{malign}\label{eq:diff_xyz} 
    \partial_t x_{u} &= \frac{J}{2^{q-2}} \left(z_{u}  y^{q-1}_u  -  y_{u}  z^{q-1}_u\right) + \frac{\jt}{2} \left(z_{u } \partial_u^2 y_u  - y_{u}  \partial_u^2 z_u \right), \\  \partial_t y_{u} &= \frac{J}{2^{q-2}}  \left(  z_{u}x^{q-1}_u  -  x_{u}z^{q-1}_u  \right) + \frac{\jt}{2} \left( z_{u} \partial_u^2 x_u - x_{u } \partial_u^2 z_u  \right), \\
    \partial_t z_{u} &= \frac{J}{2^{q-2}} \left(y_{ u} x^{q-1}_u  -x_{u} y^{q-1}_u \right) + \frac{\jt}{2} \left( y_{u} \partial_u^2 x_u - x_{u} \partial_u^2 y_u \right)\,,
\end{malign}
where we have defined the Laplacian on the chain as:
\begin{equation}
    \partial^2_u f_u \equiv \left( f_{u+1} + f_{u-1} - 2\,  f_u \right)\,.
\end{equation}

As discussed below equation~\eqref{eq:g_sigma}, $\Tr(g^2_u)$ is conserved. For $n = 2$, $\Tr(g^2_u)  \propto x_u^2 + z_u^2 - y_u^2$. Therefore, $x_u^2 + z_u^2 - y_u^2$ is conserved. As discussed around equation~\eqref{eq:boundary_constraints}, the boundary conditions satisfy $x_u^2 + z_u^2 -y_u^2 = 1$ at $t = 0$ and $t = T$. Therefore, \begin{equation}\label{eq:xyz_constraint}
    x_u^2 + z_u^2 -y_u^2 = 1 \quad \forall u, t\,.
\end{equation}
Hence the effective dynamics on the Brownian SYK chain is described by two independent variables at every site. It is convenient to  use the following parametrization of the variables:
\begin{equation}
    x = \sqrt{r^2+1}  \sin \theta,\qquad z=\sqrt{r^2+1} \cos \theta, \qquad y = r\,.
\end{equation}
\subsection{Continuum limit}\label{sec:continuum}
We will be interested in the continuum limit of the Brownian SYK chain. To see what scaling of parameters corresponds to the continuum limit in this model, let us consider the ``Hamiltonian" part of the action in equation~\eqref{eq:g_sigma_action} for $n = 2$.
\begin{malign} 
\frac{H(t)}{N} &= \sum_u \frac{J}{2^{q-1}q} \left( 1 - x_u^q - z_u^q + y_u^q\right) + \sum_u \frac{\jt}{4} \left(1 -  x_u x_{u+1} - z_{u}z_{u+1} + y_u y_{u+1}\right) \\
&= \sum_u  \frac{J}{2^{q-1}q} \left( 1 - x_u^q - z_u^q + y_u^q\right) + \frac{\jt}{8} \sum_u \left( (x_u - x_{u+1})^2 + (z_u - z_{u+1})^2 - (y_u - y_{u+1})^2 \right)\,.
\end{malign}
In the second step, we used equation~\eqref{eq:xyz_constraint}.
The continuum limit corresponds to replacing $u$ by $\sigma = a u$ (where $a$ is the lattice spacing) and taking the $a \rightarrow 0$ limit. Under this replacement, the Hamiltonian is
\begin{malign} \label{eq:continuum_hamiltonian}
    \frac{H(t)}{N} &= a\sum_\sigma  \frac{ J_c}{2^{q-1}q} (1 - x_\sigma^q -z_\sigma^q + y_\sigma^q)+  a\sum_\sigma \frac{\jt_c}{8}\left( (\partial_{\sigma} x)^2 + (\partial_{\sigma} z)^2-(\partial_{\sigma} y)^2\right) \\ &\rightarrow \int \d \sigma\, \frac{J_c}{2^{q-1}q} \left(1 - x^q(\sigma) - z^q(\sigma) + y^q(\sigma)\right) +  \int \d \sigma\, \frac{\jt_c}{8} \left((\partial_{\sigma} x)^2 + (\partial_{\sigma} z)^2-(\partial_{\sigma} y)^2\right) \,,
\end{malign}
where, $J_c, \jt_c$ are related to  $J, \jt$ as:
\begin{equation}\label{eq:rescaling}
    J = a J_c, \quad \jt = \frac{\jt_c}{a}\,.
\end{equation} 

To make sense of the continuum limit, we must also rescale $t$. This can be checked by taking the continuum limit of equations~\eqref{eq:diff_xyz}. Consider, for example, the first equation of~\eqref{eq:diff_xyz}:
\begin{malign}
    \partial_t x_u &= a  \frac{J_c}{2^{q-2}} (z_u y_u^{q-1}- y_u z_u^{q-1}) + \frac{\jt_c}{2 a} (z_u \partial_u^2 y_u - y_u \partial_u^2 z_u) \\ \implies \partial_t x(\sigma) &= a \left[  \frac{J_c}{2^{q-2}} (z(\sigma) \,y^{q-1}(\sigma) - y (\sigma)z^{q-1}(\sigma)) +  \frac{\jt_c}{2 } \left(z (\sigma)\partial_\sigma^2 y(\sigma) - y(\sigma) \partial_\sigma^2 z(\sigma) \right) \right]\,.
\end{malign}

The above equation implies that the dynamical variables vary slowly in $t$. We can absorb the overall factor of $a$ by redefining the time variable as 
\begin{equation}
    \tau =  a t \,.
\end{equation}
In $t$ coordinate, the chain evolves in time by time $T$. The duration of evolution in $\tau$ coordinate is
\begin{equation}\label{eq:cont_t}
     T_\tau \equiv \Delta \tau = a T\,. 
\end{equation}
Another important parameter governing the time evolution on the chain is the size of $B$. If $|B|$ is the number of sites contained in $B$, then the size of $B$ in units of $\sigma$ is \begin{equation}\label{eq:cont_u}
   L_\sigma \equiv \Delta \sigma = a |B| \,.
\end{equation} 

Now, we can re-write the saddle point equations in $\tau, \sigma$ coordinates as follows:
\begin{malign}\label{eq:diff_xyz_cont}
    \partial_\tau x = \frac{J_c}{2^{q-2}} (z\, y^{q-1} - y \,z^{q-1}) +  \frac{\jt_c}{2 } \left(z\,\partial_\sigma^2 y - y \,\partial_\sigma^2 z\right) \\ 
     \partial_\tau y = \frac{J_c}{2^{q-2}} (y \,x^{q-1} - x\, y^{q-1}) +  \frac{\jt_c}{2 } \left(y\,\partial_\sigma^2 x - x \,\partial_\sigma^2 y\right)\\
      \partial_\tau z = \frac{J_c}{2^{q-2}} (z\, \,x^{q-1} - x \,z^{q-1}) +  \frac{\jt_c}{2 } \left(z\,\partial_\sigma^2 x - x \,\partial_\sigma^2 z\right)
\end{malign}
In the continuum limit, we can always rescale $\sigma$ and $\tau$ to rescale $J_c$ and $\jt_c$. We will leverage this property to set $\jt_c = 2$ throughout this paper. We conclude the discussion of the continuum limit by rewriting the relation between $\sigma$ and $u$ in terms of the original couplings of the Brownian SYK chain:
\begin{equation}\label{sigmadef}
    \sigma = a u\,, \quad a = \sqrt{\frac{\jt_c J }{  J_c  \jt} } = \sqrt{\frac{2 J }{ J_c \jt } } \,.
\end{equation}

In the $r,\theta$ variables, equations~\eqref{eq:diff_xyz_cont} can be written as 
\begin{malign}  \label{eq:rtheta_eqations}
   \partial_\tau r &= \frac{J_c}{2^{q-1}} (r^2+1)^{q/2} \sin 2\theta \left( \sin^{q-2} \theta - \cos^{q-2} \theta  \right) + \partial_\sigma ((r^2+1) \partial_\sigma \theta ) \\\; 
   \partial_\tau \theta &= \frac{J_c}{ 2^{q-2}} \left( r^{q-1} -  r (r^2+1)^{q/2-1}  \left( \cos^q \theta + \sin^q \theta \right) \right) + \frac{\partial_\sigma^2 r}{r^2+1} - \frac{r (\partial_\sigma r)^2}{(r^2+1)^2} + r (\partial_\sigma \theta)^2\,,
\end{malign}
Notice that the above equations can be derived from the following action: 
\es{eq:hamiltonian}{
    -S =\frac{1}{2 a}&\int_0^{T_\tau} \d \tau \, \d \sigma \left( r \,\partial_\tau \theta  - h(\tau,\sigma)\right) \\ 
    h(\tau,\sigma) =&  \frac{J_c}{q 2^{q-2}} \left( 1 + y^q - (x^q + z^q)\right) +\frac{1}{2} \left((\partial_\sigma x)^2 + (\partial_\sigma z)^2 -(\partial_\sigma y)^2 \right)\\ =&\frac{J_c}{q 2^{q-2}} \left(1 + r^q -(r^2+1)^{q/2}  (\sin ^q \theta + \cos^q \theta) \right) + \frac{1}{2} \left((r^2+1)(\partial_\sigma \theta)^2 - \frac{(\partial_\sigma r)^2}{r^2+1}\right)\,,}
In section~\ref{sec:action}, we will show that this is indeed the action of the Brownian SYK chain in the continuum limit. $h(\tau,\sigma)$ is the Hamiltonian density of the chain in the continuum limit. Note that $h(\tau,\sigma)$ can be negative for large enough $\partial_\sigma r$. In particular, there exist classical configurations $(r(\sigma), \theta(\sigma))$ for which the Hamiltonian is negative.  However, this does not contradict the fact that the Hamiltonian defined in equation~\eqref{eq:pos_hamiltonian} is a positive operator. This simply means that classical configurations which can emerge under Euclidean evolution in the large $N$ limit must have a positive Hamiltonian.

Importantly, all derivations in this section  go through without change for higher dimensional lattices and the corresponding lattice Laplacian would appear in~\eqref{eq:diff_xyz}. In the continuum limit, we obtain a rotationally symmetric kinetic term in~\eqref{eq:hamiltonian} and the continuum Laplacian in the equation of motion. Thus the resulting membrane effective theory will be isotropic.\footnote{This is far from guaranteed in general: the lattice anisotropy can imprint itself on growing operators~\cite{Nahum2018OperatorSpreading} and the membrane~\cite{vardhan_moudgalya2024}.}

\section{Solving the saddle point equations}\label{sec:solitons}

In this section, we solve the saddle point equations for the boundary conditions in~\eqref{eq:boundary_conditions_xyz0},~\eqref{eq:boundary_conditions_xyzT}. An important parameter for this discussion is the membrane velocity defined in equation~\eqref{eq:membrane_velocity} as
\begin{equation}
    v = \frac{|B|}{T} =  \frac{L_\sigma}{T_\tau},
\end{equation}
where $|B|$ is the number of sites contained in region $B$ and $L_\sigma, T_\tau$ are defined in equations~\eqref{eq:cont_t},~\eqref{eq:cont_u} respectively. 

According to equations~\eqref{eq:diff_xyz}, the dynamical variables  vary at the time scale $t \sim 1/J$. Therefore, $1/J$ sets the time scale of thermalization. The membrane picture emerges at times much longer than the time scale of thermalisation. Therefore, we must set $T \gg {1/J}$. Similarly, the scale of spatial variation is set by $\sqrt{\jt/J}$ and  we must demand that $|B| \gg \sqrt{{\jt/J}}$. In the continuum limit, we can use~\eqref{sigmadef} to determine the conditions on $T_\tau$ and $L_\sigma$:
\begin{malign}
    T_\tau \gg \frac{1}{J_c} \,, \quad L_\sigma \gg \sqrt{\frac{1}{J_c}}\,.
\end{malign}

We split our analysis of the saddle point equations into two regimes:
\begin{enumerate}
    \item $v > v_B$ 
    \item $v < v_B$
\end{enumerate}
where $v_B$ is the butterfly velocity, which we determine shortly for our model. Let us first consider the case $v>v_B$.

\subsection{\texorpdfstring{$v > v_B$}{u to pminfty}} \label{sec:vgvb}
For $v > v_B$, it is simpler to describe the solution in terms of the $(x_u,y_u,z_u)$ variables. Consider the solution at $T = 0$ which can be found by simultaneously solving the boundary conditions at $t = 0$ and $t = T$:
\begin{equation}
    (x_u,y_u,z_u) =  \begin{cases}
        (0,0,1) \quad  u \in A \\
        (1, 1,1) \quad  u \in B \\
        (1,0,0) \quad  u \in C\,.
\end{cases}
\end{equation}
Notice that the $T = 0$ solution in the three subregions are at the fixed points of the saddle point equation. The solution has domain walls at $A|B$ and $B|C$, where $X|Y$  denotes the boundary between $X$ and $Y$. When $T$ is nonzero but small, we expect nontrivial dynamics only near the domain walls $A|B$ and $B|C$. If $ |B|$ is large, we can ignore $C$ (respectively $A$) to study the dynamics near $A|B$ (respectively $B|C$). This allows us to find an approximate solution to the saddle point equations. 

\subsubsection{Traveling wave near $B|C$}
\begin{figure}
    \centering
    \includegraphics[width=0.8\linewidth]{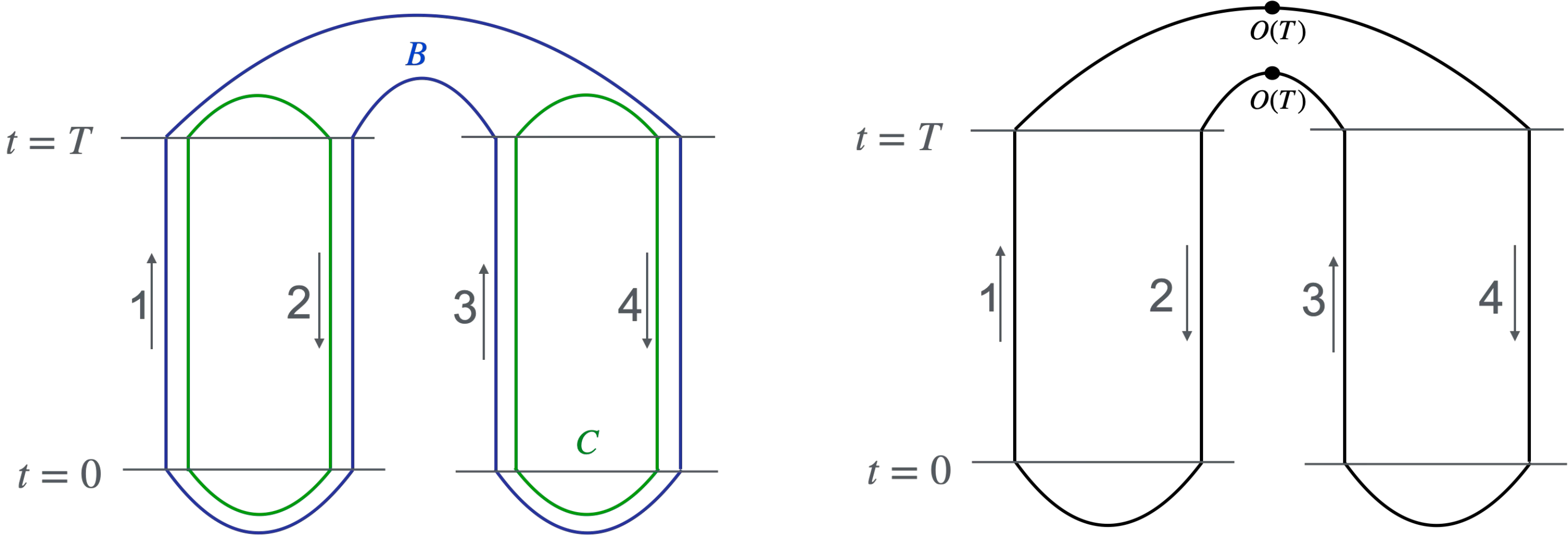}
    \caption{\textbf{Left:} Contour for the second R\'enyi entropy ignoring $A$. \textbf{Right:} The OTOC contour. The second R\'enyi contour resembles the OTOC contour when $A$ is ignored.}
    \label{fig:contour_wo_A}
\end{figure}
To solve the equations near $B|C$ approximately, let us ignore $A$. The contour for the second R\'enyi entropy (ignoring  $A$) is shown in figure~\ref{fig:contour_wo_A}. Notice that the contour at $t= 0$ corresponds to insertion of a maximally entangled state at $t  = 0$. Since the maximally entangled state does not evolve in time, the boundary conditions at $t = 0$ are satisfied for all $t$.   Therefore, according to the boundary conditions for $B\cup C$ in equations~\eqref{eq:boundary_conditions_xyz0}, we can set $x_u =1$ and $ y_u=z_u$ everywhere. The resulting equation is
\begin{equation}\label{eq:past_fkpp}
        -\partial_t y_u = -\frac{J}{2^{q-2}}\left( y_u- y^{q-1}_u \right) + \frac{\jt}{2} \, \partial_u^2 y_u\,.
\end{equation}
Since the boundary conditions at $t= 0$ are satisfied everywhere, they are also satisfied at $t = T$. This allows us to fix the boundary conditions for $y_u$ explicitly:
\begin{align}\label{eq:fkpp_bdy}
    y_u(T) = \begin{cases}
        1 \quad u \in B\,, \\
        0 \quad u \in C\,.
    \end{cases}
\end{align}

If we flipped the sign of the $\partial_t y_u$ term in equation~\eqref{eq:past_fkpp} we would get the FKPP equation \cite{Fisher, kolmogorov1937, BramsonLectures,TongLectures}. This flipped sign implies that we should be thinking about backward time evolution, which is appropriate to the setting where the initial condition was specified at the future boundary in~\eqref{eq:fkpp_bdy}.

It is well known that when the initial condition of the FKPP equation is a step function, the solution approaches a traveling wave with velocity\footnote{This formula for $v_B$ in terms of the coupling is valid only in the continuum limit. On the lattice the dependence on the couplings is more nontrivial.}
\begin{equation}\label{eq:butterfly} v_B^{(u)} = \sqrt{\frac{2(q-2)J \jt}{2^{q-2}}}\,.
\end{equation} 
We sketch the derivation of the butterfly velocity in Appendix~\ref{sec:fkpp}. The superscript in $v_B^{(u)}$ indicates that this velocity is in the $u$ spatial coordinate, which needs to be converted to $v_B^{(\sigma)}$ when working in the continuum $\sigma$ coordinates defined in~\eqref{sigmadef}.

For us the resulting traveling wave solution propagates towards region $B$ under backward time evolution and it has a wavefront of width $O\left(\sqrt{J/\jt}\right)$. Away from the wavefront, the solution approaches $y = 0$ to the right and  $y = 1$ to the left. As we explained above, the boundary conditions at the $t=0$ boundary allow for an arbitrary $y_u(t=0)$ profile, hence the traveling wave can end its journey there without any further consistency requirement: indeed the FKPP equation is a parabolic differential equation, so it should only come with an intial condition and not a final one. 

$v_B$ in equation~\eqref{eq:butterfly} is indeed the butterfly velocity. The reason the wavefronts propagate with the butterfly velocity goes as follows: Notice that the contour in figure~\ref{fig:contour_wo_A} resembles an OTOC contour. On the OTOC contour shown in figure~\ref{fig:contour_wo_A},  $y_u(t)$ is equal to the following OTOC:
\begin{malign}
    y_u(t) &= \frac{1}{N} \sum_{I} \langle \psi_{I,u}(t) \mathcal{O}(T) \psi_{I,u}(t) \mathcal{O}(T)\rangle \\&= \frac{1}{N} \sum_{I} \langle \psi_{I,u}(0) \mathcal{O}(T-t) \psi_{I,u}(0) \mathcal{O}(T-t)\rangle\, .
\end{malign}
Expanding $\mathcal{O}(T-t)$ in the basis of products of Majorana fermions and using the anticommutation relation, we can express the OTOC in terms of the average number of Majorana fermions in $\mathcal{O}(T-t)$ at site $u$ (denoted as $n_u(T-t)$ below):
\begin{equation}
    y_u(t) = 1 - 2\frac{ n_u(T-t)}{N}\,.
\end{equation}
In the large $N$ limit, we can replace the $\frac{n_u(t)}{N}$ by a continuous number $\phi_u(t)$. Then, the equation of $y_u(t)$ in~\eqref{eq:past_fkpp} is related to the operator growth equation by the above change of variables \cite{Xu_2019_PRX}: 
\begin{equation}
    \partial_t \phi_u = \frac{J}{2^{q-1}} \left( 1 - 2\phi_u(t)\right)(1 - (1- 2 \phi_u(t))^{q-2}) + \frac{\jt}{2} \partial_u^2 \phi_u\,.
\end{equation}
The boundary condition in~\eqref{eq:fkpp_bdy} implies the following boundary condition for operator size $\phi_u(t)$:
\begin{malign} 
       \phi_u(T) = \begin{cases}
        0 \quad &u \in B \\
        1/2 \quad &u \in C\,. \end{cases}
\end{malign}
The above boundary condition corresponds to an operator that contains  $N/2$ fermions at every site $C$ but no fermions in $B$.  Under time evolution, such an operator spreads towards region $B$ with the butterfly velocity $v_B$.

The coincidence of $v_B$ defined through the OTOC of simple local operators with the maximal slope of the entanglement membrane has an intriguing explanation in holographic theories that is distinct from the argument we presented above. The argument goes as follows: Let us denote the OTOC velocity by $v_B$ and the velocity corresponding to the maximal slope of the entanglement membrane by $v_m$. Consider the entanglement membrane of a spherically symmetric subregion $S$ in the boundary CFT, for which the entanglement growth has saturated.  In this case, the entanglement membrane of $S$ is a cone of slope $v_m$. In holography, this membrane is determined by projecting the Ryu-Takayanagi surface corresponding to $S$ to the boundary along the constant infalling time \cite{Jiang:2024tdj}. As a consequence of this fact, $v_m$ also dictates the rate of operator growth on the boundary as measured using entanglement wedge reconstruction \cite{Mezei_Stanford_2017}. The authors of \cite{Chua:2025vig} showed in gravity that the shockwave solution used to compute the OTOC is the imaginary part of the metric of the replica manifold computing the R\'enyi entropy linearised in $(n-1)$. Hence, $v_B = v_m$. See~\cite{Mezei_Stanford_2017,Dong:2022ucb} for earlier work explaining various aspects of the equality between the two velocities.

\begin{figure}[t]
    \centering
    \includegraphics[width =0.8\textwidth]{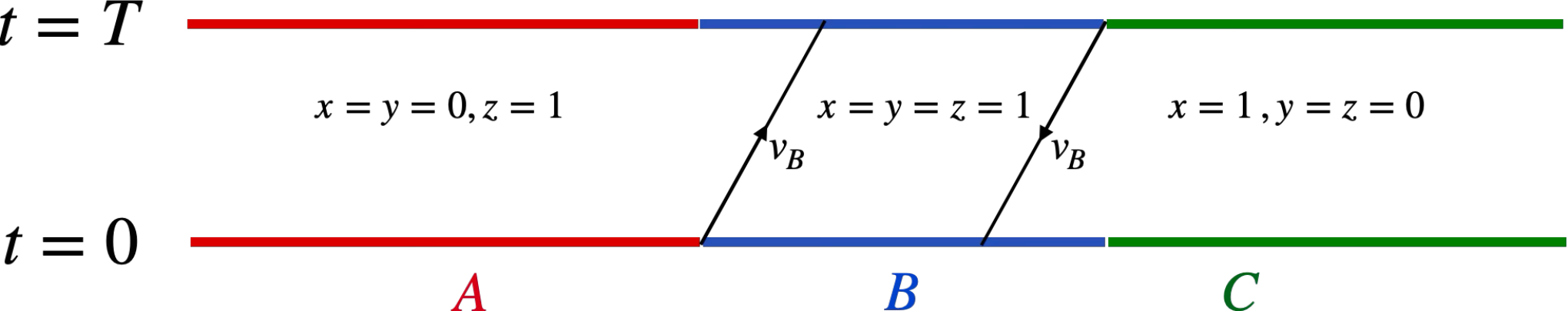}
    \caption{{Saddle point solution for $v >v_B$. The solution consists of two fronts: (i) originating from the domain wall $A|B$ at $t = 0$}, (ii) originating from $B|C$ at $t = T$. Both fronts propagate with the butterfly velocity towards region $B$, one forward and one backward in time as indicated by the arrows. }
    \label{fig:sol_vgvb}
\end{figure}

\subsubsection{Traveling wave near $A|B$}
In this case, the $x,y,z$ variables in $A$ and $B$ satisfy the same boundary conditions at $t = T$. If we ignore $C$, then it is easy to check that the boundary conditions at $t= T$ are preserved for all $t \in [0,T]$ under backward time evolution. Therefore, to solve the differential equations near $A|B$, we can set $x_u = y_u$ and $z_u =1$. The equations reduce to the following partial differential equation:
\begin{equation}\label{eq:future_fkpp}
        \partial_t y_u = -\frac{J}{2^{q-2}} \left( y_u- y^{q-1}_u \right) +  \frac{\jt}{2} \partial_u^2 y_u\,,
    \end{equation}
This is the again the FKPP equation but with the conventional sign for $\partial_t y_u$, implying that the equation naturally describes forward time evolution. This is the opposite behaviour than what we saw for the traveling wave near $B\vert C$, see the discussion around~\eqref{eq:past_fkpp}.
Since the boundary conditions at $t = T$ are always preserved, they are also preserved at $t = 0$. This fixes the boundary condition of $y_u$ at $t = 0$.
\begin{align}\label{eq:bc_operatorgrowth}
    y_u(0) = \begin{cases}
        0 \quad u \in A \\
        1 \quad u \in B\,.
    \end{cases}
\end{align}
The solution is a traveling  wave that propagates towards region $B$ under forward time evolution. 

The two traveling wave solutions obtained above approach the fixed point $(x,y,z) = (1,1,1)$ in the interior of $B$. Therefore, we can obtain the global solution by patching the two traveling wave solutions in $B$. This is shown schematically in figure~\ref{fig:sol_vgvb}. The solution can be implicitly represented in $r,\theta$ variables as follows:
\begin{malign}  \label{eq:rtheta_vb}
   r(\theta) =  r_c(\theta) \equiv \begin{cases}
        \tan \theta \quad 0\leq \theta \leq \pi/4 \\
        \cot \theta \quad \pi/4 \leq \theta \leq \pi/2 \,.
    \end{cases}
\end{malign}

Although we began this section with the assumption that $T$ is small, it is clear that our construction is valid as long as $v > v_B$. This is because the two traveling waves are separated by a distance $(v - v_B)T \gg \sqrt{J/\jt}$. Since the wavefronts near $A|B$ and $B|C$ have a width of $O(\sqrt{J/\jt})$, we can consistently patch the two traveling wave solutions in the bulk of $B$. We illustrate a rough sketch of the solution in figure~\ref{fig:sol_vgvb}. 

Although the saddle point solution for $v > v_B$ is a traveling wave, it contains a macroscopic region where the solution is away from the domains $x = 1, y = z = 0$ and $x = y =0, z = 1$. Since the entanglement membrane is expected to have an $O(1)$ thickness, we conclude that the membrane ceases to exist
 for $v > v_B$.

In appendix~\ref{app:vgvb}, we show that we can extend the above solution to arbitrary replica index. 

\
\subsubsection{$1/N$ corrections}
\label{sec:1_on_correction}
In the previous section, we constructed a traveling wave solution that approaches $(x,y,z) = (0,0,1)$ when $u \rightarrow -\infty$, remains close to $(x,y,z) = (1,1,1)$ in a macroscopic region of size $(v - v_B)T$ in $B$, and approaches $(x,y,z) = (1,0,0)$ as $u \rightarrow \infty$.  Although $(x,y,z) = (1,1,1)$ is a fixed point of the saddle point equations, it only emerges in $N \rightarrow \infty$ limit.  At large but finite $N$, it is important to consider the effects of quantum fluctuations around the fixed point. We can estimate fluctuations around the fixed point by computing two point functions of the $x,y,z$ variables. Consider, for example, the connected correlation function between $x_u(T)$ and $y_u(0)$ at some site $u \in B$ where the classical solution is close to the fixed point $(1,1,1)$: 
\begin{malign}   \langle x_u(T) y_u(0)\rangle_c &=  \langle x_u(T) y_u(0)\rangle -\langle x_u(T)\rangle \langle y_u(0) \rangle \\&= \langle x_u(T) y_u(0)\rangle -1\,. \end{malign}
The connected correlator can be written in terms of an OTOC as:
\begin{equation}
     \langle x_u(T) y_u(0)\rangle_c   = -\frac{1}{N^2} \sum_{I,J} \langle \psi_{I,u}(T) \psi_{J,u} (0) \psi_{I,u}(T) \psi_{J,u}(0) \rangle  -1\,.
\end{equation}
For $T = O(1)$, the OTOC is close to 1 and the connected correlator is small.  However, when $ T = O(\log N)$, the  scramblon modes \cite{Gu_2022, Stanford:2021bhl,Choi_2023,StanfordVardhanYao2024ScramblonLoops} contribute to the decay of the OTOC and the connected correlator becomes $O(1)$. Therefore, to describe the dynamics of the $x,y,z$ variables for $T \sim \log N$ we cannot ignore quantum fluctuations. We conclude that the classical solution discussed in this section is valid only when $T = O(N^0)$. 

On the other hand, $1/N$ corrections are always small for the fixed points in regions $A$ and $C$. This is because connected correlators of $x,y,z$ in $A \cup C$ are related to time-ordered correlators between fermions. Hence, they are always small.

Ref.~\cite{Xu_2019_PRX,StanfordVardhanYao2024ScramblonLoops} considered the effects of $1/N$ corrections in the operator growth problem. The authors showed that such corrections lead to a diffusive broadening of the operator wavefront when $T \sim \log^{3} N$. Since the saddle point equations near the domain walls map to the operator growth problem, we believe that $1/N$ corrections would lead to a similar broadening of the wave-fronts at late times.  It would be interesting to analyze the details of dynamics of the $(x,y,z)$ variables in this regime. We leave this for future work.

\subsubsection{Comparison with random quantum circuits and holography}

Since the splitting of the membrane into two wavefront is a phenomenon previously not seen before it may be useful to contrast it to membrane configurations computing operator entanglement for $v>v_B$ in other solvable models. 

\begin{figure}[h]
    \centering
\includegraphics[width=0.7\linewidth]{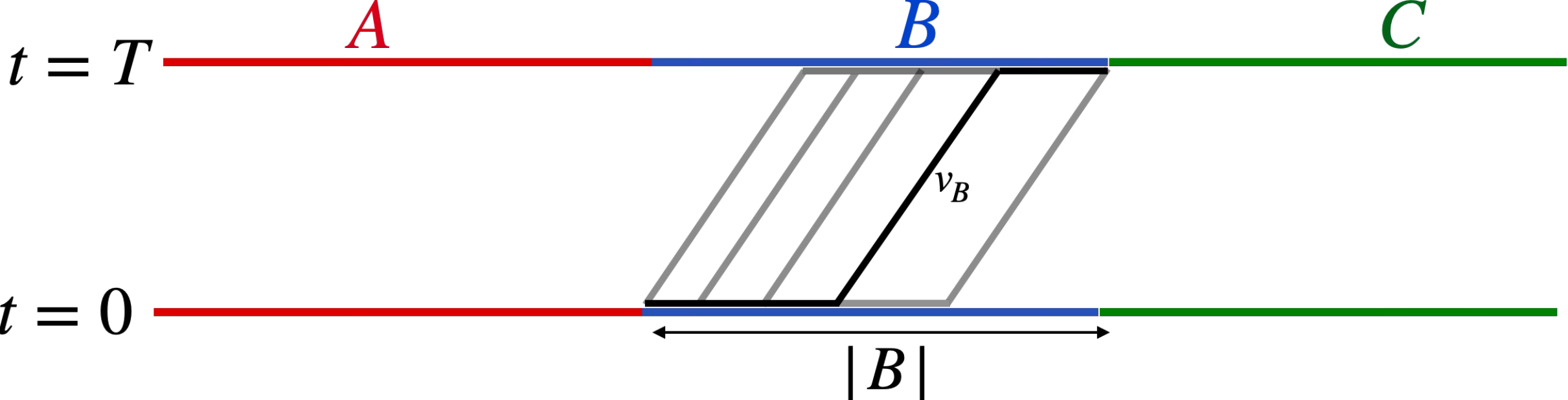}
\caption{The family of piecewise membranes that gives the same entropy for $v>v_B$ in random quantum circuits. We show five membranes and highlight one.  }
    \label{fig:rqc_membranes}
\end{figure}

In random quantum circuits, in the minimisation~\eqref{eq:entanglement_growth_conjecture} over membranes of slope $v$ one is also allowed to consider piecewise membranes with horizontal segments on the lower and upper boundaries~\cite{JonayHuseNahum2018,Zhou:2019pob}. A horizontal segment of spatial length $L$ costs $s_\text{eq}\, L$ entropy. This is different from taking the $v\to\infty$ limit of regular membranes, which would include an additional $\lim_{v\to\infty}\mathcal{E}(v)/v>1$ enhancement factor.
We can draw a family of membranes with horizontal segments adding up to $L=(v-v_B)T$ and a $v_B$ slope bulk segment, see figure~\ref{fig:rqc_membranes}. These all give the same value for the entropy:
\es{S_piecewises}{
S^{(n)}=s_\text{eq}\le((v-v_B)T+v_BT\ri)=s_\text{eq} v\, T=s_\text{eq}\,\abs{B}\,,
}
reproducing the effective membrane tension $\mathcal{E}^{(n)}_{\text{eff}}(v)=v$ for $v>v_B$. The ensemble of membranes fill the same spacetime region that the new unstable fixed point in between the two wavefront fills in the Brownian SYK chain.

One key difference in the physics of random quantum circuits and SYK chains is that the former has $O(1)$ scrambling time, hence the membrane only emerges after the scrambling time, whereas in the  Brownian SYK chain the membrane at $v<v_B$ and the spacetime region with the new fixed point for $v>v_B$ emerge well before the scrambling time, and the latter domain becomes highly fluctuating after the scrambling time. Could these fluctuations be related to the multiple membranes giving the same entropy in  figure~\ref{fig:rqc_membranes}? We leave this question for the future.

\begin{figure}[!h]
    \centering
\includegraphics[width=0.5\linewidth]{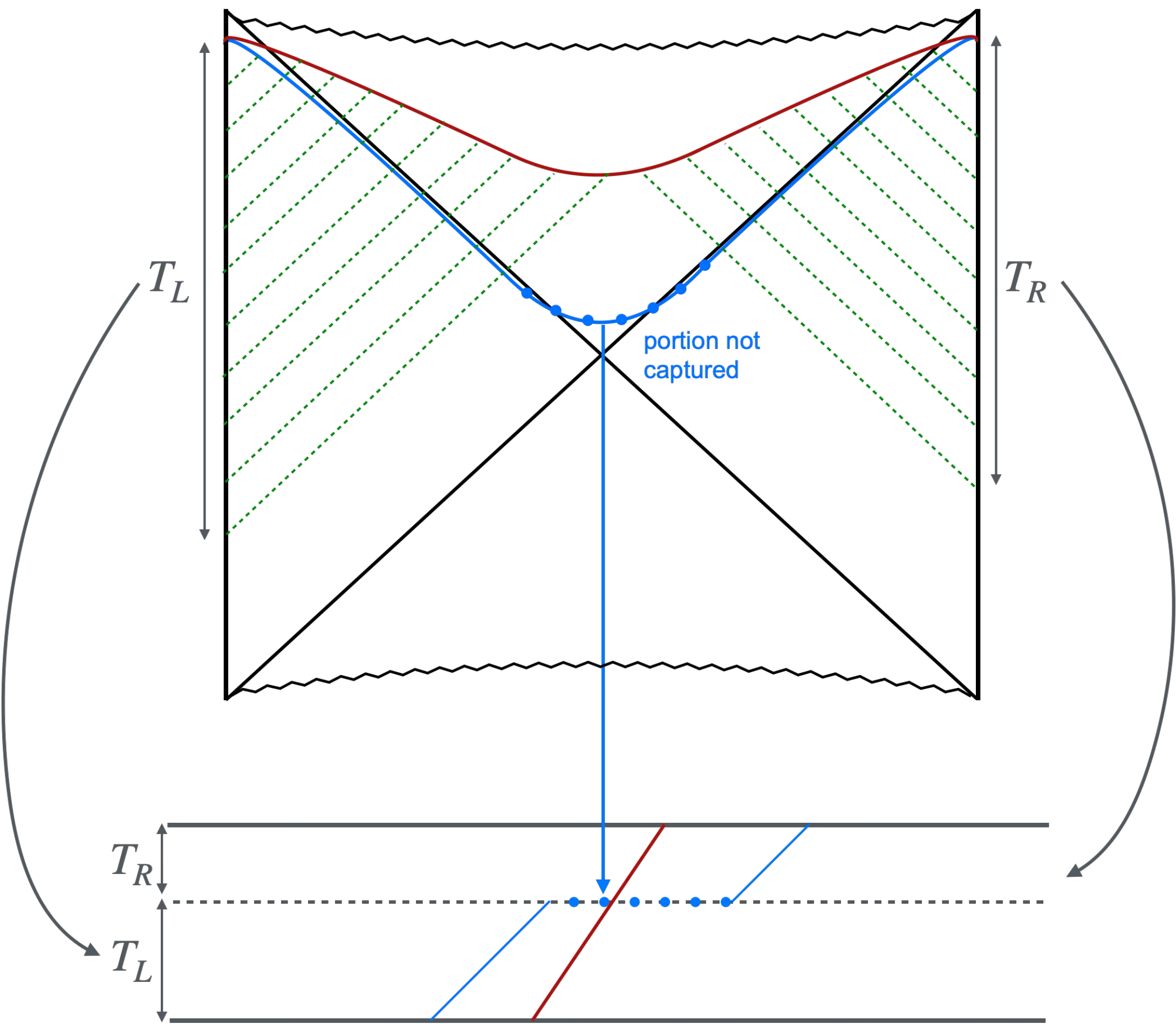}
\caption{The holographic mapping of HRT surfaces to entanglement membranes. On the top we show the eternal black hole Penrose diagram with two HRT surfaces. Note that the boundary spatial directions are suppressed and only time and the bulk radial coordinate is included. The two HRT surfaces are projected to the boundary along radial null geodesics and the projections become the entanglement membranes displayed on the bottom. Here the time and spatial directions are shown. If we stop the projections at the same time for the red and blue HRT surfaces, we get that $T_L+T_R=T$, but a portion of the blue HRT surface is not captured. We can account for its area by including a horizontal membrane segment.  }
    \label{fig:holo}
\end{figure}

In holographic theories, the membrane is obtained by projecting the HRT surface~\cite{Ryu:2006bv,Ryu:2006ef,Hubeny:2007xt} in the gravitational description to the conformal boundary~\cite{Mezei_Membrane_Theory_Holography}. The projection is done along infalling null geodesics. The image is holographic: the HRT surface can be reconstructed from the projection, as bulk depth is tied to the local membrane velocity $v$. In the operator entanglement setup the gravity dual is a two-sided eternal black hole and some portion of the HRT surface needs to be projected to the left and the other to the right boundary, see figure~\ref{fig:holo} for illustration. The two projections  are subsequently glued together. Together they give a spacetime slab of time width $T$ with a continuous membrane through it. There is a simple recipe for choosing where to switch from left to right projection for $v<v_B$,\footnote{In the case when the subregions are half spaces, the corresponding HRT surface spends most of its life at a constant radial coordinate $z(v)$ behind the black brane horizon. We can choose an arbitrary switching point between the left and right and we will get the same boundary membrane after gluing. For more complicated subregions the HRT surface exhibits nontrivial evolution in $z$, but it can never surpass a barrier surface $z_*$~\cite{Hartman:2013qma}. We switch projection from left to right whenever $z_*$ is reached~\cite{Mezei_2017,Mezei_Membrane_Theory_Holography,Jiang:2024tdj}. This point projects to a membrane patch with $v=0$ locally.} however for $v>v_B$ these projections would naturally give a spacetime slab of time width $\abs{B}>T$~\cite{Jiang:2024tdj}. Since this would be hard to interpret from the boundary quantum field theory perspective, we may stop the projection at some point, where $T_L+T_R=T$ and connect the disconnected membrane segments with a horizontal line that costs $s_\text{eq}\, L$ entropy for length $L$~\cite{Mezei_Membrane_Theory_Holography,Jiang:2024tdj}. This prescription is reminiscent of how membranes work in random quantum circuits, note however that for that case the horizontal membrane segments are confined to the time boundaries.\footnote{One may contemplate replacing the horizontal line with a $v_B$ zig-zag, but we have not found a very natural prescription to do this.} The holographic $v>v_B$ membrane lives in the same spacetime domain, where the Brownian SYK chain has a new unstable fixed point.

A shared feature between the Brownian SYK chain and holographic theories is an $O(\log N)$ scrambling time. For $v>v_B$ the HRT surface spends most of its life exponentially close to the horizon. If $T\sim \log N$, we expect that off-shell shock waves on the horizon (not sourced by matter stress tensor) have large fluctuations~\cite{Stanford:2021bhl}, making the HRT surface and its membrane projection also highly fluctuating without changing the entropy. This is in close analogy with the behaviour of the Brownian SYK chain in the same regime $v>v_B$ described in the previous section.

\subsection{$v < v_B$} \label{sec:vlvb}

When $v < v_B$, we must simultaneously solve the nonlinear partial differential equations for $x_u,\,y_u,\,z_u$. However, the boundary conditions in~\eqref{eq:boundary_conditions_xyz0} and~\eqref{eq:boundary_conditions_xyzT} are only implicitly known at $t = 0$ and $t = T$. As we increase the length of the chain, the number of unknown boundary conditions grows linearly with the length. Moreover, the saddle point equations admit exponentially growing and decaying solution in time. The existence of such modes can be  seen by considering the linearised saddle point equations around a fixed point. Consider, for example the fixed point $(x,y,z) = (0,0,1)$. The linearised equations are:
\begin{malign}
    \partial_t x_u &= -\frac{J}{2^{q-2}} y_u + \frac{\jt}{2} \partial_u^2 y_u, \quad
     \partial_t y_u = -\frac{J}{2^{q-2}} x_u + \frac{\jt}{2} \partial_u^2 x_u,
   \quad  \partial_t z_u = 0\,.
\end{malign}
One can check that the solutions to these equations either grow or decay exponentially in time. 
The resulting instabilities make it hard to shoot for the right initial condition. 
\begin{figure}[h]
    \centering
\includegraphics[width=0.7\linewidth]{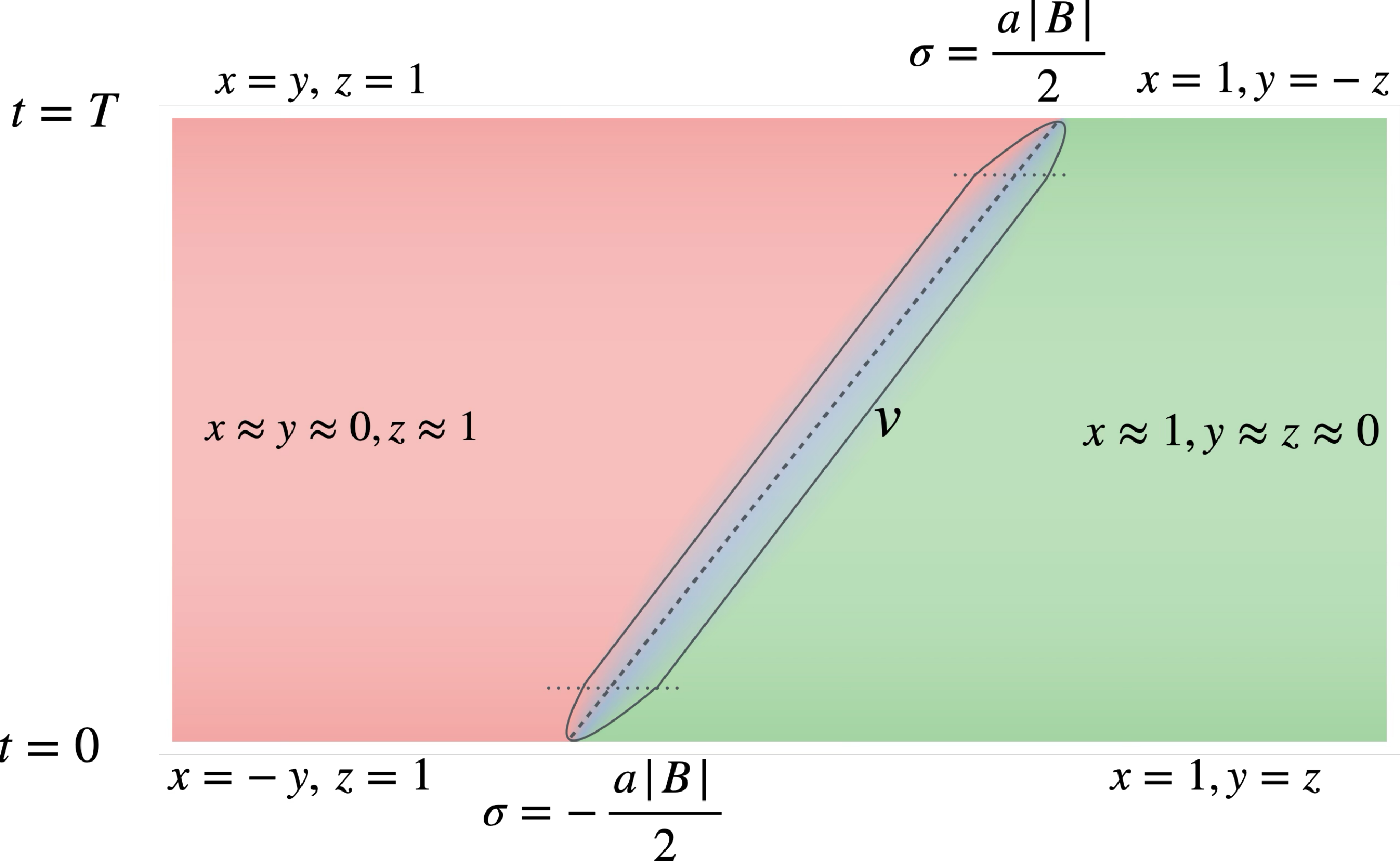}
    \caption{A rough sketch of a traveling wave solution with velocity $v$ that emerges from a sharp boundary at $t= 0$ and ends on a sharp boundary at $t = T$. The boundary layers end at the horizontal dotted lines and the traveling wave we construct captures the bulk of the solution, where it is of constant width and has an enhanced translation symmetry. }
    \label{fig:sketch}
\end{figure}

It is however very plausible that in the large $T$ and $|B|$ limit the growing modes are localised in boundary layers of spacetime of fixed size ${1/ J}$ near $t=0$ and $t=T$. Away from these layers the solution should be stationary and have an emergent translation symmetry characterised by the velocity $v$: it should be a traveling wave propagating with velocity $v$. Near $t = 0$ and $ t = T$, the exponentially growing and decaying modes are important. We assume that there exist boundary field configurations at $t = 0,\,T$ satisfying~\eqref{eq:boundary_conditions_xyz0},~\eqref{eq:boundary_conditions_xyzT} such that the solution approaches the traveling wave solution in the bulk.
%In this section, we will make the simplifying assumption that for large $T$ and $|B|$, the saddle point solution can be approximated by a traveling wave propagating with velocity $v$ for the time interval $t \in [0,T]$ as long as $t \gg \frac{1}{J}$ and $|T -t| \gg  \frac{1}{J}$. Near $t = 0$ and $ t = T$, the exponentially growing and decaying modes are important. We assume that there exist boundary conditions at $t = 0,T$ satisfying~\eqref{eq:boundary_conditions_xyz0},~\eqref{eq:boundary_conditions_xyzT} such that the solution approaches the traveling wave solution in the bulk. 
In figure~\ref{fig:sketch}, we show a rough sketch of a traveling wave emerging from an initial condition with sharp boundaries at $t = 0,\,T$. It would be interesting to construct the full solution numerically; we leave this problem for future work.

Throughout this section, we will work in continuum limit. We will use the variables $r,\theta$ and coordinates $\sigma, \tau$ to solve the differential equations. We will also set $q = 4$. It will be useful to note from equation~\eqref{eq:butterfly} and the relation~\eqref{eq:rescaling} that the butterfly velocity in $\sigma, \tau$ coordinates is\begin{malign} 
    v_B =  \sqrt{J_c \jt_c}\, = \sqrt{2 J_c}.
\end{malign}

\subsubsection{Symmetries}
Before attempting to solve the differential equations it is useful to consider their  symmetries. Let us denote the center of subregion $B$ by $\sigma = 0$. The equations are invariant under the following transformation
\begin{enumerate}
    \item $\tau \rightarrow T_\tau-\tau$ and $\theta \rightarrow \frac{\pi}{2} - \theta$ 
    \item $\sigma \rightarrow -\sigma$.
\end{enumerate}
The boundary conditions in~\eqref{eq:boundary_conditions_xyz0},~\eqref{eq:boundary_conditions_xyzT} are also invariant under the joint transformation:
\begin{align} \theta  \rightarrow \frac{\pi}{2} -\theta, \quad \tau \rightarrow T_\tau - \tau, \quad \sigma \rightarrow -\sigma\,.
\end{align}
Hence we demand that the solution to equations~\eqref{eq:rtheta_eqations} is symmetric under the above transformation:
\begin{equation}
    r(\sigma, \tau) = r(-\sigma, T_\tau-\tau), \quad 
    \theta(\sigma, \tau) = \frac{\pi}{2} - \theta(-\sigma, T_\tau-\tau)\,.
\end{equation}
For a traveling wave solution, this condition implies
\begin{malign}
    r(\sigma) = r(-\sigma),\quad  \theta(\sigma) = \frac{\pi}{2} - \theta(-\sigma)\,.
\end{malign}
Therefore, at $\sigma = 0$, the traveling wave solution must satisfy
\begin{malign} \label{eq:symmetry}
    r'(\sigma)|_{\sigma = 0} = 0,\quad \theta(\sigma)|_{\sigma = 0} = \frac{\pi}{4}\,.
\end{malign}
where $'$ denotes derivative w.r.t $\sigma$. 

\subsubsection{Traveling waves for $v < v_B$}\label{sec:vlvb_waves}

To find the traveling wave solutions with velocity $v$, we  replace $\partial_\tau$ by $- v \partial_\sigma$ in the equations~\eqref{eq:rtheta_eqations}. With this replacement, our task of solving the nonlinear partial differential equations has reduced to solving nonlinear ordinary differential equations. 

Rather than solving the~\eqref{eq:rtheta_eqations} directly, consider the action in~\eqref{eq:hamiltonian} restricted to traveling wave configurations with velocity $v$:
\begin{equation}
    r(\sigma,\tau)= r(\sigma - v\tau, 0), \quad \theta(\sigma,\tau) = \theta(\sigma - v\tau, 0)\,.
\end{equation}
For such configurations the time integral produces a constant factor of $T_\tau$ and the action is given by the integral of a spatial Lagrangian $\mathcal{L}(\sigma)$: 
\begin{malign}
-\frac{S}{T_\tau} &= \frac{1}{2a} \int \d \sigma \,\mathcal{L}\,, \\ 
 \mathcal{L} &= - v r \,\theta' - \frac{1}{2} \left( (r^2+1) \theta'^2 - \frac{r'^2}{r^2+1}\right) - V\,,
\end{malign}
where $V$ is 
\begin{malign}
    V &= \frac{J_c}{16} (1 + r^4 - (r^2+1)^2 (\sin^4 \theta + \cos^4 \theta)) = \frac{J_c}{16} \left( \frac{(r^2 + 1)^2 \sin^2 2 \theta}{2} - 2 r^2\right)\,.
\end{malign}

By construction, the Euler-Lagrange equations derived from the above Lagrangian must agree with the equations in~\eqref{eq:rtheta_eqations} under the replacement $\partial_\tau \rightarrow - v\partial_\sigma$. The Lagrangian description is useful because it allows us to find a ``spatial Hamiltonian'' $\hat h(\sigma)$ that is conserved along $\sigma$:
\begin{malign}  \label{hath}
    \hat h = (r^2+ 1) \frac{p_r^2}{2} - \frac{(p_\theta+ vr)^2}{2(r^2+1)} + V\,,
\end{malign}
where $p_r$ and $p_\theta$ are the canonical conjugates of $r$ and $\theta$:
\begin{malign}
    p_r = \frac{\partial \mathcal{L}}{\partial r'}\,, \quad p_\theta = \frac{\partial \mathcal{L}}{\partial \theta'} \,.
\end{malign}
We can rewrite the equations of motion in~\eqref{eq:rtheta_eqations} as Hamilton's equations:
\begin{malign} \label{eq:hamiltons_equation}
     r' &= p_r (r^2+1)\,,&\quad 
    \theta' &= - \frac{p_\theta + v r}{r^2+1}\,, \\
    p_\theta' &= -\partial_\theta V\,, &\quad 
    p_r' &= -r\, p_r^2  + \frac{\partial_r}{2}\left(\frac{p_\theta + v r}{r^2+1}\right)^2 - \partial_r V\,.
\end{malign}
The equations are supplemented with the following boundary conditions at $\sigma \rightarrow \pm \infty$
\begin{malign}\label{eq:bc_sigma}
  &\sigma \rightarrow -\infty: \, r = 0\,,\, \theta = 0\,, \qquad
  \sigma \rightarrow \infty: \, r = 0\,,\, \theta = \frac{\pi}{2}\,.
\end{malign}

The above boundary conditions determine the asymptotic form (the tail region) of the traveling wave solution. In the $\sigma \rightarrow \infty$ limit we can linearise the equations in $(r ,\, \delta \theta \equiv \theta - \pi/2,\, p_r,\, p_\theta)$. We get 
\begin{malign}
    r' &= p_r\,, &\quad \delta \theta' &= - (p_\theta + v r)\,,\\ 
    p_\theta' &= -\frac{J_c}{4} \delta \theta\,, &\quad p_r' &= \left(v^2  + \frac{J_c}{4}\right) r + v\, p_\theta \,.
\end{malign}
The linearised equations have two exponentially growing and two exponentially decaying modes. A general solution that decays as  $\sigma \rightarrow \infty$ is the following linear combination:
\begin{malign}
(r, \delta \theta, p_r, p_\theta) = \alpha_+ \, e^{-\lambda_+ \sigma } \psi_+ + \alpha_- \, e^{-\lambda_- \sigma } \psi_- \,,    
\end{malign}
where 
\begin{equation}
    \lambda_\pm = \frac{\pm v - \sqrt{v^2  + J_c}}{2}
\end{equation}
and 
\begin{malign}
    \psi_+ &= (4 \lambda_+ , -4 \lambda_+, \lambda_+^2 , J_c)\,, \\  \quad \psi_- &= (4 \lambda_- , 4 \lambda_-, \lambda_-^2 , -J_c)\,.
\end{malign}
$\alpha_\pm$ are undetermined infinitesimal parameters that must be fixed by solving the nonlinear equations.  The solution to the linearised equations imply that the traveling wave solution has an exponential tail at infinity.

Now, we consider solutions to the full non-linear equation. The boundary conditions in equation \eqref{eq:bc_sigma} imply that $\hat h = 0$ as $\sigma \rightarrow \pm \infty$. Since $\hat h$ is conserved, it vanishes everywhere. Therefore,
\begin{equation} \label{eq:vanishing_hamiltonian}
    \hat h = 0\,.
\end{equation}
By evaluating $\hat h=0$ at $\sigma=0$ and by imposing the  symmetry constraints, we can fix $p_\theta$ at the origin:
\begin{malign} 
    p_\theta(0) &= - vr(0) - \sqrt{2(r(0)^2+1)V_0}\,.
\end{malign}
Thus, we have fixed three out of the four unknown variables at $\sigma = 0$. The problem of solving the Hamilton's equations has now been reduced to a shooting problem in the variable $r_v(0)$. 

\begin{figure}[h]
    \centering
\includegraphics[width=0.49\textwidth]{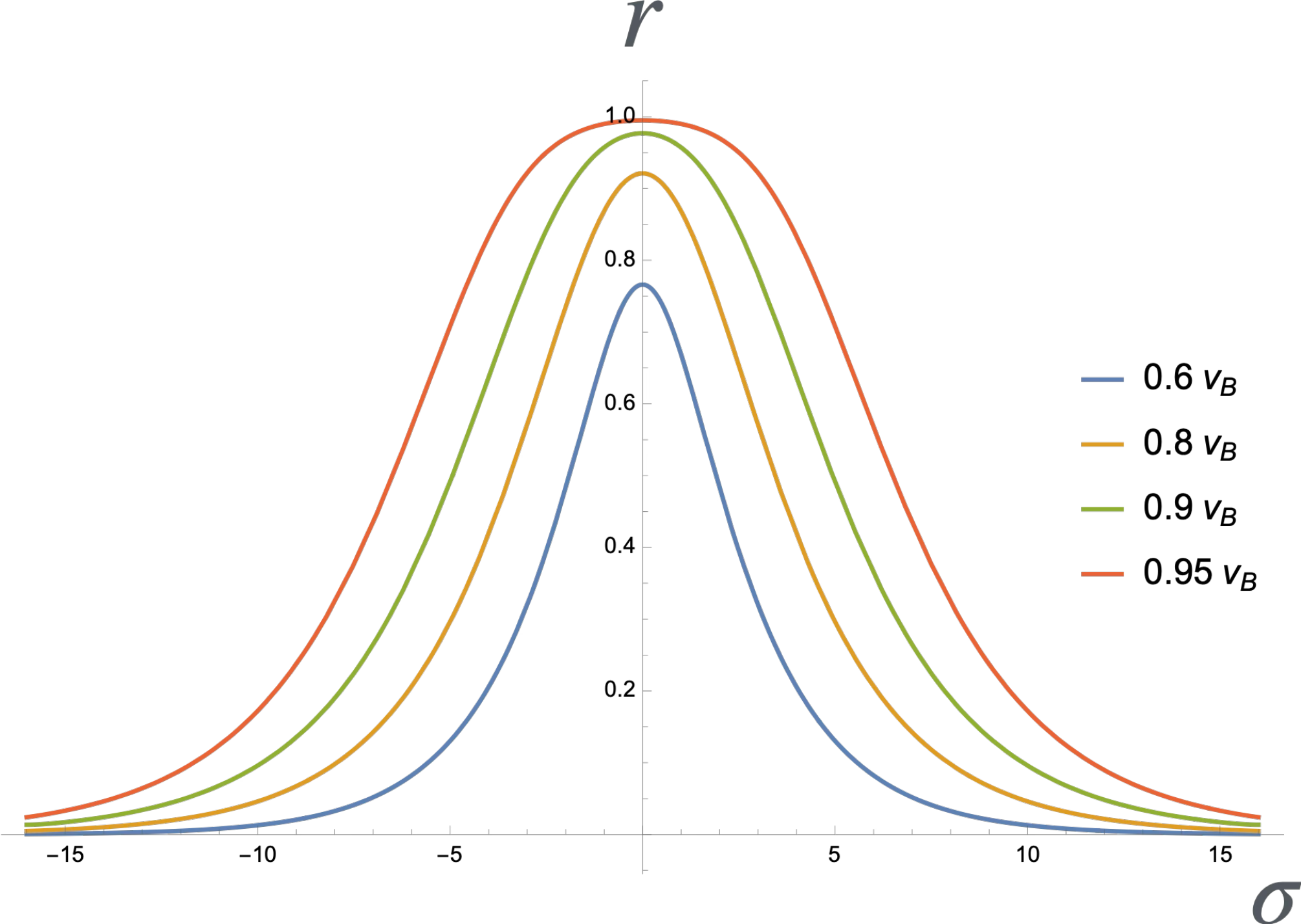} 
\includegraphics[width=0.49\textwidth]{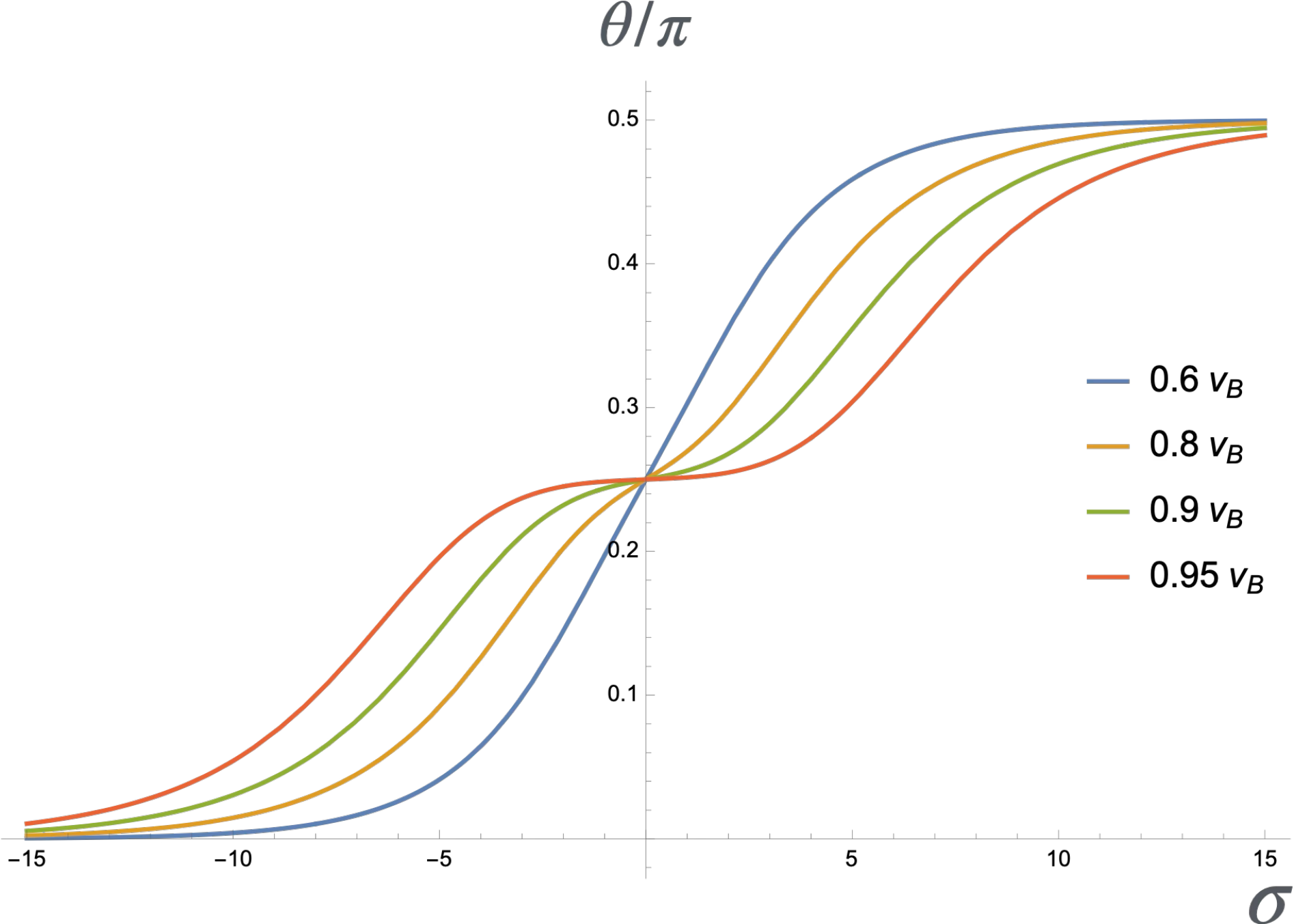}
    \caption{Traveling wave solution (soliton) for $v < v_B$. {\bf Left:} $r(\sigma)$. {\bf Right:} $\theta(\sigma)$. The width of the soliton increases as $v$ approaches $v_B$. } 
    \label{fig:membrane_thickness}
\end{figure}

In figure~\ref{fig:membrane_thickness}, we show the profile of the traveling wave solution for some values of $v < v_B$.\footnote{In this section, we have only considered traveling wave solutions for $v > 0$. Traveling wave solution with $v < 0$ are related by the simultaneous transformation $v\rightarrow -v, \,r \rightarrow -r$ and $p_r \rightarrow -p_r$.} It follows from these numerical solutions combined with the linearised analysis above that the traveling waves are finite in size and are exponentially localised.

Let us denote a traveling wave solution with velocity $v$ by $(r_v(\sigma),\theta_v(\sigma))$. We find that this family of solutions has the following properties established in appendix~\ref{app:wavesize}: 
\begin{enumerate}
    \item  $-1 < r_v(0) < 1$ for $-v_B < v < v_B$ and $r_v(0)$ increases monotonically with $v$. In the limit $v\rightarrow  v_B$,  $r_v(0) \rightarrow 1 $.\footnote{See appendix \ref{app:wavesize} for more details on the scaling near $ v_B$.}
    \item As $v \rightarrow v_B$: \begin{equation} \label{rc_sol}
        r_v(\sigma) \rightarrow r_c(\theta(\sigma)) = \begin{cases}
            \tan(\theta(\sigma))  \quad 0< \theta(\sigma) < \frac{\pi}{4} \\
            \cot(\theta(\sigma)) \quad \frac{\pi}{4} < \theta(\sigma) < \frac{\pi}{2}\,.
        \end{cases} 
    \end{equation}
    \item Let $\Delta_v $ denote the size of interval around $\sigma = 0$ where $|\theta_v(\sigma) - \frac{\pi}{4}| < \frac{\pi}{8}$.  $\Delta_v$ monotonically increases with $v$ and it diverges in the limit $v \rightarrow v_B$.
\end{enumerate}

To summarise this section, we found the traveling wave solution to our saddle point equations~\eqref{eq:diff_xyz}. For $v < v_B$, the traveling wave solution is a soliton that propagates from the boundary $A|B$ at $ \tau  = 0$ to the boundary $B|C$ at $ \tau = T_\tau$. The soliton has an $O(1)$ width that increases with the velocity.  As  $v \rightarrow v_B$, the width diverges. For $ v> v_B$, the soliton splits into two fronts, each front propagating with the butterfly velocity. The distance between the fronts is $(v-v_B) T_\tau$. In turn, we may think of the $v < v_B$ traveling wave as a bound state of the two fronts.

\subsection{Analytic solution at large $q$}\label{sec:largeq}

One may wonder if there exists a simplifying regime of model parameters, where an analytic solution for the traveling wave can be found. Since $J_c$ can be absorbed by rescaling $\sigma$, the only parameter we can adjust is $q$. In the $q\rightarrow \infty$ limit, in the ``core region" of the wave, to be defined more precisely below,  we find that the potential terms (that have a $J_c$ prefactor) in~\eqref{eq:rtheta_eqations} are negligible. To get a complete description of the wave, we would  need to solve the equations in the tail region as well, and match the core and tail solutions. However using the spatial Hamiltonian allows us to circumvent treating the tail in detail. 

\begin{figure}[!h]
    \centering
 \includegraphics[width=0.25\textwidth]{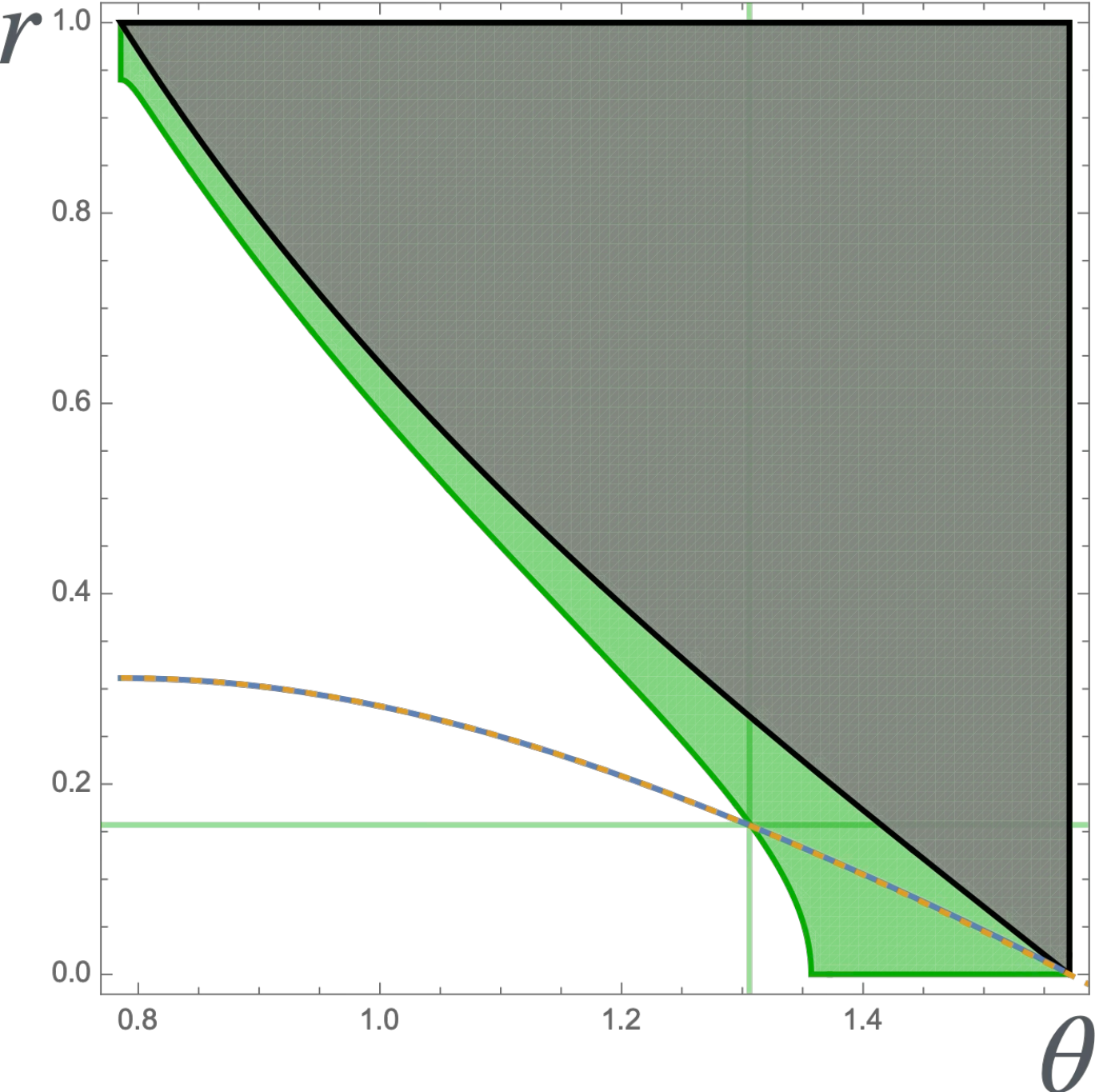}\hspace{0.3cm}
\includegraphics[width=0.3\textwidth]{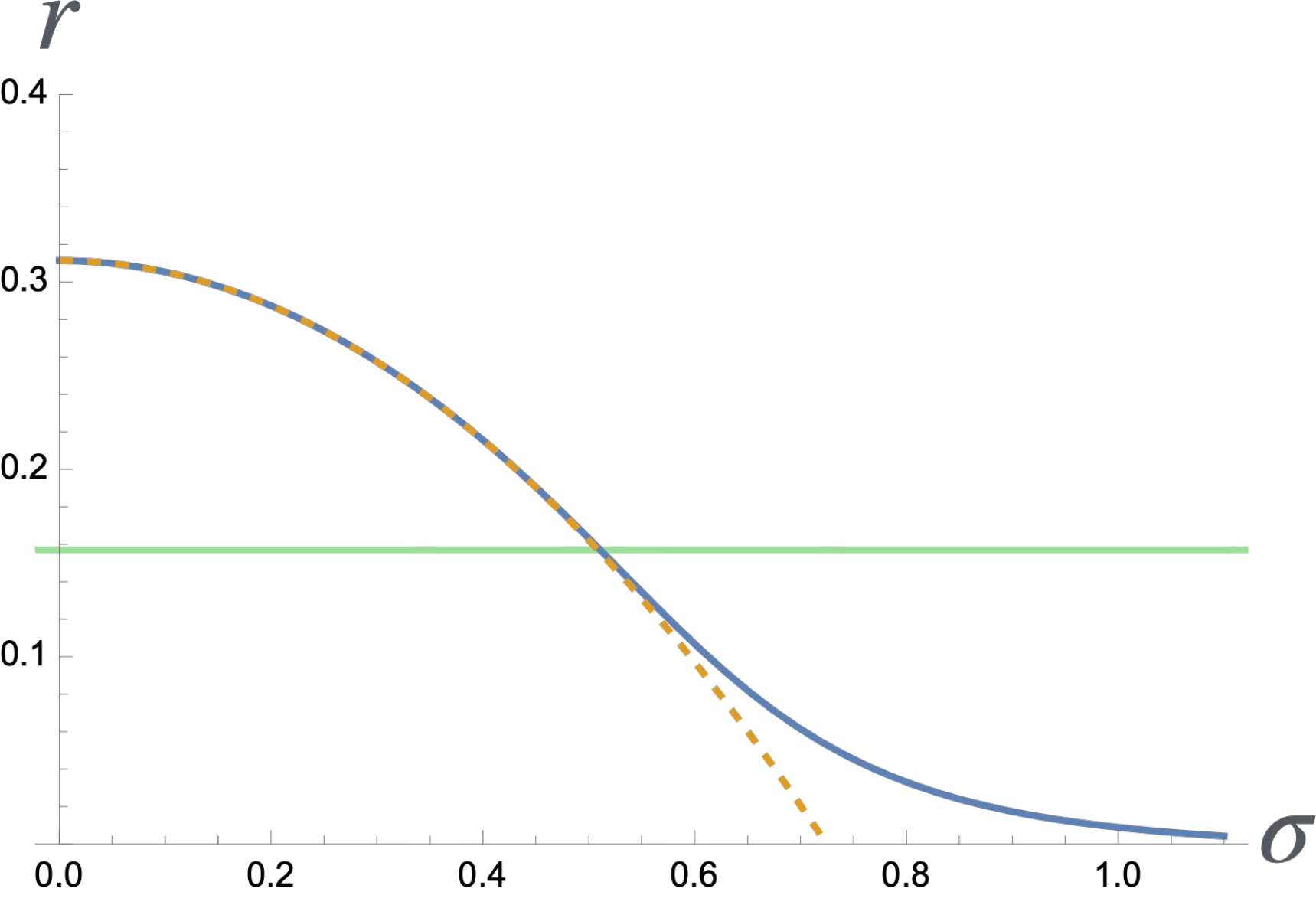}\hspace{0.3cm}
\includegraphics[width=0.3\textwidth]
{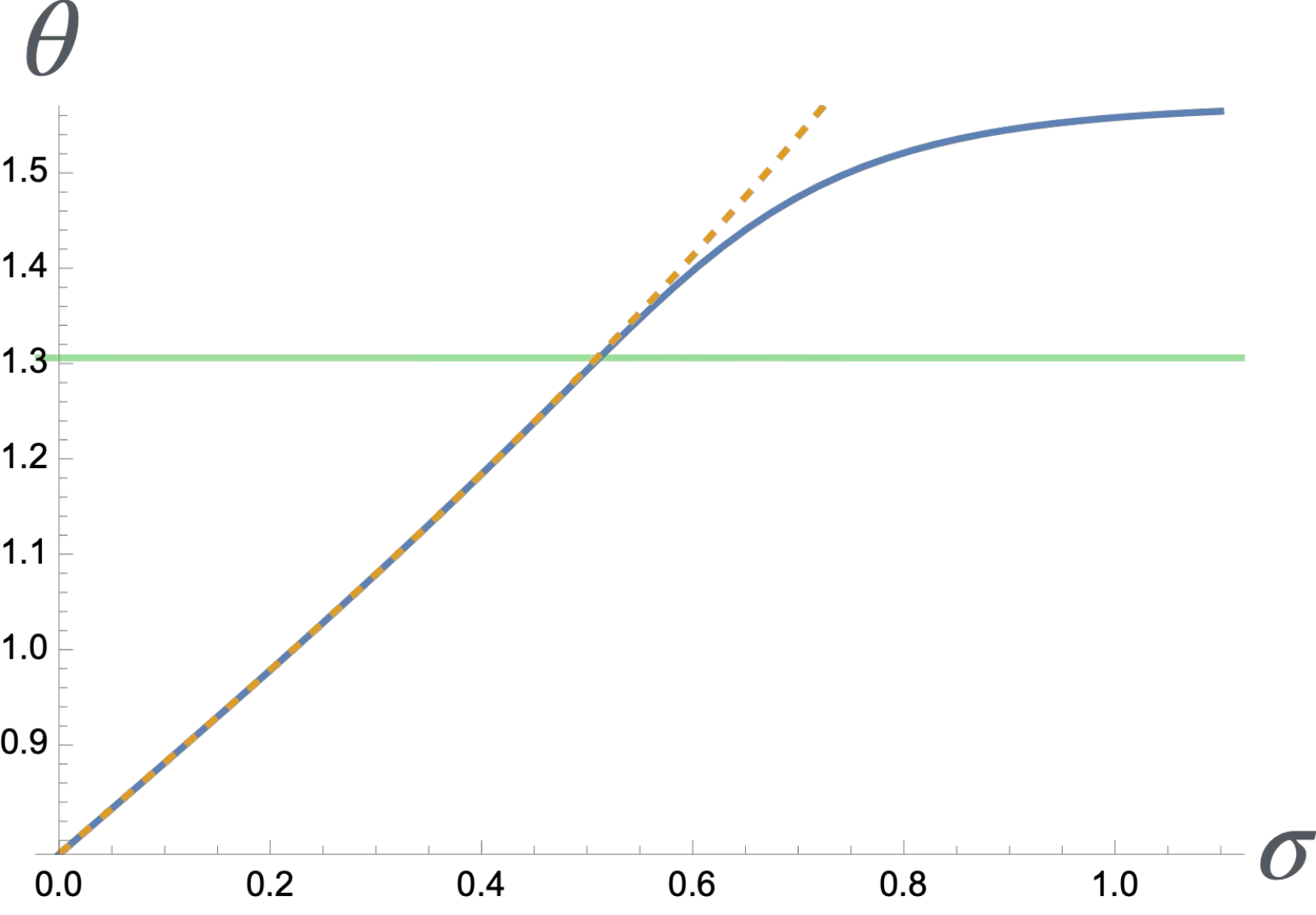}
    \caption{Match in the core region between numerical solutions for $q=100$ (blue) and the analytic prediction from the large $q$ expansion (orange dashed). Note that there is no free parameters involved in these plots. {\bf Left:} Saddle point solution in the $r,\theta$ plane. The gray exclusion region is $r > \cot \theta$, where the potential blows up as $q\to \infty$. The green exclusion region is where $V$ cannot be approximated by a constant for $q=100$. The point where the large $q$ solution enters this region is where the core ends and the tail region starts, over which we do not have analytic control. {\bf Middle and right:} The green lines indicate the value of $r,\theta$ where the core ends. These are exactly the points where the core solution starts to deviate from the numerical solution. } 
    \label{fig:largeqMatch}
\end{figure}

To obtain the core solution, we would like to neglect the non-constant potential terms form~\eqref{eq:hamiltonian}
\es{other_pot_terms}{
1&\gg r^q\,,\\
1&\gg (r^2+1)^{q/2}  \left( \cos^q \theta + \sin^q \theta \right)\,.
}
For simplicity we will focus on the $\sigma>0$ half line, hence $\pi/4<\theta<\pi/2$. The solution for $\sigma < 0$ follows from symmetry. For $q\to \infty$
the first inequality is automatically satisfied since $r<1$. However the second inequality can be violated:  we use that $\sin\theta>\cos\theta$ for our range of $\theta$ to simplify the inequality to $(r^2+1)\sin^2 \theta<1$, which simplifies to $r<\cot \theta=r_c(\theta)$. We have argued before that for $v<v_B$ all traveling waves satisfy this inequality. We conclude that the large $q$ approximation is self-consistent. However, for finite but large $q$ the core solution may enter a region, where the inequalities~\eqref{other_pot_terms} are not obeyed. This provides a prediction for the end of the core region and the start of the tail region. These predicitions are verified by numerics, see figure~\ref{fig:largeqMatch} (especially the left plot).  In numerics we will detect violation if the right hand sides in~\eqref{other_pot_terms} exceed $0.1$.  

In the core region we set the potential term to $V={{\cal J}\ov 2}\equiv \frac{J_c}{q 2^{q-2}}$, which
reduces Hamilton's equations~\eqref{eq:hamiltons_equation} to
\begin{malign} \label{eq:hamiltons_equation_largeq}
     r' &= p_r (r^2+1)\,,&\quad 
    \theta' &= - \frac{p_\theta + v r}{r^2+1}\,, \\
    p_\theta' &= 0\,, &\quad 
    p_r' &= -r\, p_r^2  + \frac{\partial_r}{2}\left(\frac{p_\theta + v r}{r^2+1}\right)^2 \,,
\end{malign}
while the spatial Hamiltonian~\eqref{hath} takes the form
\begin{malign} \label{eq:hamiltons_equation_largeq2}
0&=2\hat{h}= (r^2+ 1) {p_r^2} - \frac{(p_\theta+ vr)^2}{r^2+1}+{\cal J}\,.
\end{malign}

From considering the first equation at the origin, we get that
\begin{malign} 
    p_\theta(\sigma) &= - vr_v - \sqrt{(1+r_v^2){\cal J}}
\end{malign}
is constant and where we introduced the notation $r_v\equiv r(0)$. This allows us to express $p_r$ as a function of $r$ and to obtain a first order equation for $r(\sigma)$ with the solution
\es{rsigLargeq}{
r(\sigma)=r_v-{2\le(r_v+\sqrt{1+r_v^2}\,\tilde v\ri)\ov 1-\tilde v^2}\,\sin^2\le({\sqrt{1-\tilde v^2}\ov 2}\,\sqrt{{\cal J}}\sigma\ri)\,,\qquad \tilde v\equiv {v\ov \sqrt {\cal J}}\,.
}
We note that the solution holds both for $\tilde v<1$ and $\tilde v>1$, where in the latter case the sine simply turns into sinh. The butterfly velocity is $\tilde v_B\sim q$, which is effectively infinite for our purposes.

\begin{figure}[h]
    \centering
\includegraphics[width=0.3\textwidth]{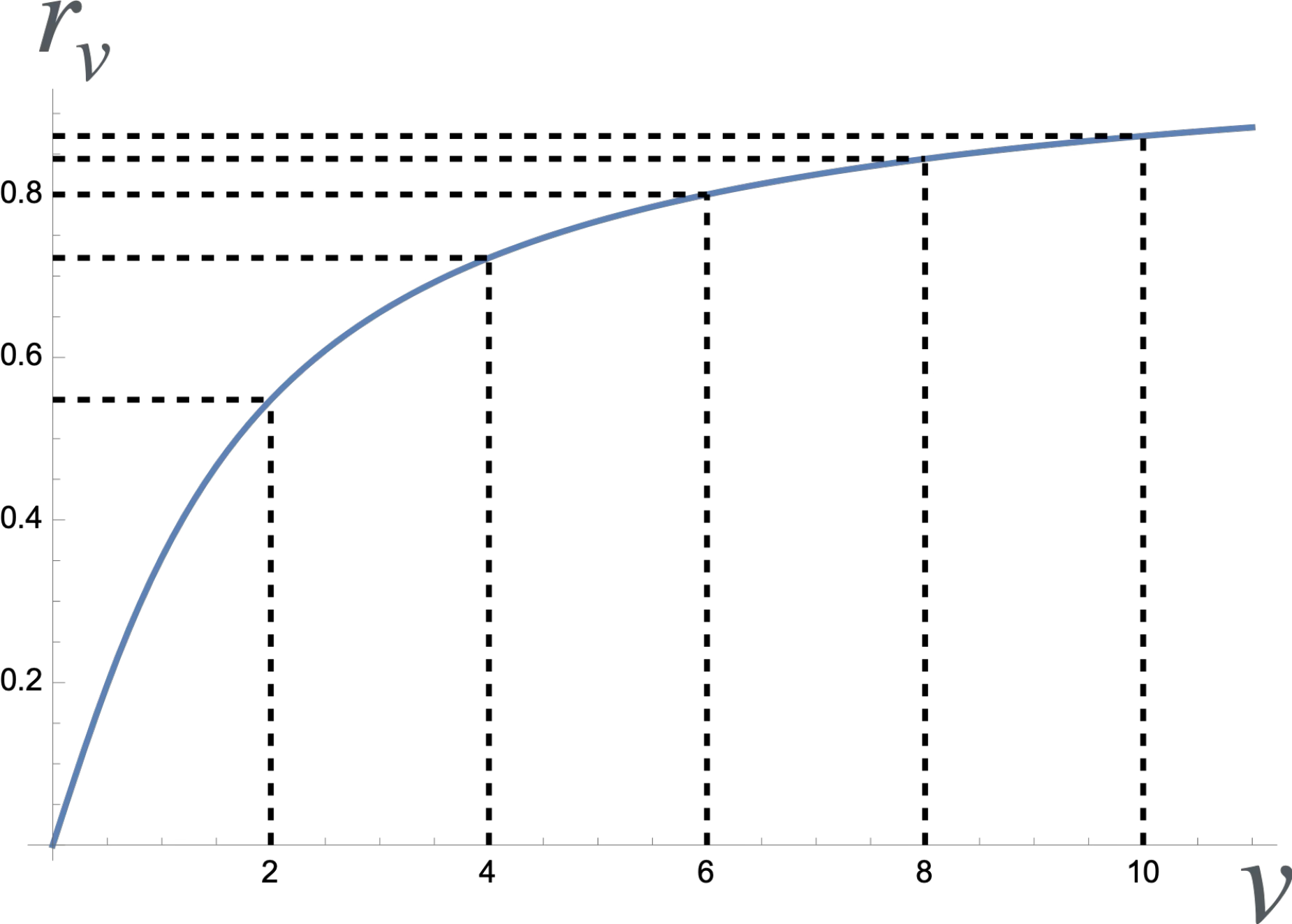}\hspace{0.3cm}
\includegraphics[width=0.3\textwidth]{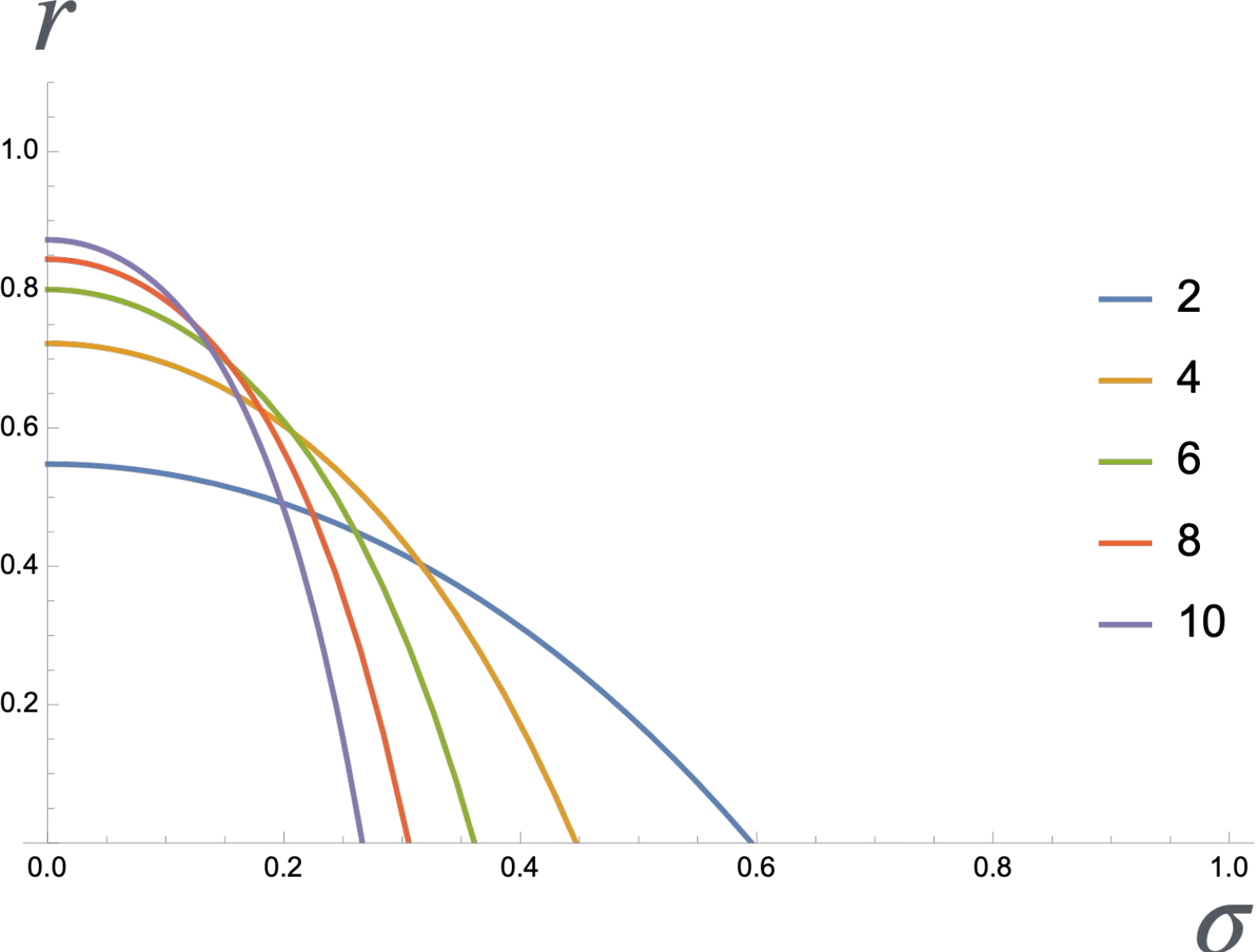}\hspace{0.3cm}
\includegraphics[width=0.3\textwidth]{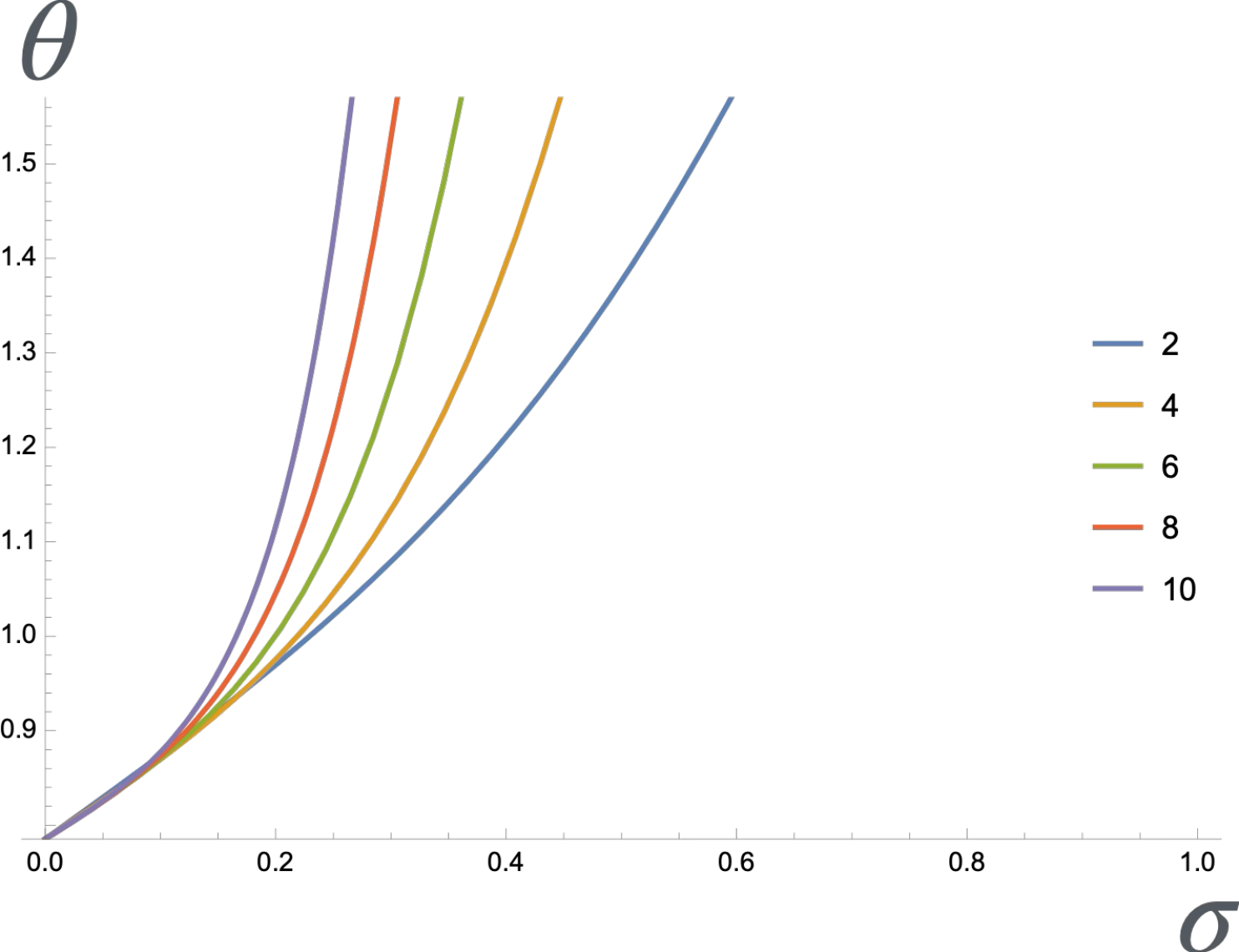}
    \caption{{\bf Left:} $r_v$ at large $q$. We read off the $r_v$ values of interest from this graph, which we use to construct the core of traveling waves at large $q$. {\bf Middle and right:}  
    The core of traveling waves for different choices of $v$ at large $q$.} 
    %\label{fig:manysols}
    \label{fig:r0v}
\end{figure}

We change variables from $\theta(s)$ to $\theta(r)$ to obtain the separable equation
\es{DthetaDr}{
{d\theta\over d r}=-{\sqrt{1+r_v^2}+ \tilde v(r_v-r)\ov (1+r^2)\sqrt{(r_v-r)\le(2\le(r_v+\sqrt{1+r_v^2}\,\tilde v\ri)+(\tilde v^2-1)(r_v-r)\ri)}}\,.
}
This equation can be integrated in closed form (using the initial condition $\theta(r_v)=\pi/4$), but the answer is tedious. Requiring that $\theta(r=0)=\pi/2$, which should hold to leading order in $q$, determines $r_v$ as a function of $v$, hence giving an analytic solution to the shooting problem, see the left plot in figure~\ref{fig:r0v} for $r_v$.\footnote{We write down the transcendental equation whose solution is $r_v$ in \eqref{rv0sol}. } As expected,  $\lim_{v\rightarrow v_B} r_v=1$, which for large $q$ is the asymptotic behaviour, since $v_B$ diverges as commented above.

On figure~\ref{fig:r0v} we show the analytic profiles of the traveling waves for multiple choices of $v$ in figure~\ref{fig:r0v}. The plots are qualitatively different from the $q=4$ case shown if figure~\ref{fig:membrane_thickness}, e.g.~the $r(\sigma)$ graphs intersect and the size of the traveling wave decreases with increasing $v$, unlike in the $q=4$ case. Note however that our large $q$ treatment does not access the near $v_B$ behaviour (since $v_B=O(q)$, while we assumed $v=O(q^0)$); we expect that the traveling wave diverges in size as $v\rightarrow v_B$ for any $q$ and that $r(\theta)$ approaches~\eqref{rc_sol} (so that $x=y$ and $y=z$ is imposed on the left and right respectively).

Using the left plot on figure~\ref{fig:r0v}, we can predict the shape of the traveling wave cores and contrast them with numerics at large but finite $q$. We get a perfect match as shown in figure~\ref{fig:largeqMatch} without fitting any parameters. We also provide a perfect prediction for where the core region ends, as demonstrated by the right plot on figure~\ref{fig:largeqMatch}.

\section{Saddle point action}
\label{sec:action}

In this section, we will evaluate the $g,\sigma$ action in~\eqref{eq:g_sigma_action} for $n =2$ in the continuum limit. Since the original action of the Brownian SYK was written on the lattice, we work with the lattice coordinate $u$ and time coordinate $t$. However, it will be important to note from our discussion in section~\ref{sec:continuum} that the continuum limit corresponds to $J \rightarrow 0, \jt \rightarrow \infty$, keeping $J \jt$ fixed. Therefore, it is sufficient to evaluate the action to first order in the ratio $J/\jt$. 

For $n = 2$, the $g,\sigma$ action is given by
\begin{malign} 
- S &=  \frac{N}{2}\int \d t \sum_{u} \left[ 2 \log \text{Pf}(\partial_t - \sigma_u) - \sum_{i,j} \sigma^{i,j}_u g^{i,j}_u - \frac{J}{q}\left( \sum_{i,j}  s_i s_j (g^{i,j}_u)^q + \frac{1}{2^{q-2}}\right) \right. \\ &\quad\left.-   \frac{\jt}{2} \left(\sum_{i,j} g_{u}^{i,j} g_{u+1}^{i,j} + 1\right)\right] \\ &=  \frac{N}{2}\int \d t \sum_{u} \left(  2\log \text{Pf}(\partial_t - \sigma_u) - \sum_{i,j} \sigma^{i,j}_u g^{i,j}_u -h_u(t) \right)\,. 
\end{malign}

where $h_u(t)$ is the Hamiltonian density defined as:
\begin{malign}  \label{eq:hamiltonian_density}
   \frac{ h_u(t)}{2} & =\frac{J}{2q}\left( \sum_{i,j}  s_i s_j (g^{i,j}_u)^q +  \frac{1}{2^{q-2}}\right) - \sum_{i,j} \frac{\jt}{4}(  g_{u}^{i,j} g_{u+1}^{i,j} +1)\,. 
\end{malign}
In equation~\eqref{eq:continuum_hamiltonian}, we have already computed the continuum limit of $\sum_u h_u(t)$.  For completeness, we will reproduce the result here:
\es{}{
    \sum_u \frac{h_u(t)}{2} &= H(t)\\&=\frac{J_c}{2^{q-1}q} \int \d \sigma\,  \left(1 - x^q - z^q+ y^q\right) + \frac{1}{4}  \int \d \sigma\, \left((\partial_{\sigma} x)^2 + (\partial_{\sigma} z)^2-(\partial_{\sigma} y)^2\right)\\ &= \frac{J_c}{2^{q-1}q} \int \d \sigma \left(1 - r^q(\cos^q \theta + \sin^q \theta-1) \right) + \frac{1}{4} \int \d \sigma\,  \left((r^2 +1) (\partial_\sigma \theta)^2 - \frac{(\partial_\sigma r)^2}{r^2+1}\right)\,.
}
\subsection{The $g\cdot \sigma$ term and the Pfaffian}
Let us first consider the $g\cdot \sigma$ term. From the saddle point equation in~\eqref{eq:g_sigma}, $\sigma_u$ is:
\begin{malign} \label{eq:sigma} 
    \sigma_u^{i,j} &= - J s_i s_j (g_u^{i,j})^{q-1} - \frac{\jt}{2}  (g_{u+1}^{i,j}+ g_{u-1}^{i,j}) \\&= - J s_i s_j (g_u^{i,j})^{q-1} - \frac{\jt}{2} g_u^{i,j} \partial_u^2 g_u^{i,j} - \jt g_u^{i,j}\,.
\end{malign}
In the continuum limit the dominant contribution to $\sigma_u$ comes from $-\jt g_u^{i,j}$, hence we split it in two parts
\es{eq:sigma2}{
 \sigma_u^{i,j} &=- \jt g_u^{i,j}+ \delta \sigma^{i,j}_u\,, \\
 \delta \sigma^{i,j}_u &\equiv - J s_i s_j (g_u^{i,j})^{q-1}- \frac{\jt}{2} g_u^{i,j} \partial_u^2 g_u^{i,j}\,.
}

We can write the $g\cdot \sigma$ term in the action as:
\begin{malign}\label{gsig_writtenout}
   - S_{g,\sigma} &= \frac{N}{2} \int \d t \sum_{u}  \Tr(\sigma_u g_u) \\ &=  \frac{N}{2} \int \d t \, \left( -\jt \Tr(g_u^2) + \Tr(\delta \sigma_u g_u )\right)\,.
\end{malign}
The Paffian is 
\begin{malign} 
    \text{Pf}(\partial_t - \sigma_u) = \int \mathcal{D} \psi_i \exp\left(-\frac{1}{2}\int_0^T\sum_k (\psi_k \partial_t \psi_k - \sigma^{i,j}_u(t) \psi_i \psi_j)\right)\,.
\end{malign}
The boundary conditions of the path integral depend on $u$ as shown in~\eqref{eq:initial_state},~\eqref{eq:final_state}. We can think of the Pfaffian as the following transition amplitude~\cite{Stanford:2021bhl}:
\begin{malign}  \label{eq:Pfaffian_transition_amplitude}
    \text{Pf}(\partial_t - \sigma_u) =\frac{\sqrt{2}}{\langle{\Gamma_{f}^u,\Gamma_i^u}\rangle}  \bra{\Gamma_f^u} \mathcal T\exp\left(-\frac{1}{2}\int_0^T \d t \, \mathcal{O}_u(t)\right)\ket{\Gamma_i^u}\,,
\end{malign}
where $\ket{\Gamma_{f}^u}$ and $\ket{\Gamma_i^u}$ are defined by restricting $\ket{\Gamma_f}$ and $\ket{\Gamma_i}$ on site $u$ and 
\begin{malign} 
    \mathcal{O}_u(t) \equiv - \sum_{i,j}\sigma_u^{i,j}(t) \hat \psi_i \hat \psi_j\,.
\end{malign}

According to equation~\eqref{eq:sigma}, $\sigma \sim O(\jt)$. Therefore, the eigenvalues of $\mathcal{O}_u(t)$ differ by $O(\jt)$ and $  \mathcal{\dot O}_u(t)$ is of order  $O(J \jt)$.  Therefore, in the continuum limit, we can use the adiabatic approximation to evaluate the Pfaffian. 

To evaluate the Pfaffian in the adiabatic approximation, we need to understand the ``ground state" of $\mathcal{O}_u(t)$, the eigenstate with smallest real part of its eigenvalue. Since $\sigma_u(t)$ is a $2n\times 2n$ complex
antisymmetric matrix, its eigenvalues come in pairs $(\alpha_k,-\alpha_k)$, where we choose $\text{Re}(\alpha_k)\geq0$, and its left and right eigenvectors are not equal. We can choose the left and right eigenvectors corresponding to $\alpha_k$ to be $l_k$ and $r_k$ respectively, such that $r_k \cdot l_{k'} = \delta_{k,k'}$. Then we can can decompose $\sigma_u(t)$ as
\es{sigma_decomp}{
\sigma^{i,j}=\sum_{k=1}^n \alpha_k\le(r_k^i \,l_k^j - r_k^j \,l_k^i\ri)\,,
}
where the $u,\, t$ dependence has been suppressed in every quantity to avoid clutter. We can write $\mathcal{O}_u(t)$ as
\begin{malign}\label{eq:O_decomposition}
    \mathcal{O} = - \sum_{k=1}^n\alpha_k \left( \bar \chi_k \chi_k -\chi_k \bar \chi_k\right),
\end{malign}where $\chi_k =\sum_i l_k^i \hat \psi_i$ and $\bar \chi_k =\sum_i r_k^i \hat \psi_i$. The fermions $\chi_k,\, \bar \chi_k$ satisfy the usual anticommutation relation, but they are not related by Hermitian conjugation.

To find the eigenstates of $\mathcal{O}$, consider a ``vacuum" state $\ket{0}$ that is annihilated by $\chi_k$
\begin{malign}
    \chi_k \ket{0} = 0\,.
\end{malign}
By construction, $\ket{0}$ is an eigenstate of $\mathcal{O}$ with eigenvalue $\sum_k \alpha_k$.  Other eigenstates of $\mathcal{O}$ can be constructed by acting with $\bar \chi_k$ on $\ket{0}$. Now, it is straightforward to check that the spectrum of $\mathcal{O}$ is given by $\sum_k \pm \alpha_k$.

In the adiabatic approximation, the Pfaffian gets the dominant contribution from the eigenvalue of $\mathcal{O}_u(t)$ with smallest real part. This is given by the sum $\sum_k (-\alpha_k)$. Thus, we can write the Pfaffian as
\begin{malign} 
    \text{Pf}(\partial_t - \sigma_u) = 2 \exp\left( \frac{1}{2} \int \d t \sum_{k=1}^n \alpha_{u,k}(t) + \int \d  t \, \phi_u(t) \right)\,.
\end{malign}
The second term in the exponential is a consequence of the  non-Hermicity of $\mathcal{O}_u(t)$ and we will refer to it as the Berry phase.  It is given by
\begin{malign} 
    \int \d t \, \phi_u(t) = \int \d t\, \frac{r_u \,\partial_t \theta_u}{2}\,.
\end{malign}
We refer the reader to appendix~\ref{sec:adiabatic} for the derivation.

At leading order in $\jt \rightarrow \infty$ limit, $\sigma_u \approx - \jt g_u$. We argued below equation~\eqref{eq:g_sigma} that the eigenvalues of $g_u$ are constant in time. Since we are evaluating the eigenvalues for a traveling wave solution (that relates time derivatives to space derivatives), they are also $u$ independent.  Therefore, we can evaluate them for $g_u$ in the limit $u \rightarrow \infty$, where $g_u$ approaches the following block diagonal matrix
\begin{malign}\label{gu_asymp}
    g_u = \frac{i}{2}  \begin{pmatrix} 0 & 1 & 0 & 0 \\ -1 & 0 & 0 & 0 \\ 0 & 0& 0&  1 \\ 0 & 0 & -1 & 0 \end{pmatrix}\; .  
\end{malign}
We conclude that the eigenvalues of $g_u$ are $\pm {1/2}$ everywhere. We will use this to compute $\sum_k \alpha_k$. Let us rewrite $\sum_k \alpha_k$ as
\begin{malign}
    \sum_k \alpha_k =  \sum_k \sum_{i,j} l_k^i \, \sigma_u^{i,j} \, r_k^j = \Tr(\sigma_u \sum_k r_k l_k^T) \,. 
\end{malign}
At leading order in $J/\jt$, we can use the relation $\sigma_u = - \jt g_u$ and we get
\begin{malign}
\sum_{k} \alpha_k  = - \jt \,\Tr( g_u \sum_k r_k(g_u)  l^T_k(g_u))  = \jt \, .
\end{malign}
where $l_k(g_u)$ and $r_k(g_u)$ are the left and right eigenvectors of $g_u$ respectively with eigenvalue $-1/2$. Using perturbation theory, we can find the first subleading correction to $ \alpha_k$. This is given by \begin{malign} l_k^i(g_u) \, \delta\sigma_u^{i,j} \, r_k^j(g_u) = \Tr( \delta \sigma_u  r_k(g_u)  l^T_k(g_u))\,.
\end{malign}
Hence, we can write $\sum_k \alpha_k$ as \begin{equation}\label{eq:sum_eigenvalues}
    \sum_k \alpha_k = \jt + \Tr (\delta \sigma_u \sum_u r_k(g_u) l^T_k(g_u))\,, 
\end{equation}
We can think of $\sum_k r_k(g_u) l_k(g_u)$ as the ``projection" operator onto the eigenstates of $g_u$ with eigenvalue $-1/2$. It has a simple representation in terms of $g_u$ as
\begin{malign}
    \sum_k r_k(g_u)  l^T_k(g_u) = \frac{1}{2} -  g_u \,.
\end{malign}
Using the relation in equation~\eqref{eq:sum_eigenvalues}, we get
\begin{malign} 
    \sum_k  \alpha_k &= \jt +  \Tr \left( \,\delta\sigma_u \left(\frac{1}{2} - g_u \right) \right) \\ &= \jt - \Tr ( \delta \sigma_u g_u)\,. 
\end{malign}
Summing up the Pfaffian and the $g\cdot \sigma$ terms from~\eqref{gsig_writtenout}, we get,
\begin{malign} 
   &N\sum_u \log \text{Pf}(\partial_t -\sigma_u) - S_{g,\sigma}\\ =& N \sum_u \log 2 + N\sum_u \int \d t \left[  \, \phi_u(t) + \frac{1}{2} \left(\jt - \Tr(\delta \sigma_u g_u) \right)\right] + \frac{N}{2}\sum_u \int \d t \,  \left( -\jt \Tr(g_u^2) + \Tr(\delta \sigma_u g_u )\right) \\
  =& N\sum_u \log 2 +  N\sum_u \int \d t \, \phi_u(t) +  \frac{N \jt}{2} \sum_u \int \d t \,\le(1 - \Tr(g_u^2)\ri) \\=& N \sum_u \log 2  + N \sum_u \int \d t \, \phi_u(t) \,,
\end{malign}
where in going to the last line we used the same argument for the constancy of the eigenvalues of $g_u$ as around~\eqref{gu_asymp}.
The total  action therefore is
\begin{malign} 
    -S = \sum_u N  \left( \log 2 +   \int \d t \, \phi_u(t) -   \int \d t\, \frac{h_u(t)}{2} \right) = S_0 -S_{\text{eff}} \,,
\end{malign}
where we have defined
\begin{malign} 
    S_0 &= \sum_u  N  \log 2\,, \\ 
    -S_{\text{eff}} &= N \sum_u \int \d t  \left(\phi_u(t) - \frac{ h_u(t)}{2} \right)\,.  \label{eq:seff_Renyi2}
\end{malign}
$S_0$ cancels the overall normalization  of $\Tr(\log  \rho^2_{A \cup B \cup \bar A})$ in equation~\eqref{eq:path_integral} and we get the following expression for the second R\'enyi entropy:
\begin{malign} 
    S^{(2)} =  \log \Tr(\rho^2) = S_{\text{eff}}\,.
\end{malign}
Now we can write $S_{\text{eff}}$ in the continuum limit
\begin{malign}\label{eq:continuum_action}
    -S_{\text{eff}} &=  \frac{N}{a} \int_0^{T_\tau} \d \tau \left( \int \d \sigma\, \frac{ r \, \partial_\tau \sigma}{2} -   H(\tau) \right)\\&=\frac{N}{2 a} \int_0^{T_\tau} \d \tau\int \d \sigma \left(   r \, \partial_\tau \sigma-   h(\tau,\sigma) \right)\,.
\end{malign}
where $h(\tau,\sigma)$ is defined in~\eqref{eq:hamiltonian} as
\es{}{
   h(\tau,\sigma) =  \frac{J_c}{q 2^{q-2}} \left(1 + r^q -(r^2+1)^{q/2}  (\sin ^q \theta + \cos^q \theta) \right) + \frac{1}{2} \left((r^2+1)(\partial_\sigma \theta)^2 - \frac{(\partial_\sigma r)^2}{r^2+1}\right)\,.
}
We can follow similar steps to evaluate the saddle point action for the third R\'enyi entropy. In appendix~\ref{app:Renyi3}, we show that the dynamics of the third R\'enyi entropy has an analogous description in terms of $r, \theta$ variables defined in equation~\eqref{eq:rtheta_Renyi3} and the saddle point action takes a similar form to~\eqref{eq:seff_Renyi2} (see equation~\eqref{eq:Renyi3action}). The third R\'enyi is related to the corresponding saddle point action $S^{(3)}_\text{eff}$ by
\begin{equation}
    S^{(3)} = \frac{1}{2} \log \Tr(\rho^3) = \frac{1}{2} S_{\text{eff}}^{(3)}\,.
\end{equation}

While the higher R\'enyi entropies are all determined by similar saddle points and their on-shell effective actions, we found that for $n\geq 4$ we need to solve for more dynamical variables, than $r,\theta$ needed for $n=2,3$. We describe initial steps in the $n=4$ case in appendix~\ref{app:highern}, but we do not complete the computation of $S^{(4)}$. It would be very interesting to compute ${\cal E}^{(4)}(v)$ and especially to do the analytic continuation in $n$ down to the von Neumann value $n=1$.

Before we end this section, we would like to clarify that we have ignored the contribution to saddle point action from the regions near the domain wall $A|B$ at $t = 0$ and the domain wall $B|C$ at $t = T$. As discussed in section~\ref{sec:vlvb}, the saddle point solution in these regions differs significantly from the traveling wave solution. Since the size of these regions does not scale with $T$, their contribution to the saddle point action is $O(1)$, which can be ignored in the large $T$ limit.

\subsection{Evaluating $S_\text{eff}$ on the traveling wave solution} 
To evaluate the action on a traveling wave solution with velocity $v$, we can replace $\partial_\tau$ with $-v \partial_\sigma$ in equation~\eqref{eq:continuum_action}. Then, $S_\text{eff}$ can be written as
\es{}{ 
    -\frac{S_{\text{eff}}}{N T_\tau} &= -\frac{1}{2a} \int \d \sigma \, \left(v r\partial_\sigma \theta + h(\sigma) \right)  \\ &= -\frac{1}{2a} \int \d \sigma \, \left[v r\partial_\sigma \theta +\frac{J_c}{16} \left((r^2+1)^2 \frac{\sin^2 2 \theta}{2} - 2r^2 \right)  +\frac{1}{2}(r^2+1) (\partial_\sigma \theta)^2 - \frac{(\partial_\sigma r)^2}{2(r^2+1)}  \right] .
}
In equation~\eqref{eq:vanishing_hamiltonian}, we found that the following quantity vanishes for all $\sigma$
\begin{equation}
    \hat h =   -\frac{1}{2}(r^2+1) (\partial_\sigma \theta)^2 + \frac{(\partial_\sigma r)^2}{2(r^2+1)}+ \frac{J_c}{16} \left((r^2+1)^2 \frac{\sin^2 2 \theta}{2} - 2r^2 \right) = 0\,.
\end{equation}
Using the above equation, we can write the total action as
\begin{malign} \label{eq:seff}
    -\frac{S_{\text{eff}}}{N T_\tau} = - \frac{1}{2a} \int \d \sigma \left(v\, r \,\partial_\sigma \theta + (r^2+1) (\partial_\sigma \theta)^2 -\frac{ (\partial_\sigma r)^2}{r^2+1}\right)\,.
\end{malign}

\begin{figure}[h]
    \centering
\includegraphics[width=0.6\textwidth]{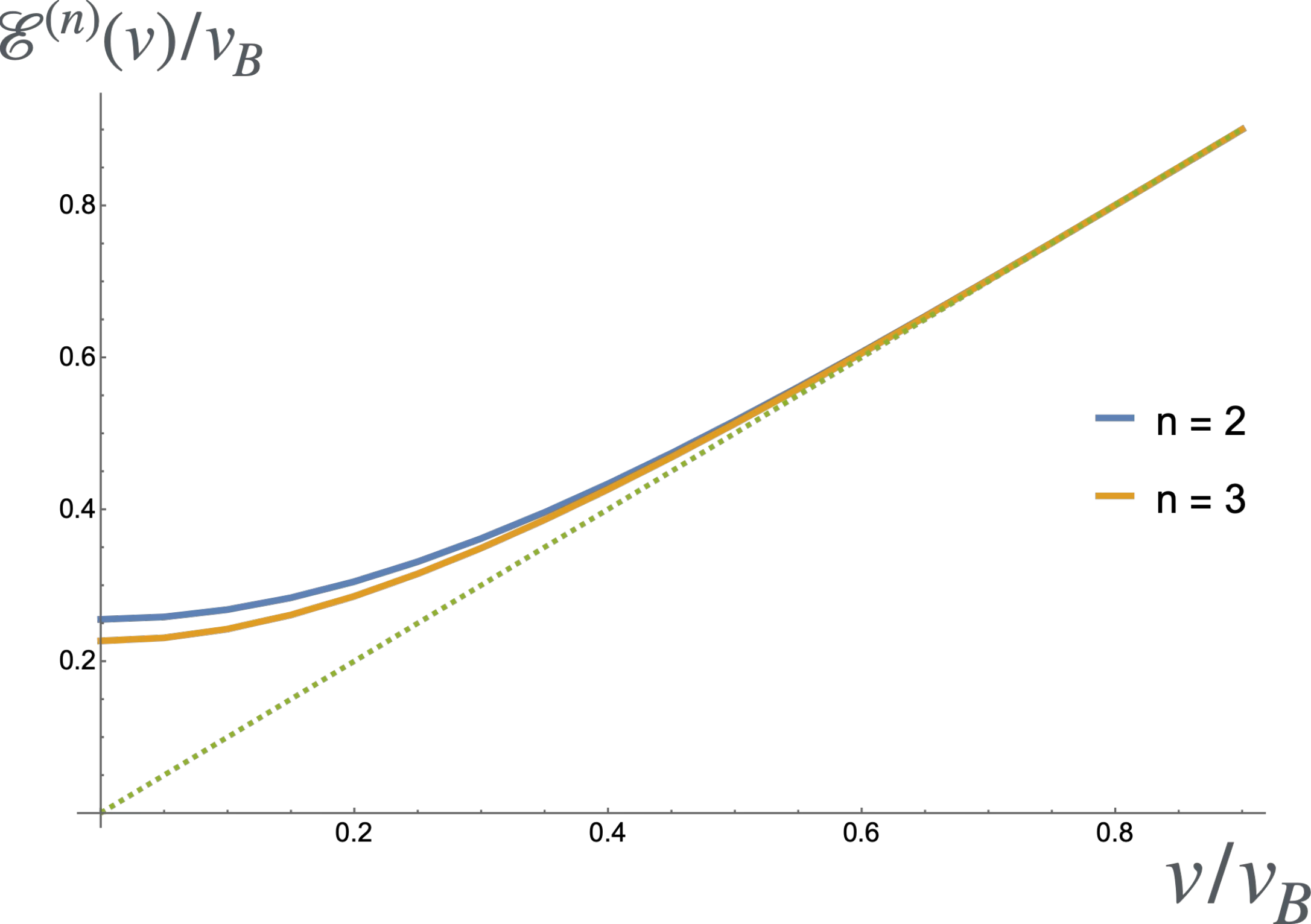} 
    \caption{Membrane tension for R\'enyi indices $n=2,3$ for the $q=4$ Brownian SYK chain. The dashed line is at $45^\circ$ and $\mathcal{E}^{(n)}(v)$ touches it at $v_B$.}
    \label{fig:membrane_tension}
\end{figure}

In figure~\ref{fig:membrane_tension} we plot the membrane tension for the second and third R\'enyi entropies for $q=4$.\footnote{See appendix~\ref{app:Renyi3} for the calculation of the third R\'enyi entropy.} It is evident from the plot that the membrane tension satisfies the conjectured properties in equation~\eqref{eq:tension_membrane}. Since the R\'enyi entropies are non-increasing functions of $n$, it follows that 
${\cal E}^{(3)}(v)\leq {\cal E}^{(2)}(v)$ should hold. Indeed, we find that our results obey this inequality, with equality at $v=v_B$.

\begin{figure}[h]
    \centering
\includegraphics[width=0.6\textwidth]{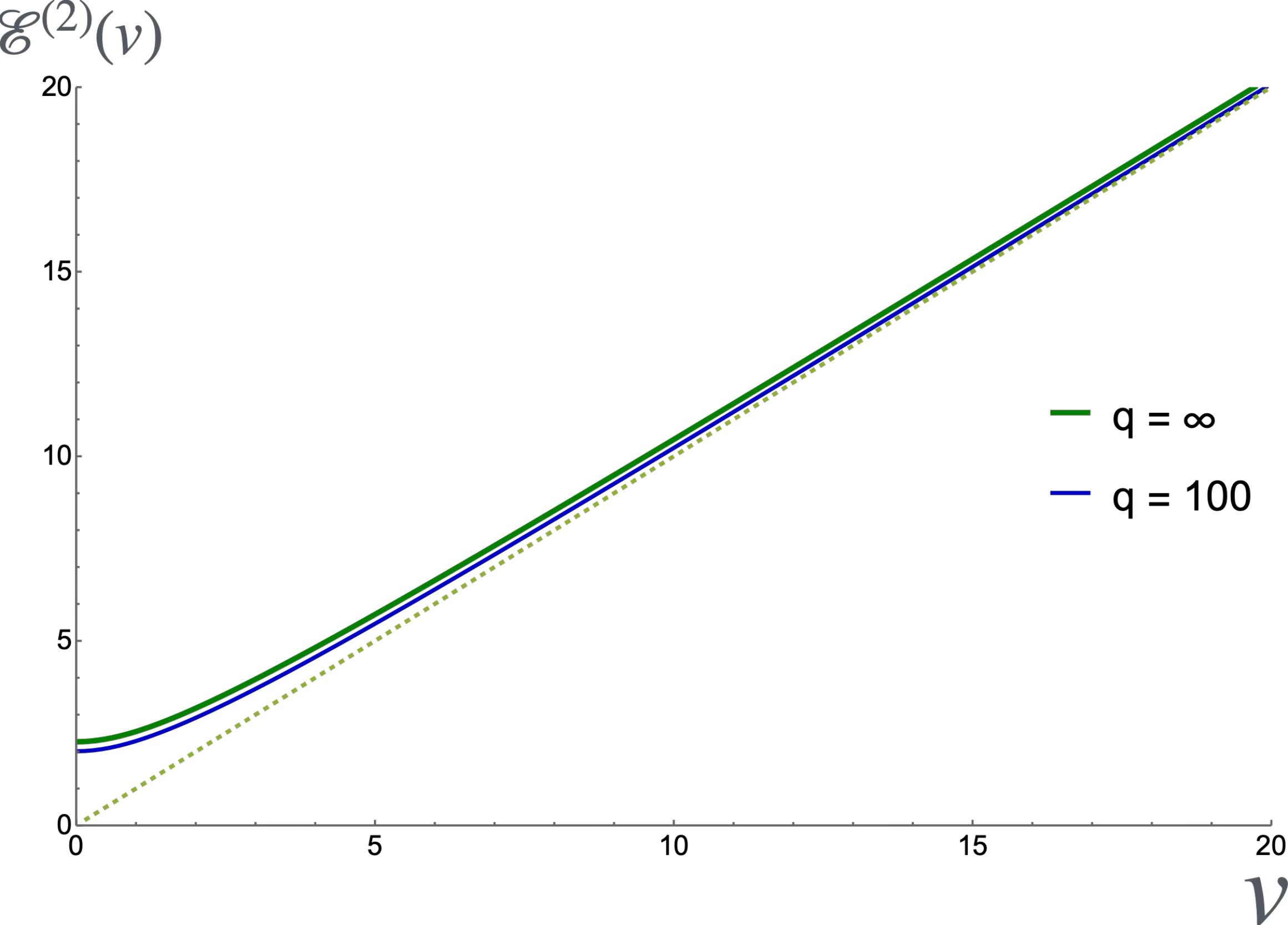} 
    \caption{Membrane tension for the second R\'enyi entropy in the large $q$ limit and $q = 100$. At $q = 100$ and $\mathcal{J} = 1$, $v_B = 140$. For large $q$, $v_B \sim q$. However, $\mathcal{E}^{(2)}(v) \approx v$  already when $1\ll v \ll q$ as shown in~\eqref{EvTaylor}.}
    \label{fig:membrane_tension_largeq}
\end{figure}

We have also evaluated~\eqref{eq:seff} analytically for the large $q$ solutions found in section~\ref{sec:largeq}. We find that for ${\cal J}=1$\footnote{For this choice $v=\tilde v$. We can restore ${\cal J}$ using dimensional analysis, $\mathcal{\widetilde E}^{(2)}(v)=\sqrt{{\cal J}} \,\mathcal{E}^{(2)}(v/\sqrt{{\cal J}})$.}
\es{E2v_largeq}{
\mathcal{E}^{(2)}(v)=-{1\ov \log(2)\,r_v}\,\le[\le(\sqrt{1+r_v^2}-V\ri)\mathrm{Im}(f)+\sqrt B \,\mathrm{atanh}\le(\sqrt{B\ov A}\ri)\ri]\,,
}
where we have defined the following quantities
\es{threequants}{
V&\equiv r_v\, v+\sqrt{1+r_v^2}\,,\qquad  A\equiv V^2-1\,, \qquad B=1-2\sqrt{1+r_v^2}\,V+V^2\,,\\[4pt]
f&\equiv \frac{\left(\sqrt{r_v-i}+i \sqrt{r_v+i} \,V\right) \mathrm{atanh}\left(\sqrt{\frac{A r_v+i B}{A (r_v+i)}}\right)}{\sqrt{A
   r_v+i B}}\,,
}
and $r_v\equiv r_v(0)$ is the solution of the transcendental equation
\es{rv0sol}{
\mathrm{Re}\le(f(v,r_v)\ri)={\pi\ov 8}\,.
}
%In deriving \eqref{E2v_largeq} we have used \eqref{rv0sol}.
We have plotted the solution of this equation in figure~\ref{fig:r0v}. We can solve this equation for small and large $v$,\footnote{And also for the special case $v=1$ where the solution is $r_v=1/\sqrt8$ and $\mathcal{E}^{(2)}(1)={\mathrm{asinh}(1)\ov \log(\sqrt2)}$.} hence we can evaluate \eqref{threequants} explicitly
\es{EvTaylor}{
\mathcal{E}^{(2)}(v)&={1\ov 2\log(2)}\le(\pi+{4-\pi\ov 2}\,v^2+{8-3\pi\ov 24}\,v^4+{52-15\pi\ov 240}\,v^6+O(v^8)\ri)\,,\\
\mathcal{E}^{(2)}(v)&=v+{\log(2e\,v^2)\ov 2\log(2)\,v}+O\le(1\ov v^2\ri)\,.
}
In figure~\ref{fig:membrane_tension_largeq} we plot the full curve together with numerical results for $q=100$ to demonstrate that it is indeed well approximated by the analytic large $q$ result.

Note however, as it is also clear from figure~\ref{fig:membrane_tension_largeq} that we have chosen to scale velocities such that $v_B=O({q})$, hence effectively infinite. If we instead scaled differently, such that $\nu_B=O(1)$ with $\nu$ denoting rescaled velocities, then we would have obtained
\es{Enu}{
\mathcal{E}^{(2)}(\nu)&=\abs{\nu}\,.
}
This is the simple consequence of the general properties of the membrane tension function and $v_E/v_B \to 0$. The same result holds for charged black holes in the extremal limit~\cite{Mezei_2017}. The same result was found in the large on-site Hilbert space limit (which they denote by $q$ and which is different from the SYK $q$) of the Brownian local GUE chain in~\cite{vardhan_moudgalya2024}.

\subsection{Second R\'enyi entanglement velocity }

If we consider the time evolution of a chain from a product state $\vert \omega\rangle$, the membrane minimising~\eqref{eq:entanglement_growth_conjecture} will be vertical with $v=0$. The   R\'enyi entropy growth rate is controlled by $\mathcal{E}^{(n)}(0)\equiv v_E^{(n)}$ which is referred to as the  R\'enyi entanglement velocity in the literature~\cite{Liu_2014}.

To evaluate the Pfaffian, we relied on the fact that the saddle point solution is a traveling wave. We used this assumption to find that the eigenvalues of $g_u$ are $u$ independent. In fact the saddle point solution also has this property at $v =0$. We can argue for this property as follows: The saddle point solutions form a continuous one-parameter family of solutions parametrised by $v$ and the eigenvalues of $g_u$ is $u$ independent for all $v \neq 0$. Since we can find the $v = 0$ solution by taking $v\rightarrow 0$ limit on this family of the solutions, the eigenvalues of $g_u$ at $v =0$ must be $u$ independent. 

To see where the analysis of the Pfaffian may differ, let us revisit equation~\eqref{eq:Pfaffian_transition_amplitude}. Since the $v = 0$ solution is time-independent, we can drop the time dependence in $\mathcal{O}_u(t)$. We have
\begin{malign}
\label{eq:Pfaffian_zero_velocity}
    \text{Pf}(\partial_t - \sigma_u) &=\frac{\sqrt{2}}{\langle{\Gamma_{f}^u,\Gamma_i^u}\rangle}  \bra{\Gamma_f^u} \mathcal T\exp\left(-\frac{\mathcal{O}_u T }{2} \right)\ket{\Gamma_i^u} \\ &\rightarrow \, \frac{\sqrt{2}}{{\langle{\Gamma_{f}^u, \Gamma_i^u}\rangle}} e^{ \frac{T}{2} \sum_k \alpha_k }  \times  \langle \Gamma_f^u , \Omega_r^u \rangle \langle \Omega_l^u , \Gamma_i^u \rangle\,,
\end{malign}
where $\ket{\Omega_r^u}$ and $\bra{\Omega_l^u}$ are the right and left ground states of $\mathcal O_u$. We used the large $T$ limit in the second step. 

As discussed in the previous section, the exponential term in the above equation cancels the contribution from the $g\cdot \sigma$ term and the last factor is the analogue of the Berry phase term. The Berry phase term has a non-trivial contribution near the boundary between $A$ and $C$,\footnote{Note that $B = \emptyset$ at $v = 0$.} where the saddle point solution is away from the fixed points. Since the size of this region does not scale with $T$, the Berry phase gives an $O(1)$ contribution, which can be ignored for large $T$. We conclude that the contribution to the saddle point action at $v =0$ only comes from the Hamiltonian density in~\eqref{eq:continuum_action}.

Now, we will use the action derived in~\eqref{eq:seff} to calculate the second R\'enyi entanglement velocity. At $v =0$, we can consistently set $r = 0$ everywhere. The reduced set of equations is
\begin{malign}
    \theta' &= - p_\theta, \quad
    p_\theta' = -\partial_\theta V\, = -\frac{J_c}{16} \sin 4 \theta .
\end{malign}
We can exactly solve the above equations
\begin{malign}
   \theta'' &= \frac{J_c}{16} \sin 4 \theta 
    \implies \theta' = \frac{\sqrt{J_c} }{4} \sin 2 \theta  \implies   \cos 2 \theta = -\tanh \frac{\sqrt{J_c}\, \sigma}{2} 
\end{malign}
To evaluate the saddle point action, we can set $r =0$ in the effective action~\eqref{eq:seff}.  We get
\es{vE_expl_comp}{
    -S_{\text{eff}} &= -\frac{N T_\tau}{2a} \int \d \sigma\, (\partial_\sigma \theta)^2 = -\frac{N T}{2}\int \d \sigma\, \frac{J_c}{16 \cosh^2 \frac{\sqrt{J_c} \sigma}{2}} = -\frac{NT \sqrt{J_c}}{8}  = - \frac{NT v_B}{8 \sqrt{2}} \\ &= - T\log d \, \frac{v_B}{4 \sqrt{2} \log 2}\,.
}
Now, we can extract the second R\'enyi entanglement velocity
\begin{malign}
    v_E^{(2)}=\mathcal{E}^{(2)} (0) = \frac{v_B}{4 \sqrt{2} \log 2}\,.
\end{malign}
This value agrees with the result ref.~\cite{Swann:2023vpg} found upon converting conventions.

As expected from general constraints on the membrane tension~\eqref{eq:membrane_conjecture}, the second R\'enyi entanglement velocity is smaller than the butterfly velocity. It is also greater than the third R\'enyi entanglement velocity $v_E^{(3)}=\mathcal{E}^{(3)} (0) = \frac{2v_B}{9 \sqrt{2} \log 2}$ (computed in appendix~\ref{app:Renyi3}), as required by the ordering of R\'enyi entropies by $n$.

\subsection{R\'enyi entropy for $v > v_B$} \label{subsec:Renyi_entropy_vgvb}
In section~\ref{sec:vgvb}, we found that the Brownian SYK chain does not have a membrane like solution for $v > v_B$. Therefore,  we must not associate the R\'enyi entropies for $v > v_B$ to the membrane tension in this model. We use the notation $\mathcal{E}^{(n)}_{\text{eff}}(v)$ to denote the $n$-th R\'enyi entropy for this case. 

For $v > v_B$, we found an implicit form of the solution in~\eqref{eq:rtheta_vb}:
\begin{equation}
    r(\theta) = \begin{cases} 
    \tan \theta, \quad 0 < \theta < \frac{\pi}{4}\,, \\ 
    \cot{\theta}, \quad \frac{\pi}{4} < \theta < \frac{\pi}{2} \,.
    \end{cases}
\end{equation}
This is sufficient to calculate $\mathcal{E}^{(2)}_\text{eff}(v)$ because in this case the Hamiltonian density vanishes. The nontrivial contribution comes from the Berry phase term. 
\begin{malign}
    S^{(2)} = S_{\text{eff}} = \frac{N}{2 a}   \int \d \sigma \int \d \theta\, r  = N |B| 
    \log \sqrt{2} = v T 
    \log d\,.
\end{malign}
Thus, \begin{equation}\label{eq:tension_vgvb} \mathcal{E}^{(2)}_{\text{eff}}(v) = v\quad \text{for} \quad v > v_B\,. \end{equation}

For $v > v_B$, we can calculate the Pfaffian in equation~\eqref{eq:Pfaffian_transition_amplitude} without using the adiabatic approximation; this computation is performed in appendix~\ref{app:vgvb} and gives the same result as~\eqref{eq:tension_vgvb} without the adiabatic approximation.

In section~\ref{sec:1_on_correction}, we argued that the $v>v_B$ solution receives nontrivial quantum correction when $T \sim \log N$. However, one must note that such quantum corrections do not change the classical action. The reason is that nonzero contribution to the R\'enyi entropy for $v > v_B$ comes from the path integral of fermions in region $B$. As shown in figure~\ref{fig:density_matrices}, the contour of fermionic path integral in $B$ is an OTOC contour. Without an operator insertion, the time evolution along an OTOC contour is trivial. Therefore, the partition function along an OTOC contour is also time-independent. Hence, quantum corrections do not affect the R\'enyi entropy for $v > v_B$.

We can understand the result~\eqref{eq:tension_vgvb} from the operator growth picture as follows:  We are interested in the second R\'enyi entropy of $A \cup B \cup \bar A$ in the state $\ket{\Psi(T)}$. In appendix~\ref{sec:purity_prob}, we show that $\dim \cH_B \times \Tr_{A \cup B \cup \bar A}(\rho^2)$ is the probability that a typical operator $\mathcal O$ in $A$ at $t = 0$, under time evolution by the unitary operator $U(T)$, evolves to an operator with support in $A \cup B$ \cite{Mezei_Stanford_2017}. A typical operator with support in $A$ spreads on the chain away from $A$ with the butterfly velocity $v_B$. If $|B| > v_B T$, the probability that such an operator is supported inside $A \cup B$ is close to 1. Therefore, \begin{malign}
    \dim \cH_B \times \Tr_{A \cup B \cup \bar A}(\rho^2_{A \cup B \cup \bar A}) &\approx 1  \\ 
  \implies  \Tr_{A \cup B \cup \bar A}(\rho^2_{A \cup B \cup \bar A}) &\approx \frac{1}{\dim \cH_B}\,,
\end{malign}
and we get the result in equation~\eqref{eq:tension_vgvb}.

\section{Discussion and open directions}\label{sec:discussion}

We computed the second and third Rényi membrane tensions of the Brownian SYK chain in the  $N\rightarrow\infty$ and the continuum limit. We found that the membrane has a description in terms of a soliton that moves with velocity $v$. For velocities $v<v_B$, the width of the soliton depends on $v$ and diverges as $v\rightarrow v_B$. For $v>v_B$, the soliton splits into two fronts, each propagating at the butterfly velocity $v_B$, with their separation growing linearly with time, $\Delta u \sim (v-v_B)T$. Therefore, the membrane description ceases to exist for $v > v_B$.

It is interesting to note that the absence of a membrane picture for $v > v_B$ is irrelevant for entanglement growth in the quench problem discussed at the beginning of section~\ref{sec:membrane_review}. To see this, consider the membrane formula in equation 
\eqref{eq:entanglement_growth_conjecture}:
\begin{malign} \label{eq:egc}
   & S^{(n)}(x,T) = \min_v \,(s_\text{eq}\,\mathcal{E}^{(n)}(v) T + S^{(n)}(x- v T,0))\,.
\end{malign}
Since $\mathcal{E}^{(n)}(v) = v$ for $v > v_B$, the argument of minimization in the above formula can be written as:
\begin{malign} 
    s_\text{eq} T \left(  v \,  + \frac{1}{s_\text{eq} T}S^{(n)}(x - vT) \right) \quad \text{for} \quad v> v_B\,.
\end{malign}
Since the Brownian SYK chain thermalises to an infinite temperature state, $s_\text{eq} = \log d$ (the on-site Hilbert space dimension). On the other hand, $S^{(n)}(x,T)$ cannot vary faster than $\log d$ as a function of $x$. Hence, the above function increases monotonically for $v >v_B$. We conclude that the argument in~\eqref{eq:egc} does not have a minimum in the range $v > v_B$.  Similarly, the argument does not have a minimum in the range $v < -v_B$. Therefore, it must have a minimum in the range $ -v_B < v < v_B$, where the membrane description exists. However, we recall that  the $v>v_B$ saddles, where the membrane splits into two wavefronts, are relevant for the computation of operator entanglement entropy.

As discussed at the end of section~\ref{sec:vgvb}, a new fixed point emerges at $(x,y,z) = (1,1,1)$ in the large $N$ limit.  When $v > v_B$, the saddle point solution is close to this fixed point in region $B$ for the spatial length that scales as $(v - v_B)T$. The fixed point in $B$ is unstable under fluctuations from $1/N$ corrections. The fluctuations are important when $T \sim \log N$. It would be interesting to understand their consequences at late times. The $1/N$ corrections however do not affect the membrane picture for $v < v_B$. This is because in this case the saddle point solution is away from the fixed points $(x,y,z) = (1,0,0)$ and $(x,y,z) = (0,0,1)$ only for times of $O(N^0)$ at any site in $B$. In summary, the fluctuations are only relevant in the regime where the membrane description breaks down. Indeed, we expect that the membrane picture is valid for $T\gg 1/J$ independent of how $T$ compares to $\log(N)$.

\begin{figure}[h]
    \centering
\includegraphics[width=0.5\linewidth]{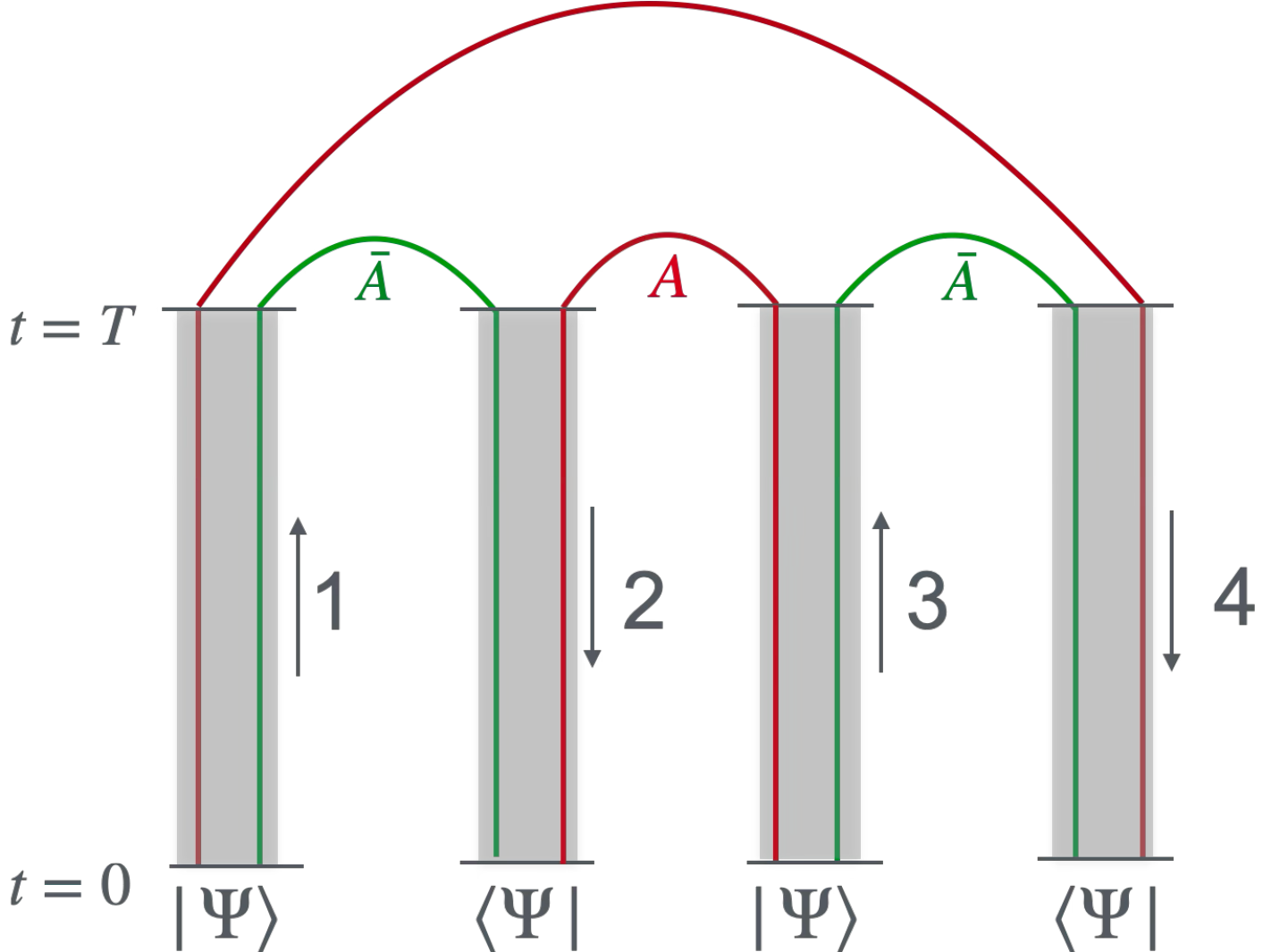}
    \caption{Schwinger-Keldysh contour for $\Tr(\rho^2_{A})$ of a pure initial state $\ket{\Psi}$ of the chain. Since the legs at $t =0$ are disconnected, there does not exist a simple description in $x,y,z$ variables.}
    \label{fig:pure_state}
\end{figure}

\begin{figure}[h]
    \centering
\includegraphics[width=0.99\linewidth]{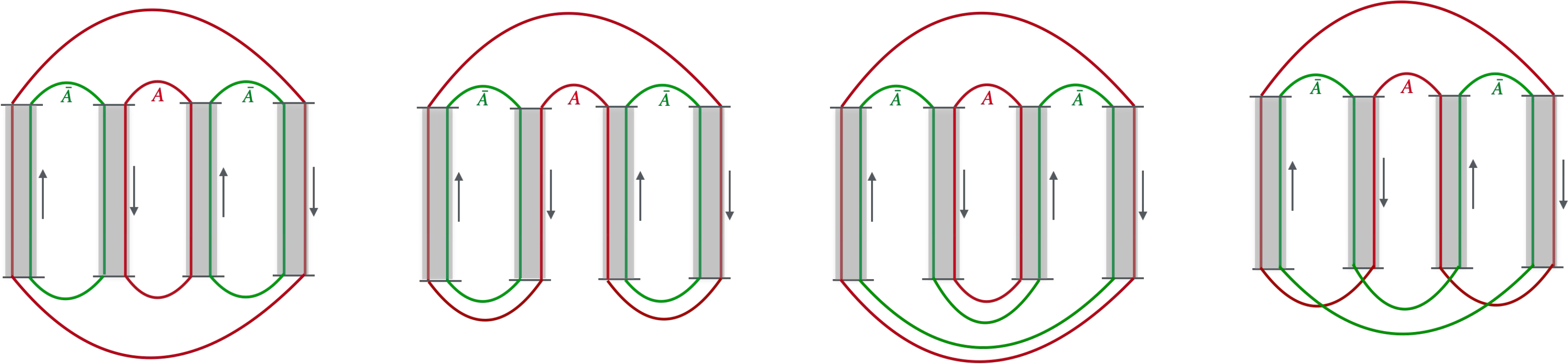}
    \caption{An example of the ensemble of states is $U_A U_{\bar A} \ket{\Psi}$ where $U_A, U_{\bar A}$ are Haar random unitaries. After averaging over $U_A, U_{\bar A}$ in $\Tr(\rho^2_{A})$, we get a sum of four contours. The boundary conditions of each contour have a simple description in $x,y,z$.}
\label{fig:pure_state_averaging}
\end{figure}
In this paper, we only focused on computing the R\'enyi membrane tension in the Brownian SYK chain. This calculation required a very special initial state, namely, a maximally entangled state between two copies of the chain. However, the membrane formula in equation~\eqref{eq:entanglement_growth_conjecture} allows us to calculate entanglement growth of subsystems in more general states (for example, a pure state of a single chain). In such problems, there may not exist a simple description of the initial state in terms of the boundary conditions of the $x,y,z$ variables used here. We illustrate this point in figure~\ref{fig:pure_state}. A simple way to deal with this problem is to consider entanglement growth in an ensemble of states related to one another by random unitary operators.\footnote{We thank Onkar Parrikar for this suggestion.}${}^,$\footnote{The choice of the ensemble of unitaries may depend on the initial pattern of entanglement between different subregions.} For the second R\'enyi entropy, an ensemble average would produce a family of contours similar to the one shown in figure~\ref{fig:pure_state_averaging}. The boundary conditions on each contour has a simple description in terms of the $x,y,z$ variables. 
Moreover, the family of contours would proliferate with the index $n$ of the R\'enyi entropy.  In the large $N$-limit, the dominant contribution to the entropy may come from one of these contours and the dominant contour may also change as the system evolves in time. It would be interesting to understand how a membrane description emerges in this scenario.

It is also interesting to consider charged Brownian SYK chains where the dynamics preserves a global $U(1)$ charge \cite{Rakovszky2019SubBallistic, ZhouLudwig2020DiffusiveScaling, Znidaric2020DiffusiveSystems,HanChen2023U1Automaton,BernardPiroli2021QSSEP,Zhang:2023vpm}. Zhang \cite{Zhang:2023vpm} found the dependence of the entanglement velocity on the local charge density in a perturbative treatment of the fermion hopping strength. Using an analogue of the $x,y,z$ variables in the charged Brownian SYK chain, we can find the large $N$ equations of motion that governs the entanglement dynamics in this model. Such a description would be nonperturbative the hopping strength. We expect that the large $N$ equations of motion provide a description of entanglement membranes coupled to the local charge density.  It would also be interesting to derive the entanglement membrane in the SYK chain (with time-independent disorder) that conserves energy, building on the works~\cite{Gu_lucas_2017,zhang2022syk}. 

\section*{Acknowledgments}

We thank Adam Nahum, Onkar Parrikar, Shreya Vardhan, and Tianci Zhou for useful discussions and comments on our draft. %MM is supported in part by the STFC grant ST/X000761/1. 
MM is supported by the ERC Consolidator Grant GeoChaos-101169611. HR acknowledges support from the Department of Atomic Energy, Government of India, under project identification number RTI 4002, and from the Infosys Endowment for the study of the Quantum Structure of Spacetime. HR also acknowledges support from grant NSF PHY-2309135 to the Kavli Institute for Theoretical Physics (KITP).

This work is funded by the European Union. Views and opinions expressed are however those of the authors only and do not necessarily reflect those of the European Union or the European Research Council Executive Agency. Neither the European Union nor the granting authority can be held responsible for them.

For the purpose of open access, the authors have applied a CC BY public copyright licence to any Author Accepted Manuscript (AAM) version arising from this submission.

\appendix
\section{Replica symmetry}\label{sec:replica_symmetry}

In section~\ref{sec:Renyi2}, we mapped the calculation of $\Tr(\rho^2)$ to a transition amplitude between initial and final states. We found that the effective dynamics is described on four copies of the Hilbert space $\cH$. In general, $\Tr(\rho^n)$ can be mapped to a transition amplitude in $2n$ copies of $\cH$, namely, $\left( \cH\otimes \cH^*\right)^{\otimes n}$. In this section, we will use this description to discuss the action of replica symmetry on the fermion two-point functions.

For this discussion, it is sufficient to consider any two flavors of Majorana fermions acting on $\cH$ at a given site in the Brownian SYK chain. Let us denote them by $\psi^1$ and $\chi^1$. We can represent $\rho$ as a state in $\cH \otimes \cH^*$ using a second copy of fermions which we denote by $\psi^2$ and $\chi^2$.  To represent $n$ copies of $\rho$ as a state, we need the Hilbert space $(\cH \otimes \cH^*)^{\otimes n}$. We label the pair of fermions acting on the $i$-th Hilbert space in  $(\cH \otimes \cH^*)^{\otimes n}$ by
\begin{equation}
    \psi^i, \chi^i \quad \text{for} \quad  1 \leq i \leq 2n \,.
\end{equation}
 
Two Majorana fermions generate the Hilbert space of qubit. Therefore, $(\cH \otimes \cH^*)^{\otimes n}$ is the Hilbert space of $2n$ qubits. We will use the following Pauli spin representation of the Majorana fermions in the Hilbert space of $2n$ qubits: 
\begin{malign}
   &\psi^{2k-1} = \frac{1}{\sqrt{2}}\left( \prod_{i = 1}^{2k-2} Y_{i} \right)X_{2k-1}\,, \qquad \psi^{2k} = \frac{1}{\sqrt{2}}\left( \prod_{i = 1}^{2k-1} Y_{i} \right) Z_{2k}\,, \\
    &\chi^{2k-1} = \frac{1}{\sqrt{2}} \left( \prod_{i = 1}^{2k-2} Y_{i} \right)Z_{2k-1}\,, \qquad \chi^{2k} =  \frac{1}{\sqrt{2}}\left( \prod_{i = 1}^{2k-1} Y_{i} \right)X_{2k} \,.
\end{malign}
The subscript on the Pauli matrices labels the Hilbert space on which they act. 
In section~\ref{sec:Renyi2}, for $n=2$ we saw that the effective dynamics generated by the ensemble average over the couplings conserves the flavour-wise parity operator. For general $n$, the following parity operators are conserved:
\begin{align}
    P_\psi = \prod_{k =1}^{n}  -2i \psi_{2k-1} \psi_{2k} = \prod_{k = 1}^{2 n} Z_{k}, \qquad P_\chi = \prod_{k =1}^{n}  -2i \chi_{2k-1} \chi_{2k} = (-1)^n\prod_{k = 1}^{2n}  X_{k}\,.
\end{align}
In particular, their product \begin{equation}
    P_y =  \prod_{k =1}^{2n}  Y_{k} 
\end{equation} is conserved. The initial and final states satisfy
\begin{malign}
    P_\psi \ket{\Gamma_i} = P_\chi \ket{\Gamma_i} = P_y \ket{\Gamma_i} = \ket{\Gamma_i},  \qquad P_\psi \ket{\Gamma_f} = P_\chi \ket{\Gamma_f} = P_y \ket{\Gamma_f} = \ket{\Gamma_f}\,. 
\end{malign}

Replica symmetry corresponds to permutations of density matrices in $\Tr(\rho^{n})$. Thus, the action of replica symmetry for $\Tr(\rho^n)$ is given by the permutation group $S_n$. Although, $\Tr(\rho^n)$ is invariant under arbitrary permutations of $\rho$, any specific choice of ordering of $\rho$'s in $\Tr(\rho^n)$ breaks replica symmetry down to cyclic permutation of $\rho$. Therefore, correlation functions of fermions are only preserved under cyclic permutations of $\rho$.  

Let us represent a state in $\left(\cH \otimes \cH^*\right)^{\otimes n}$ as 
\begin{malign}
  \ket{\alpha}  =  \prod_{k =1}^n\ket{a_{2k-1}}  \ket{a_{2k}} \,.
\end{malign}
We define the set of cyclic permutation operators $\{\pi_\sigma^{(m)}: 1 \leq m \leq n-1 \}$ whose elements act on $\ket{\alpha}$ as
\begin{malign} 
    \pi_\sigma^{(m)} \ket{\alpha} = \prod_{k = 1}^n \ket{a_{2 \sigma(k)  - 1}} \ket{a_{2 \sigma(k)}},\quad \sigma(k) = m + \le[k \pmod n\ri]\,.
\end{malign}
Another symmetry of interest is transposition of the density matrix: $\rho \rightarrow \rho^T$
\begin{malign}
    \Tr(\rho^n) = \Tr((\rho^T)^n)
\end{malign}
Transposition acts on $\ket{\alpha}$ as
\begin{align}
    T \ket{\alpha} = \prod_{k =1}^n \ket{a_{2k}} \ket{a_{2k-1}}\,.
\end{align}
We can compose transposition with the appropriate element of the permutation group to define  a reflection symmetry $R$ that acts as:
\begin{malign}
    R \ket{\alpha} = \prod_{k = 1}^n \ket{a_{2(n-k+1)}} \ket{a_{2(n-k)+1} }\,.
\end{malign}
The reflection symmetry corresponds to the transformation:
\begin{equation}
    \rho^n \rightarrow (\rho^n)^T\,.
\end{equation}
For $n = 2$, the reflection symmetry corresponds to flipping the second R\'enyi contour in figure~\ref{fig:density_matrices} horizontally about its center.  

In the large $N$ limit of the Brownian SYK chain, we assume that the dominant saddle point solution is symmetric under cyclic permutations and reflection. This assumption imposes nontrivial constraints on the two point function $\psi^i \psi^j$.  

Consider the action of cyclic permutation on the following operator:
\begin{malign}\label{PauliRep}
 -2i   \psi^{2i-1} \psi^{2j} = Z_{2i-1} \left( \prod_{k =2i}^{2j-1} Y_k\right) Z_{2j}\,, \quad \text{for} \quad i< j\,.
\end{malign}
Under a cyclic permutation, $-2i \psi^{2i-1} \psi^{2j}$ is mapped to
\begin{malign}
   \pi_{\sigma}^{(m)\,\dagger}  \left(-2i \psi^{2i-1} \psi^{2j}   \right)  \pi_{\sigma}^{(m)} = \begin{cases}
       Z_{2r-1} \left( \prod_{k =2r}^{2s-1} Y_k\right) Z_{2s}\,,  &\quad  r < s\,,\\ 
    \left(\prod_{k = 1}^{2s-1}Y_{k}\right)  Z_{2s} \, Z_{2r-1} \left( \prod_{k =2r}^{2n} Y_k\right), &\quad  s < r\,, \\ 
   \end{cases}  
\end{malign}
where to get these results we applied the permutation on the right-hand side of~\eqref{PauliRep}. We have defined $r,s$ as
\begin{malign}
r =1 + \le[(i + m-1) \pmod n\ri] \,, \qquad s = 1 + \le[(j + m-1) \pmod n\ri] \,.
\end{malign}

We can rewrite the action of cyclic permutation in terms of fermion bilinears as
\begin{malign}
   \pi_{\sigma}^{(m)\,\dagger}  \left(-2i \psi^{2i-1} \psi^{2j}   \right)  \pi_{\sigma}^{(m)} = \begin{cases}
      -2i \,\psi^{2r-1} \psi^{2s}\,,  &\quad  r < s\,,\\ 
  -2 i\, \psi^{2s} \psi^{2r-1} P_y\,, &\quad  s < r\,. \\ 
   \end{cases}  
\end{malign}
Since $P_y = 1$ on the subspace of interest, we get the following constraint by imposing invariance of the two point function under cyclic permutation:
\begin{malign}
   \langle \psi^{2i-1} \psi^{2j}\rangle = \begin{cases} \langle \psi^{2r-1}\psi^{2s}  \rangle \quad r <s\,, \\   \langle \psi^{2s} \psi^{2r-1}\rangle \quad s<r\,. \end{cases}
\end{malign}

Similarly, we can consider the action of cyclic permutation on other two-point functions. We get the following constraint:
\begin{malign}
    \langle \psi^i \psi^j\rangle = (-1)^{f(r,s)}\langle  \psi^r \psi^s\rangle\,,
\end{malign}
where \begin{malign}
    r = 1 +  \le[(i + 2 m -1) \pmod n\ri]\,,  \qquad   s = 1 +  \le[(j + 2 m -1) \pmod n\ri] \,,
\end{malign}
and \begin{malign} 
    f(r,s) = \begin{cases}
        1 \quad &r< s \,,\\
        -1 \quad  &r> s\,.
    \end{cases} 
\end{malign}
Now consider the action of reflection on
\begin{malign}
   -2i \psi^{2i -1} \psi^{2j-1} &= Z_{2i-1} \left( \prod_{k = 2i}^{2j-2} Y_k\right) X_{2j-1}\,,\\
R^\dagger \left( -2i \psi^{2i -1} \psi^{2j-1} \right) R  &=   X_{2r} \left( \prod_{k = 2r+1}^{2s-1} Y_k\right)  Z_{2s} = 2i \psi^{2r}\psi^{2s}\,.
\end{malign}
where \begin{malign} 
    r = n-j + 1\,, \qquad s = n-i+1 \,.
\end{malign}
By acting with an appropriate cyclic permutation, we can map $\psi^{2r} \psi^{2s}$ to $\psi^{2i} \psi^{2j}$. Thus, reflection symmetry imposes the following constraint:
\begin{malign}
   \langle  \psi^{2i - 1} \psi^{2j-1} \rangle = - \langle \psi^{2i} \psi^{2j} \rangle\,.
\end{malign}

\section{The butterfly velocity from the FKPP equation}\label{sec:fkpp}

In this appendix, we sketch the derivation of the butterfly velocity from the FKPP equation. It is convenient to work with equation~\eqref{eq:future_fkpp}, so that the traveling wave propagates forward in time:  
\begin{equation}
    \partial_t y_u = -\frac{J}{2^{q-2}}(y_u - y_u^{q-1}) + \jt \partial_u^2 y_u\,.
\end{equation}
We are interested in the solution to this equation when the initial condition is a step function:
\begin{equation}
    y_u(0) = \begin{cases}
        0 \quad u \leq 0\,, \\ 
        1 \quad u > 0\,.
    \end{cases}
\end{equation}
The FKPP equation has two fixed points: (i) a stable fixed point at $y_u = 0$ and (ii) an unstable fixed point at $y_u = 1$. When the initial condition is a step function, we expect that a traveling wave solution propagates towards the unstable domain.
To determine its velocity, it is sufficient to linearise the FKPP equation near $y_u = 1$. To avoid cumbersome notation, let us define $f_u = 1- y_u$. Then $f_u$ satisfies:
\begin{equation}\label{eq:fkpp}
    \partial_t f_u =  \frac{J}{2^{q-2}} (1 - f_u) (1 - (1 -f_u)^{q-2}) + \jt \partial_u^2 f_u\,.
\end{equation}
The linearised equation in $f_u$ is
\begin{equation}\label{eq:fkpplinear}
    \partial_t f_u = \frac{J (q-2)}{2^{q-2}} f_u + \jt \partial_u^2 f_u \,.
\end{equation}
and the boundary condition is
\begin{equation}
    f_u(0) = \begin{cases}
        1 \quad u \leq 0\,, \\ 
        0 \quad u > 0\,.
    \end{cases}
\end{equation}
Notice that 
\begin{equation}
    \frac{J (q-2)}{2^{q-2}} f_u > \frac{J}{2^{q-2}} (1 - f_u) (1 - (1 -f_u)^{q-2}) \quad \text{for} \quad 0< f_u <1\,. 
\end{equation}
Therefore, the solution to equation~\eqref{eq:fkpplinear} provides an upper bound on the solution to equation~\eqref{eq:fkpp}.
The solution to the linearised equation is
\begin{equation}
    f_u(t) = e^{\frac{J (q- 2)}{2^{q-2}}t} \int_{-\infty}^0 \frac{\d s}{\sqrt{2 \pi \jt t}} e^{-\frac{(u - s)^2}{2\jt t}}\,.
\end{equation}
To locate the wavefront, let us replace $u$ by $ r +  v\,t $. We get:
\begin{equation}
    f_{r + v t}(t) = e^{ \left( \frac{J (q- 2)}{2^{q-2}} - \frac{v^2}{2 \jt }\right)t}   \int_{-\infty}^0 \frac{\d s}{\sqrt{2 \pi \jt t}} e^{- \frac{ v (r-s)}{\jt} - \frac{(r-s)^2}{2 \jt t}}
\end{equation}
The exponential dependence of $f_{r+vt}$ in $t$ vanishes if and only if \begin{equation}
    v = v_B \equiv \sqrt{\frac{J \jt (q-2)}{2^{q-1}}}\,.
\end{equation} 

The above analysis only provides an upper bound on the velocity of the traveling wave solution to the FKPP equation.  To put a lower bound, we take a traveling wave ansatz:
\begin{equation}
    f_u(t) = g(u -vt)\,.
\end{equation}
Then, $g$ satisfies the following ordinary differential equation:
\begin{equation}
    - v\, \partial_u g(u)  =  \frac{J}{2^{q-2}} (1 - g(u)) (1 - (1 -g(u))^{q-2}) + \jt \partial_u^2 g(u)\,.
\end{equation}
We require that the solution satisfies the following constraint:
\begin{equation}
    g_u \rightarrow \begin{cases}
      0\,, \quad   u \rightarrow -\infty\\
      1\,, \quad u \rightarrow +\infty
    \end{cases} \qquad 0 < g_u <1,.
\end{equation}
It turns out that such a solution exists only when $v \geq v_B$. See \cite{BramsonLectures} for a detailed explanation.  

The velocity of the traveling wave solution is both upper and lower bounded by $v_B$. If the solution approaches a traveling wave, then it must propagate with the velocity $v_B$.  

\section{Traveling wave solution with $v>v_B$ for arbitrary replica index}\label{app:vgvb}

In section~\ref{sec:vgvb}, we described the traveling wave solution for the case $v> v_B$ for $n= 2$.  We found that the solution contains three  different domains separated by domain walls $A|B$ and $B|C$. In this section, we will extend the $n = 2$ solution to arbitrary replica index. 
\begin{figure}[h]
    \centering
\includegraphics[width=0.9\textwidth]{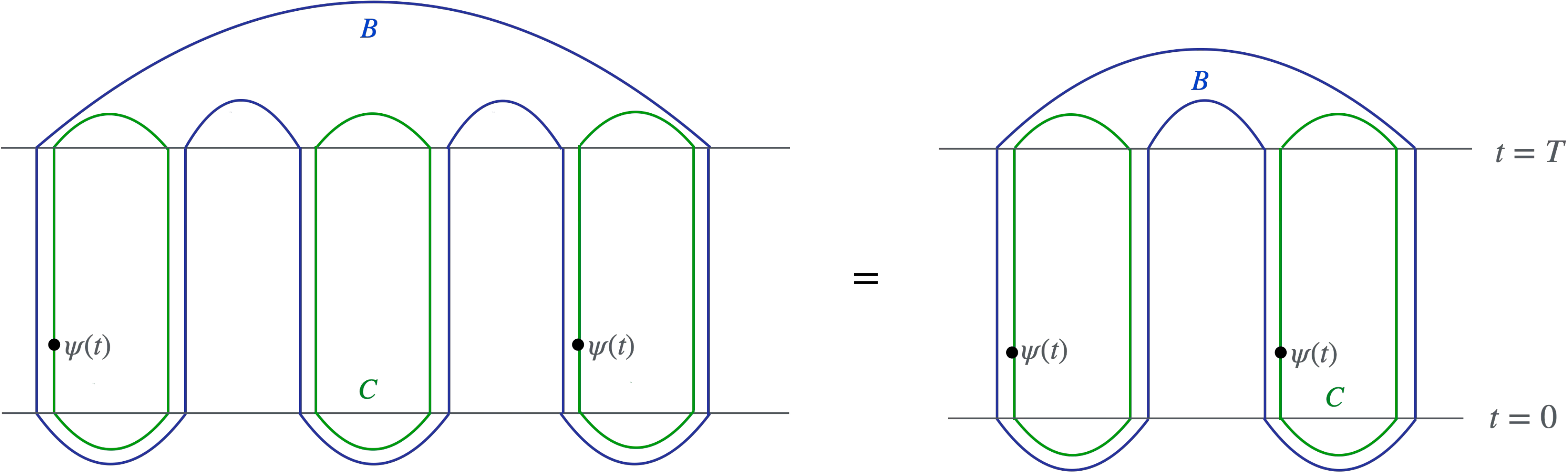} 
    \caption{Contour for $\Tr(\rho^3(T)_{B})$ ignoring region $A$. Since the operators are  inserted on legs $1$ and $5$, we can shrink legs $3$ and $4$ without changing the correlation function.} 
    \label{fig:Renyi_BC}
\end{figure}

The main idea is that when $v>v_B$, we can ignore the influence of $A$ to determine correlation functions near $B|C$ and vice versa. Now consider the situation when $A$ is completely ignored. Then, the chain $R$ (respectively  $L$) contains $B \cup C$ (respectively  $\bar B \cup \bar C$) and the calculation of $n$-th entropy would reduce to evaluating the path integral shown in figure~\ref{fig:Renyi_BC}. Consider computing the equal time two point function between operators inserted on any two legs on the contour. Notice that a contour with no operator insertion corresponds to the forward time evolution of a maximally entangled state. Since a maximally entangled state does not evolve in time, we can shrink such a contour without changing the two point function.  

The above argument allows us to compute all equal time two point functions for arbitrary $n$ in terms of equal time function for $n = 2$. Using the $t = 0$ boundary condition, we have the following relation for two point functions near the domain wall $B|C$. 
\begin{malign} 
 &\text{For}~1 \leq k < m \leq n : & \quad  g^{2k-1, 2m-1}_u &= -\frac{y_u}{2} \,, \\ 
  &&  g^{2k-1, 2m}_u &= g^{2k,2m-1}_u = -i\frac{y_u}{2} \,,\\ 
&&    g^{2k,2m}_u &= \frac{y_u}{2}\,,\\
&\text{for}~k = m: & g^{2k-1,2k}_u &= {i\ov2}\,.
\end{malign}

\begin{figure}[t]
    \centering
\includegraphics[width=0.9\textwidth]{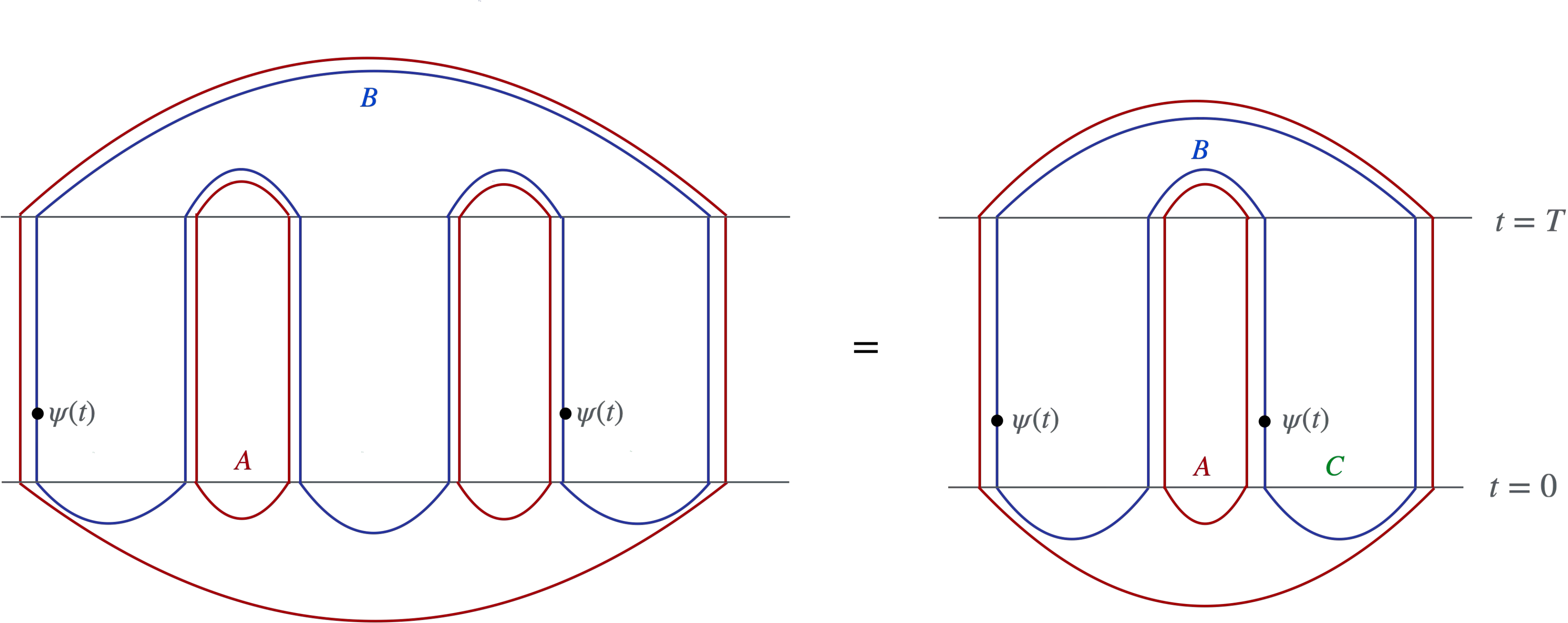} 
    \caption{Contour for $\Tr(\rho^3(T)_{A \cup B \cup \bar A})$ ignoring region $C$. Since we have inserted operators on legs $1$ and $5$, we can shrink legs $2$ and $3$ without changing the correlation function.} 
\label{fig:Renyi_AB}
\end{figure}
Now, consider the region near the domain wall $A|B$. If we ignore the influence of region $C$ in the calculation of the $n$-th R\'enyi entropy, we get the path integral contour shown in figure~\ref{fig:Renyi_AB}. In this case, the state defined at $t = T$ does not evolve under backward time evolution. Now we can use similar arguments discussed above to fix arbitrary equal time functions near the domain wall $A|B$. We get, 
\begin{malign} 
 &\text{For}~0 \leq k < m \leq n-1: & \quad    g^{2k+1, 2m+1}_u &= -\frac{y_u}{2} \, ,\\ 
&& g^{2k+1, 2m}_u &= g^{2k+2,2m+1}_u = g^{2m+1,2n} =i\frac{y_u}{2}\,, \\ 
&&    g^{2k,2m}_u &= \frac{y_u}{2}\,,\\ 
&\text{for}~k = m: &\quad  g_u^{2k,2k+1} &=  {i\ov2} \,.
\end{malign}
It can be  checked that the above solution satisfies the $g,\sigma$ equations. 

\section{Scaling of the membrane width near $v = v_B$}
\label{app:wavesize}

In section~\ref{sec:vlvb}, we found that $r_v(0) \rightarrow 1$ as $v \rightarrow v_B$. Here we provide a partial analysis of how the solutions approach this limit.

\begin{figure}[!h]
    \centering
    \includegraphics[width=0.49\linewidth]{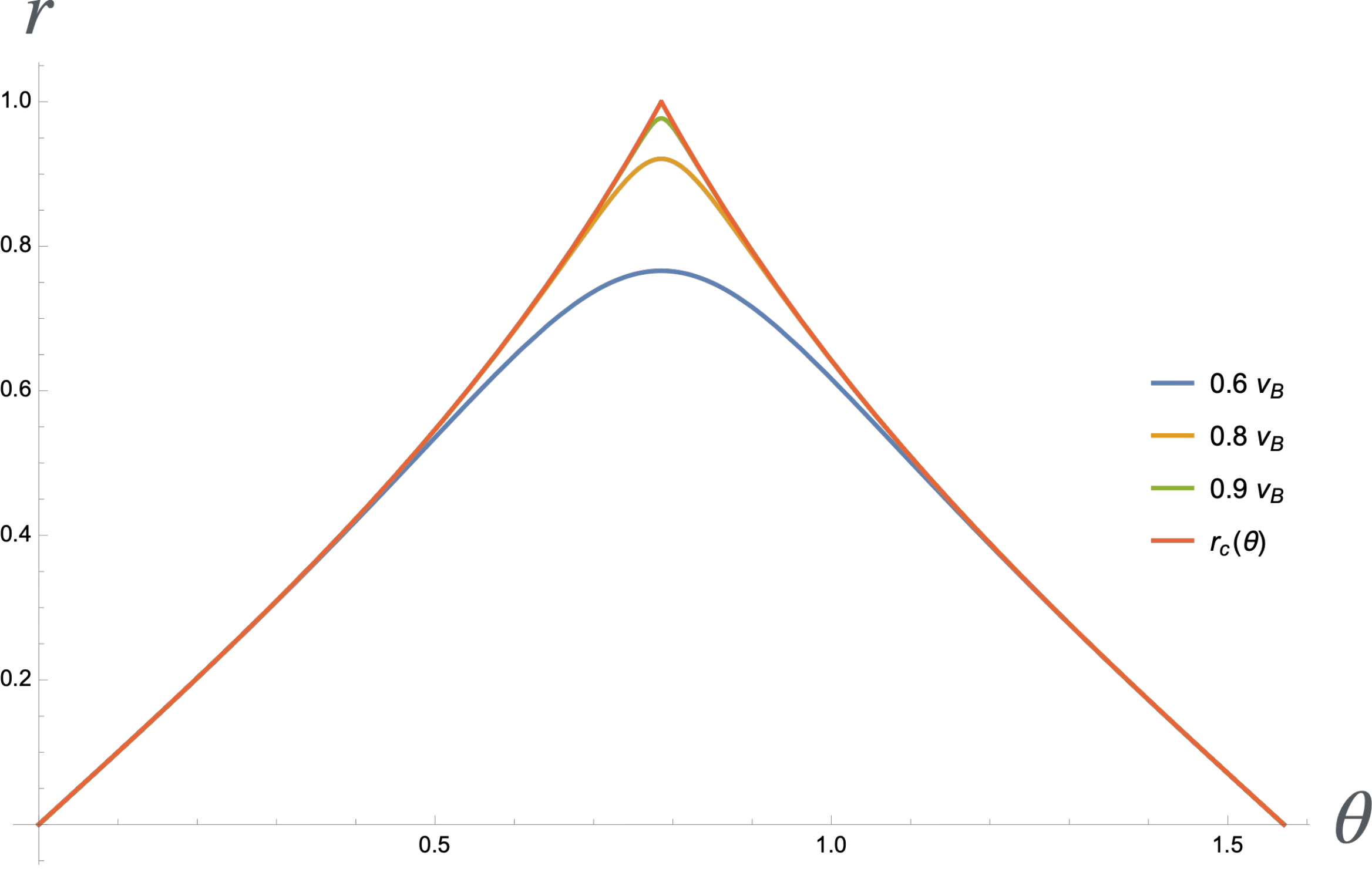}
\includegraphics[width=0.49\textwidth]{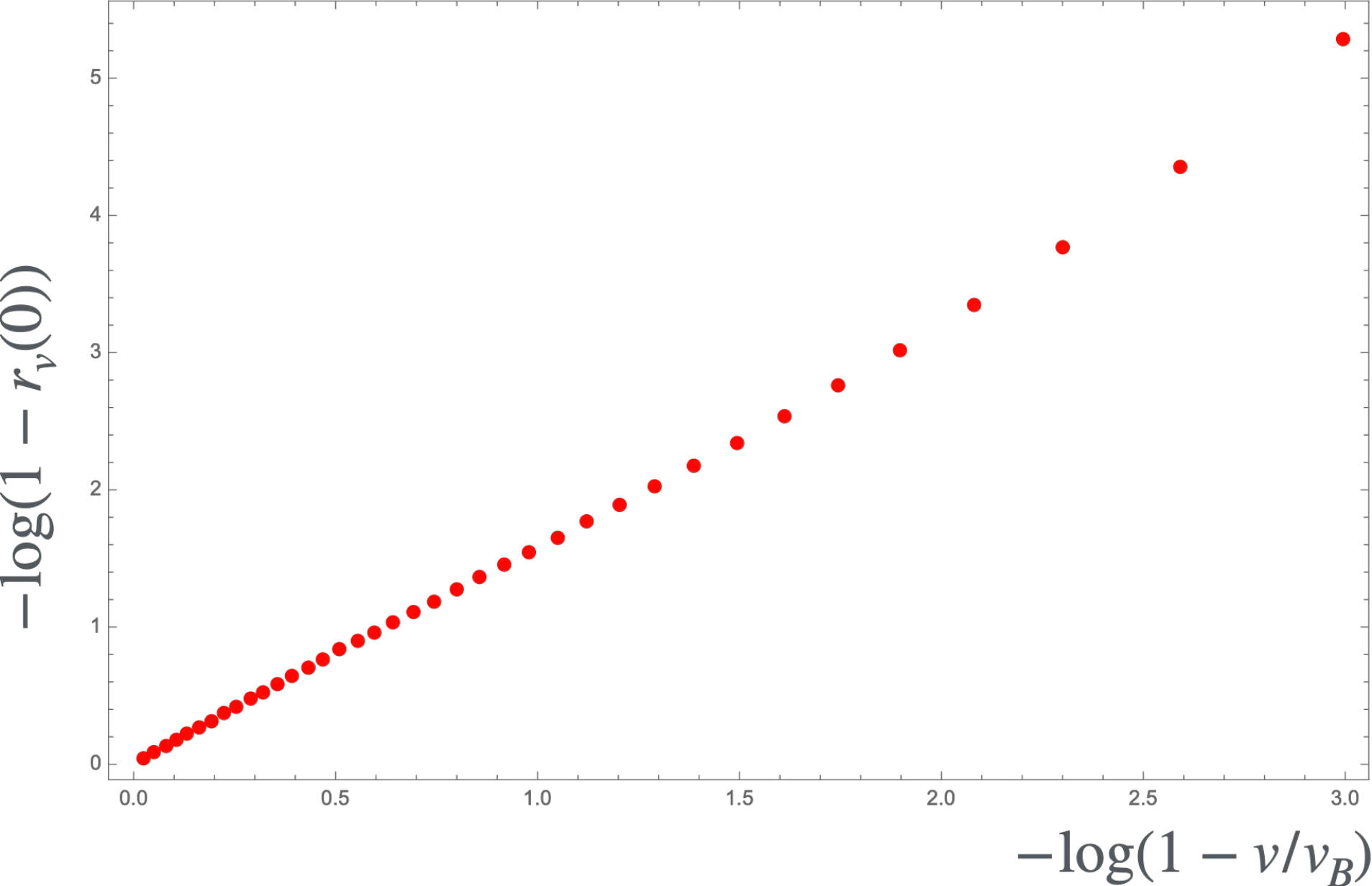} 
    \caption{{\bf Left:} $r(\theta)$ for $v = \frac{k v_B}{20}$. As $v \rightarrow v_B$, $r(\theta)$ approaches $r_c(\theta)$.
    {\bf Right:} $r_v(0)$ as a function of  $v/v_B$.} 
    \label{fig:r_theta}
\end{figure}

To guide our thinking it is useful to switch from $r(\sigma)$ to $r(\theta)$, see figure~\ref{fig:r_theta}. The critical curve is given by $r_c(\theta)$ from~\eqref{eq:rtheta_vb} and decreasing $v$ gives rise to a small deviation near $\theta=\pi/4$. We can solve the equations in this regime.

Since $(r,\theta) = (1,  \pi/4)$ is a fixed point of the differential equation, we can linearise differential the equations near $\sigma = 0$. We write \begin{equation}
    r(\sigma)  = 1 -\epsilon \,r^{(1)}(\sigma) -\epsilon^2\, r^{(2)}(\sigma) +\dots, \quad \theta(\sigma) = \frac{\pi}{4} +\epsilon\,\theta^{(1)}(\sigma)  +\epsilon^2\, \theta^{(2)}(\sigma) +\dots\,,
\end{equation}
where we choose $r(0)=1-\epsilon$ hence $r^{(1)}(0)=1$.
At leading order in $\epsilon$ the differential equations are:
\begin{malign}
    v\partial_\sigma r^{(1)} = 2(2  + \partial_\sigma^2) \theta^{(1)}, \quad  v\partial_\sigma \theta^{(1)} = \frac{1}{2} (2 + \partial_\sigma^2) r^{(1)},
\end{malign}
We must solve the differential equations with the following boundary conditions at $\sigma = 0$:
\begin{malign} 
    r^{(1)}(0) = 1,\quad \p_\sigma\theta^{(1)}(0) = \frac{1}{\sqrt{2}},  \quad \p_\sigma r^{(1)}(0) = \theta^{(1)}(0)= 0\,.
\end{malign}
The solution is
\begin{malign} 
r^{(1)}(\sigma)
&= 
\cosh(\alpha\sigma)\,\cos(\omega\sigma)
+ \frac{\gamma}{2\omega}\,\sinh(\alpha\sigma)\,\sin(\omega\sigma)
,\\[6pt]
\theta^{(1)}(\sigma)
&=\frac{1}{2}\!\left(
\,\sinh(\alpha\sigma)\,\cos(\omega\sigma)
+ \frac{\gamma}{2\omega}\,\cosh(\alpha\sigma)\,\sin(\omega\sigma)
\right)\,,
\end{malign}
where \begin{equation}
  v_B = 2\sqrt{2},\qquad  \alpha = \frac{v}{2},\qquad
\omega = \frac{\sqrt{\,v_B^2- v^2}}{2},\qquad
\gamma = v_B - v\,.
\end{equation}
We have also solved the problem at $O(\epsilon^2)$, but we will not write down the answer for general $v$. Instead, we also expand in $\gamma=v_B-v$, and write down the large $\sigma\gg 1$ (but $\epsilon e^{\sqrt2\sigma}\ll 1 $) behaviour of the solutions
\es{largesig}{
    r(\sigma) & = 1 -\epsilon\le(-\frac12 e^{\sqrt2\, \sigma}+{\gamma\over 2\sqrt2}\,\sigma^2\,e^{\sqrt2\, \sigma}+O\le(e^{-\sqrt2\, \sigma}\ri)\ri)  \\ & \quad -\epsilon^2\, \le(-\frac38 e^{2\sqrt2\, \sigma} + {\sqrt2 \sigma+1\over 2}\,e^{\sqrt2\, \sigma}-\frac54+O\le(\gamma,e^{-\sqrt2\, \sigma}\ri)\ri) +O(\epsilon^3)\,,\\
    \theta(\sigma) &= \frac{\pi}{4} +\epsilon\le(\frac14 e^{\sqrt2\, \sigma}-{\gamma\over 4\sqrt2}\,\sigma^2\,e^{\sqrt2\, \sigma}+O\le(e^{-\sqrt2\, \sigma}\ri)\ri) \\ &\quad -\epsilon^2\, \le(\frac14 e^{2\sqrt2\, \sigma}-{\sqrt2 \sigma+1\over 4}\,e^{\sqrt2\, \sigma}+O\le(\gamma,e^{-\sqrt2\, \sigma}\ri)\ri) +O(\epsilon^3)\,.
}
We can use the full expressions and make the parametric plot $(r(\sigma),\theta(\sigma))$. We can also use the large $\sigma$ behaviour given in~\eqref{largesig} to obtain the curve $r(\theta)$, which we find approaches the critical curve
\es{critcurveapproach}{
r_c(\theta)-r(\theta)=\epsilon\le( e^{-\sqrt2\, \sigma}-{\gamma\over \sqrt2}\,\sigma^2\,e^{-\sqrt2\, \sigma}+O\le(e^{-2\sqrt2\, \sigma}\ri)\ri)+\epsilon^2\, \le(1+O\le(\gamma,e^{-\sqrt2\, \sigma}\ri)\ri) +O(\epsilon^3)\,,
}
meaning the two curves approach closely.

The membrane width in the $v \rightarrow v_B$ limit is controlled by $\eps$ as \begin{equation}
    \Delta_v \sim \frac{\log(\eps^{-1})}{\sqrt{2}}
\end{equation}

In figure \ref{fig:membrane_thickness}, we provide a numerical plot of $r_v(0)$ as a function of $v/v_B$. We expect that we can derive this dependence analytically using the asymptotic series expansion techniques but we have not found a way to fix it completely.  We found that the two estimates $\Delta_v \sim \abs{\log(v_B - v)}$ and $\Delta_v \sim {1/\sqrt{(v_B - v)}}$\footnote{We thank Adam Nahum for the latter suggestion.} fit reasonably well with the numerical data. We leave the correct velocity dependence of the membrane width undetermined.

\section{Adiabatic evolution by a non-Hermitian operator} \label{sec:adiabatic}

Consider the time evolution by a time dependent non-Hermitian operator $H(s):  0 \leq s \leq 1$ operator with real eigenvalues. Assume further that there is a unique lowest eigenvalue that we fix to be zero. 

Consider the following evolution operator:
\begin{malign}
    O(T) = \mathcal{T} \exp \left( -\int_0^{T}  \d t \, H\left(\frac{t}{T}\right)\right)\,.
\end{malign}

We can break $O(T)$ into a product of $N$ operators $O_k(T)$ as follows:
\begin{align}
    O(T) &= \prod_{k = 1}^N O_k(T)\,, \\ 
    O_k(T) &= \mathcal{T} \exp \left(-\int_0^{T/N} \d t\,  H \left( s_k +  \frac{t}{T} \right)\right),\quad s_k = \frac{k-1}{N}\,.
\end{align}
Thus, $O_k(T)$ is an evolution operator between $t_k$ and $t_{k+1}$. Consider taking the limit $N \rightarrow \infty$ and $T \rightarrow \infty$ such that 
\begin{malign}
    \frac{T}{N} \rightarrow \infty, \quad \text{but} \quad \frac{T}{N^2} \rightarrow 0\,.
\end{malign}

In this limit, we can approximate $O_k(T)$ as follows:
\begin{malign} 
    O_k(T) &= \exp\left(- \int_0^{T/N} \d t\, H\left(s_k + \frac{t}{T}\right) \right)  \\&=\exp\left(- \frac{T}{N} H(s_k) - O\left(\frac{T}{N^2} \right)\right)  \\ &\approx \exp\left(-\frac{T}{N} H(s_k) \right)\,.
\end{malign}

A non-Hermitian operator admits a linear decomposition into its left and right eigenstates:
\begin{equation}
    H = \sum_k f_k \ket{r_k} \bra{l_k}, \quad \langle l_m,r_n\rangle = \delta_{m,n} \,.
\end{equation}
If $\ket{r_g},\bra{l_g}$ are the left and right eigenstates corresponding to the zero eigenvalue, then in the limit discussed above, we have,
\begin{equation}
  \lim_{\frac{T}{N} \rightarrow \infty}  \exp\left(-\frac{T}{N} H(s_k)\right) = \sum_{m} \exp\left(-f_m(s_k) \frac{T}{N}\right) \ket{r_m(s_k)}\bra{l_m(s_k)} \rightarrow \ket{r_g(s_k)}\bra{l_g(s_k)}\,.
\end{equation}
Therefore, in the limit $T \rightarrow \infty$, we can represent $O(T)$ as 
\begin{equation}
    \lim_{T \rightarrow \infty} O(T) = \lim_{T \rightarrow \infty} \prod_{k =1}^N O_k(T) = \prod_{k= 1}^N \ket{r_g(s_k)} \bra{l_g(s_k)} \approx \ket{r_g(1)} \bra{l_g(0)} \prod_{k=1}^{N-1} \langle l_g(s_k), r_g(s_{k-1})\rangle\,.
\end{equation}
The $T \rightarrow \infty$ limit of $O(T)$ maps the ground state of $H(0)$ to the ground state of $H(1)$ up to an extra ``Berry phase" contribution. If $H(s)$ is a Hermitian operator, then the eigenstates of $H$ are normalised to $1$. Therefore,
\begin{equation}
    \langle{l_g(s_{k+1})}, r_g(s_k) \rangle = \langle{r_g(s_{k+1})}, r_g(s_k) \rangle = 1 + O(1/N^2)\,.
\end{equation}
Thus, the extra Berry phase is trivial under evolution by a Hermitian Hamiltonian.   

\subsection{Example: $\dim \cH = 2$}

In the continuum limit of the Brownian SYK chain, we are interested in the Berry phase contribution  from the non-Hermiticity of the Hamiltonian in the following transition amplitude. 
\begin{align}
   \bra{\Gamma^u_f} \mathcal T \exp(-\int \d t\,\mathcal{O}_u(t) \ket{\Gamma^u_i} = \bra{\Gamma^u_f} \exp\left(-\jt\int_0^T \d t\,g_u(t)^{i,j} \hat \psi_i \hat \psi_j\right) \ket{\Gamma^u_i}\,.
\end{align}
For $n = 2$, note that $\ket{\Gamma_i^u}$ and $\ket{\Gamma_f^u}$ are invariant under the action of the parity operator defined as $P  = -4 \prod_{i} \hat \psi_i $.
Moreover, the evolution operator preserves parity throughout the dynamics. Therefore, the time evolution effectively happens in a two dimensional sector of the four dimensional Hilbert space. 

It is convenient to map the fermion bilinear operators to following Pauli operators~\cite{Stanford:2021bhl}:
\begin{malign} 
    2 i \hat \psi_1 \hat \psi_2 = X,\quad -2i \hat \psi_1 \hat \psi_3 = Y,\quad  2i \hat \psi_1 \hat \psi_4 = Z \,.
\end{malign}
Since $P = 1$ we also have \begin{malign} 
    2 i \hat \psi_3 \hat \psi_4 = X,\quad 2i \hat \psi_3 \hat \psi_4 = Y,\quad  2i \hat \psi_2 \hat \psi_3 = Z \,.
\end{malign}

The initial and final states are eigenstates of $Z$ and $X$ operators respectively. 
\begin{malign} 
    X \ket{\Gamma_i} = - \ket{\Gamma_i} ,\quad Z \ket{\Gamma_f} = -\ket{\Gamma_f}\,.
\end{malign}
Now we have a representation of the evolution operator in terms of the Pauli matrices:
\begin{malign} 
 H(t) = \sum_{i\neq j}  g_u(t)^{i,j} \hat \psi_i \hat \psi_j = x_u(t) X  + i y_u(t) Y + z_u(t) Z  = \begin{pmatrix}
     z & x + y \\ x-y & -z
 \end{pmatrix}\,.
\end{malign}
where,
\begin{equation}
    x_u^2 + z_u^2 - y_u^2 = 1\,.
\end{equation}
We will make use of the following definition:
\begin{align}
    \frac{y}{\sqrt{1 + y^2 }} = \sin \phi, \quad x \ = \sqrt{1+y^2} \sin \theta, \quad z = \sqrt{1 + y^2} \cos \theta\,.
\end{align}
The left and right eigenstates of $H$ satisfy
\begin{equation}
    \bra{l_g} H= -\bra{l_g}, \quad H \ket{r_g} = -\ket{r_g}\,.
\end{equation}
Using the above definition, we can represent the left and right instantaneous eigenstates of $H(t)$ as
\begin{equation}
    \bra{l_g} = \left(\sin\frac{\theta -\phi}{2}, -\cos \frac{\theta-\phi}{2}\right) , \quad \ket{r_g} = \frac{1}{\cos \phi}\left(\sin\frac{\theta+\phi }{2}, -\cos \frac{\theta+\phi}{2}\right)\,.
\end{equation}
we have normalised the eigenvectors so that they satisfy:
\begin{equation}\label{ortho}
    \langle l_g,r_g \rangle = 1\,.
\end{equation}
Now we can compute the inner product:
\begin{malign} 
    \langle l_g(s_{k+1}) , r_g(s_k)\rangle &\approx \langle l_g(s_{k}) , r_g(s_k)\rangle + \frac{1}{N} \langle \frac{\d}{\d s} l_g(s_k), r_g(s_k)\rangle \\ &= 1+ \frac{1}{2 N} \tan \phi \left(- \frac{\d  \phi}{\d s} +\frac{\d \theta}{\d s}  \right) \,.
\end{malign}
In the $N \rightarrow \infty$ limit, the product of all such inner products is replaced by the following integral:
\begin{equation}\label{BerryRes}
   B(T)=\lim_{N \rightarrow \infty} \prod_{k=1}^{N-1} \langle l_g(s_k), r_g(s_{k-1})\rangle = \exp \left[ \int_0^1 \d s \,\frac{\tan \phi}{2}  \, \le(- \frac{\d \phi}{\d s}  + \frac{\d  \theta} {\d s}\ri)\right] = \exp\left[\int_0^T \frac{\d t}{2}\, y\,  \dot \theta\right]\,.
\end{equation}
In the final step, we have ignored the contribution from the total derivative term and replaced the variable of integration from $s$ to $t$.

The same result can be obtained by solving the Schr\"odinger equation. The Euclidean evolution will project any initial state to the ground state in $O(1)$ time. Subsequently, the ground state evolves adiabatically.
We write 
\es{psit_adiabatic}{
\vert \psi(t)\rangle&\approx B(t)e^{t} \vert r_g(t)\rangle\,,
}
and plug the Ansatz into the Schr\"odinger equation projected onto the ground state:
\es{psit_adiabatic2}{
0&=\langle l_g(t)\vert\p_t+H(t)\vert \psi(t)\rangle=\langle l_g(t)\vert\le[\p_t\le(B(t)e^{t}\vert r_g(t)\rangle\ri)-\vert r_g(t)\rangle\ri]\\
&=e^{t}\le(\dot B(t)+B(t)\langle l_g(t)\vert \dot r_g(t)\rangle\ri)\,,\\
&\implies {\dot B(t)\ov B(t)}=\frac{1}{2} \tan \phi \left(-\dot \phi + \dot \theta \right)\,,
}
where we used the orthogonality property~\eqref{ortho} and the explicit form of the instantaneous eigenstates (or reused~\eqref{BerryRes}). The resulting differential equation is indeed solved by~\eqref{BerryRes}.

\section{Third R\'enyi entropy}\label{app:Renyi3}

The calculation of the third R\'enyi membrane tension is quite similar to the second R\'enyi membrane tension. In this appendix we work out some of the steps in the calculation of the third  R\'enyi entropy.

For $n=3$, $g_u$ is a $6\times 6$ matrix that takes the following form after imposing replica symmetry discussed in appendix~\ref{sec:replica_symmetry}. 
\begin{equation}
   -2i g_u^{i,j} = \begin{pmatrix}
       0 & a_u & i b_u & c_u & ib_u & e_u \\
       -a_u & 0 & e_u & -ib_u &  c_u& -ib_u \\
       -ib_u & -e_u & 0 & a_u & ib_u & c_u \\
       -c_u &i b_u & -a_u & 0 & e_u & -ib_u \\ 
       -i b_u & -c_u & -ib_u & -e_u & 0 & a_u \\ -e_u & i b_u & -c_u & ib_u & -a_u & 0
    \end{pmatrix}\,.
\end{equation}
The $g,\sigma$  equations again imply that $\Tr(g_u^{2k})$ is conserved. Not all such traces are independent. We find the following two independent conserved quantities:
\begin{malign}
 Q_u &= a^2 - 2b^2 + c^2 + e^2 = 1\,, \\
   H_u &= a_u + e_u - c_u = 1\,.
\end{malign}
Therefore, the dynamics is restricted to two independent variables. Define the variables $r,\theta$ as follows:
\begin{malign}  \label{eq:rtheta_Renyi3}
    r = \sqrt{3}\, b,\qquad \sqrt{r^2 +1} \cos \theta = \frac{3 c +1}{2}, \qquad \sqrt{r^2 +1} \sin \theta = \frac{\sqrt{3}}{2} (a - e)\,,
\end{malign}
In the continuum limit, the equations of motion can be derived from the following effective action:
\begin{malign}
    S_{\text{eff}}^{(3)} &= \int \d t\,\d \sigma \left(r\,  \partial_t \theta - h(\tau,\sigma) \right)\,, \\
    h(\tau,\sigma) &=  6\left(\frac{J}{64}\left( 1-a^4 - c^4 - e^4 - 2b^4 \right) + \frac{1}{8}\left(( \partial_\sigma a)^2 + ( \partial_\sigma c)^2 + ( \partial_\sigma e)^2- 2( \partial_\sigma b)^2\right)\right)\,.
\end{malign}
In terms of $r,\theta$ the Hamiltonian density is 
\begin{malign}
    h(\tau,\sigma)  &= \frac{J_c}{36}\left( 1-3 r^2  + (1 + r^2)^{3/2} \cos 3 \theta \right) + \frac{1}{2}\left( (r^2+ 1) (\partial_\sigma \theta)^2-\frac{(\partial_\sigma r)^2}{r^2+1}\right)\,.
\end{malign}
\subsection{Traveling wave solution}
We restrict the action to traveling wave configurations with velocity $v$ to define an effective spatial action:
\begin{align} 
    S = -\int \d \sigma\, \mathcal L \,,\qquad 
    \mathcal L =  v r \,\theta' + h(\sigma)\,.
\end{align}
Translational symmetry in $\sigma$ implies that the following spatial Hamiltonian is conserved:
\begin{malign} 
    \hat h =  -\frac{1}{2}\left( (r^2+ 1) (\partial_\sigma \theta)^2-\frac{(\partial_\sigma r)^2}{r^2+1}\right)+\frac{J_c}{36}\left( 1-3 r^2  + (1 + r^2)^{3/2} \cos 3 \theta \right) \,.
\end{malign}
In the $\sigma \rightarrow -\infty$ limit, the only nonzero correlation function is  $ e = -2i g_u^{2k,2k+1} = 1$ while as $\sigma \rightarrow +\infty$, the only nonzero correlation function is $a = - 2ig_u^{1,2}  = 1$. The asymptotic behaviour at $\sigma \rightarrow \pm \infty$ fixes the boundary condition for the traveling wave equation:
\begin{malign} 
    \sigma \rightarrow -\infty:& \quad r = 0,\, \theta =- \frac{\pi}{3}\,, \\ 
    \sigma \rightarrow \infty:& \quad r = 0,\, \theta = \frac{\pi}{3}  \,.
\end{malign}
From the boundary conditions, it follows that $\hat h$ vanishes everywhere. 
The traveling wave equations in the $r,\theta$ representation are the same as~\eqref{eq:hamiltons_equation}, with the only difference coming from the potential term \begin{equation}
    V = \frac{J_c}{36} (1 - 3r^2 + (1+r^2)^{3/2} \cos 3 \theta)\,.
\end{equation}
Demanding that the solution is symmetric about $\sigma = 0$, we get the following constraints:
\begin{align}
    r(\sigma) = r(-\sigma), \quad  \theta(\sigma) = -\theta(-\sigma)\,.
\end{align}
The reflection symmetry imposes following constraints on the solution at $\sigma = 0$:
\begin{align}
    r'(0) = 0,\qquad  \theta(0) = 0\,.
\end{align}
From the fact that the spatial Hamiltonian $\hat h$ vanishes everywhere, we can constrain $p_\theta$:
\begin{equation}
    p_\theta = -v r -\sqrt{2 (r^2+1) V}\,.
\end{equation}

\subsection{Membrane tension}

We can simplify the on-shell action using the fact that $\hat h$ vanishes everywhere. The on-shell action for the third R\'enyi entropy takes a similar form as~\eqref{eq:seff}.
\begin{malign}  \label{eq:Renyi3action}
    -\frac{S_{\text{eff}}^{(3)}}{NT_\tau}
    &= -\frac{1}{2a}\int \d \sigma \left( vr \, \partial_\sigma \theta + (r^2+ 1) (\partial_\sigma \theta)^2 -\frac{(\partial_\sigma r)^2}{r^2+1}\right)\,.
\end{malign}
Following similar steps from section~\ref{sec:action}, for the third  R\'enyi entropy, we find that it has the following relation to the effective action:
\begin{equation}
    S^{(3)}_{A \cup B \cup \bar A} = \frac{S_{\text{eff}}^{(3)}}{2}\,.
\end{equation}
\subsection{Entanglement velocity}
At $v = 0$, the saddle point equation can be solved exactly. Setting $r = 0$ everywhere, we find the following equation:
\begin{malign} 
  & \hspace{1cm}\theta'' = -\frac{J_c}{12} \sin 3 \theta \\&\implies \theta'^2 = \frac{2J_c }{36} (1 + \cos 3 \theta  )\\ 
    &\implies \theta' = \frac{\sqrt{J_c} }{3} \cos \frac{3 \theta}{2} \\ &\implies  \sin \frac{3\theta}{2} = \tanh\frac{ \sqrt{J_c}\, \sigma}{2}\,.
\end{malign}
The contribution to the action at $v =0$ is only from the Hamiltonian density:
\begin{malign} 
    -S_{\text{eff}}^{(3)} &=- \frac{NT_\tau}{2a}  \int \d \sigma\, \theta'^2   = - \frac{N T}{2} \times \frac{4 \sqrt{J_c}}{9} =  - \frac{\sqrt{2} N v_B T}{9 }\,.
\end{malign}
Hence the third R\'enyi entanglement velocity is:
\begin{malign} 
   v_E^{(3)}=\mathcal E^{(3)}(0)  = \log{d} \times \frac{ 2\,v_B}{9 \sqrt{2}\log 2}\,.
\end{malign}

\subsection{Third R\'enyi entropy at $v > v_B$}

For $v > v_B$, the solution is:
\begin{equation}\label{eq:rtheta_vgvb}
    r(\theta) = \begin{cases} 
    \cot(\frac{\pi}{6} - \theta), \quad -\frac{\pi}{3} < \theta < 0 \\ 
    \cot{(\frac{\pi}{6} + \theta)}, \quad 0< \theta < \frac{\pi}{3} \,.
    \end{cases}
\end{equation}
In this case, the action has contribution only from the Berry phase:
\begin{equation}
    \int_{-\pi/3}^{\pi/3} r\, \d \theta = - 2 \log  \sin \frac{\pi}{6} = \log 4\,.
\end{equation}

Plugging this result into the action yields:
\begin{malign}
S^{(3)}_{A \cup B \cup \bar A} & = \frac{S_{\text{eff}}^{(3)}}{2}
  = \frac{1}{2}N \sum_u \int \frac{ r\, \d \theta }{2} = N v T \frac{\log 4}{4} = N v T\log \sqrt{2} \\ &= v T\,\log d \,.
\end{malign}

Notice that the solution in \eqref{eq:rtheta_vgvb} can be expressed in terms of $a, b, c$ and $e$ as:
\begin{malign}\label{eq:vgvb_renyi3}
    b &= c\,, \\ 
    (a, e) &= \begin{cases}
        (b ,1), \quad - \frac{\pi}{3} < \theta < 0 \,,\\ 
        (1, b), \quad 0 < \theta < \frac{\pi}{3}\,.
    \end{cases}
\end{malign}

The above solution is consistent with the discussion 
in appendix \ref{app:vgvb} of the saddle point solution for $v >v_B$ and arbitrary $n$. We can see this once we make the following identification
\begin{equation}
    b_u =  - 2 g_u^{1,3} = y_u\,. 
\end{equation}
Indeed the membrane for the third R\'enyi entropy splits into two fronts for $v > v_B$. Since $\theta$ monotonically increases with the spatial coordinate $\sigma$, the solution for $\theta < 0$ corresponds to the wavefront near $A|B$ while the solution for $\theta > 0$ corresponds to the wavefront near~$B|C$. 

The result for the third R\'enyi entropy for $v>v_B$ is consistent with operator growth argument discussed at the end of section \ref{subsec:Renyi_entropy_vgvb} and the arguments in appendix \ref{app:tension_vgvb} used to fix the R\'enyi entropies for arbitrary $n$.  

\section{R\'enyi entropy for $n \geq 4$}
\label{app:highern}

While the membrane dynamics for $n = 2,3$ was described by two independent variables, we must solve for more than two independent variables for $n \geq 4$. 

Imposing replica symmetry for $n =4$ from appendix~\ref{sec:replica_symmetry}, $g_u^{i,j}$ takes the following form:

\begin{equation}
    g_u^{i,j} = \begin{pmatrix}
       0 & a_u & b_u & c_u & d_u & e_u & b_u & f_u \\
       -a_u & 0 & f_u & -b_u &  e_u& -d_u &c_u & -b_u\\
       -b_u & -f_u & 0 & a_u & b_u & c_u & d_u & e_u \\
       -c_u & b_u & -a_u & 0 & f_u & -b_u &e_u &-d_u\\ 
       -d_u & -e_u & -b_u & -f_u & 0 & a_u &b_u & c_u\\
       -e_u &  d_u & -c_u & b_u & -a_u & 0 & f_u & -b_u\\ 
      -b_u & -c_u & -d_u & -e_u & -b_u & -f_u & 0 &  a_u \\
     -f_u   &b_u & -e_u & d_u & -c_u &b_u & -a_u & 0
    \end{pmatrix}\,.
\end{equation}
We get the following constraints from conservation of $\Tr(g^{2k}_u)$. 
\begin{malign}
     Q_u &= a_u^2 + 2 b_u^2 + c_u^2 + d_u^2 + e_u^2 + f_u^2 \,, \\ 
     D_u &= 2 b_u d_u + a_u (c_u -f_u) + e_u (c_u + f_u)\,.
\end{malign}

Since $g_{u}^{i,j}$ has six variables and there are two nontrivial constraints, the dynamics is described by four independent variables. For $v=0$ we have $b=d=0$, which leaves two independent variables. This is unlike the $n=2,3$ cases where for $v=0$ only $\theta$ was active and the saddle was possible to obtain analytically.

Due to the increased difficulty of solving the saddle point equations for four variables, we leave the computation of ${\cal E}^{(4)}(v)$ to the future.

\section{$n$-th R\'enyi entropy for $v > v_B$} \label{app:tension_vgvb}

In this section, we will evaluate the $n$-th R\'enyi entropy of the region $A \cup B \cup \bar A$ for the case $v> v_B$. Let us first consider the second R\'enyi entropy. As we discussed in section \ref{sec:Renyi2}, we can map the second R\'enyi entropy to a transition amplitude under a Euclidean time evolution by a positive Hamiltonian $H$ defined in equation \eqref{eq:pos_hamiltonian}.
\begin{malign} \label{eq:sec_renyi_trans_amp}
    \Tr(\rho^{2}_{A \cup B \cup \bar A}) =  \bra{\Gamma_f}  e^{ - H T} \ket{\Gamma_i}\,,
\end{malign}
where $\ket{\Gamma_i}$ and $\bra{\Gamma_f}$ are the initial and final states defined in equations \eqref{eq:initial_state} and \eqref{eq:final_state} respectively. 

\begin{figure}[h]
    \centering
\includegraphics[width=0.9\textwidth]{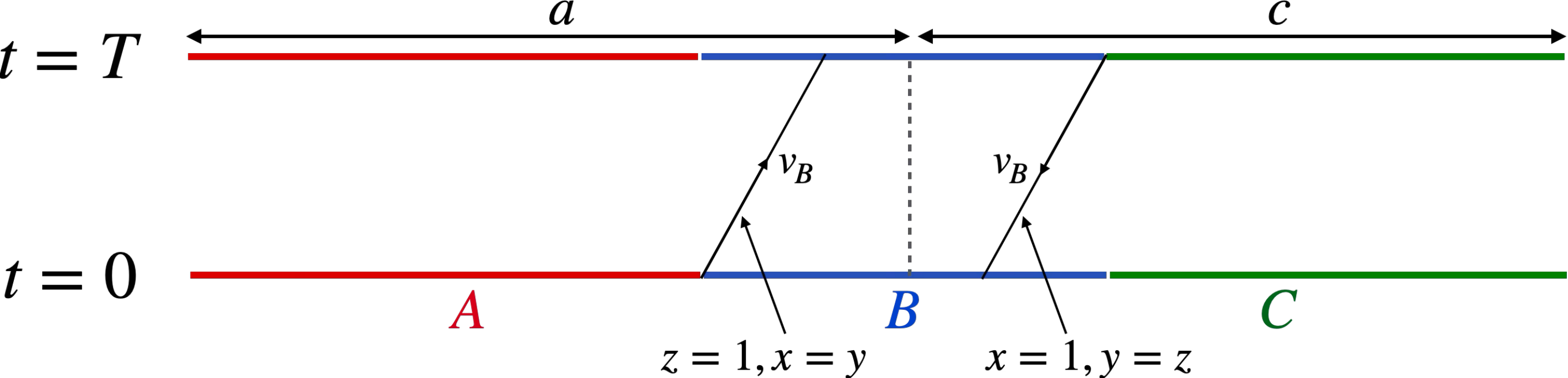} 
    \caption{Saddle point solution for $v > v_B$. We have divided the chain into two regions $a$ and $c$. The saddle point solution in $a$ and $c$ is unaffected by regions $C$ and $A$ respectively. } 
    \label{fig:schematic_sol}
\end{figure}

We will find it convenient to divide the chain into two regions denoted by $a$ and $c$ in figure \ref{fig:schematic_sol}. The initial and final states in the above transition amplitude are tensor products over pure states at different sites. Therefore, we can write them as tensor products over pure states in $a$ and $c$. 
\begin{malign}
    \ket{\Gamma_i} = \ket{\Gamma_i^a} \otimes \ket{\Gamma_i^c} \,, \quad \ket{\Gamma_f} = \ket{\Gamma_f^a} \otimes \ket{\Gamma_f^c}\,.
\end{malign}
\begin{figure}[h]
    \centering
    \includegraphics[width=0.8\linewidth]{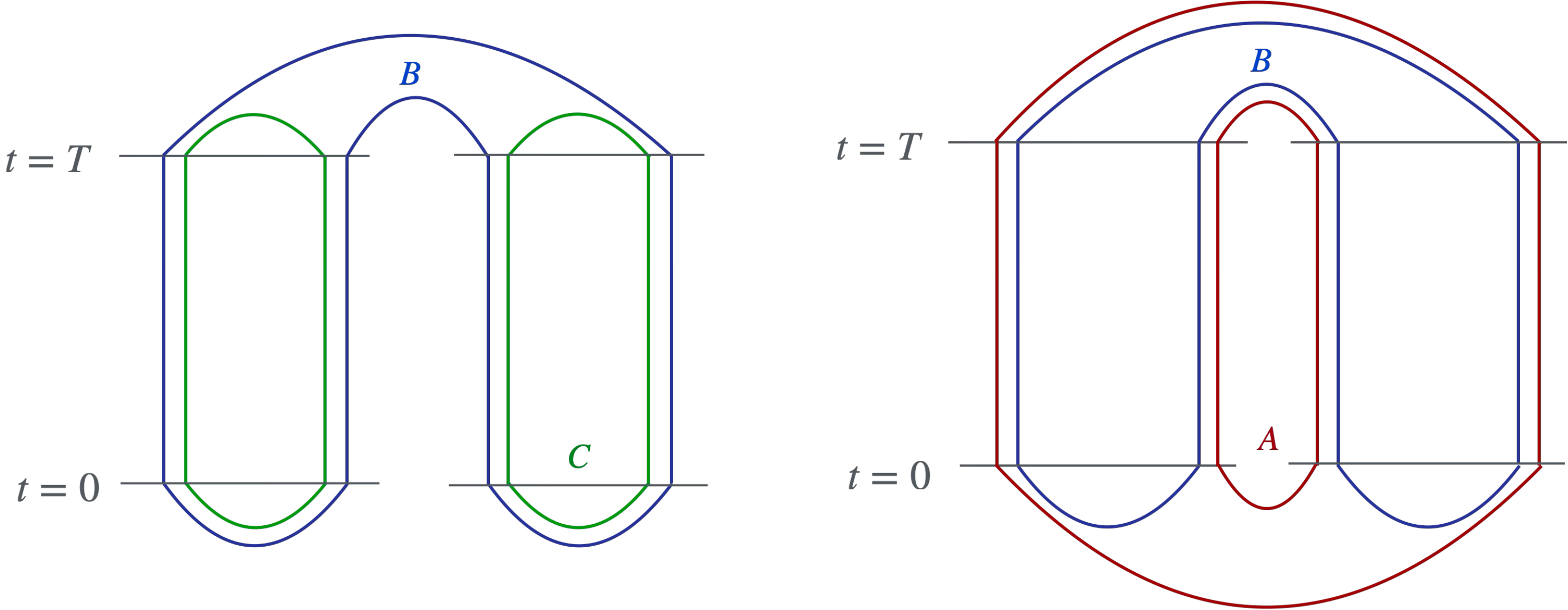}
    \caption{ \textbf{Left:} Path integral contour for $c$ ignoring region $A$. The initial state is a maximally entangled state which does not evolve in time.  \textbf{Right:} Path integral contour for $a$ ignoring region $C$. The final state is a maximally entangled state that does not evolve in time.}
\label{fig:trivial_evolution}
\end{figure}
We want to evaluate the transition amplitude in the saddle point approximation. In \ref{sec:vgvb}, we noted that to solve the saddle point equations in the range $v > v_B$ in the region $c$ we can ignore the influence of $A$. Since the saddle point solution in $c$ is not affected by $A$, we can also evaluate the saddle point action in $c$ ignoring $A$. This would correspond to computing a path integral along the contour shown in figure \ref{fig:trivial_evolution} in the saddle point approximation. This path integral computes a transition amplitude between $\bra{\Gamma_f^c}$ and $\ket{\Gamma_i^c}$. Since $\ket{\Gamma_{i}^c}$ is a maximally mixed state which (ignoring $A$) does not evolve in time, the saddle point action in $c$ is related to the inner product $\langle \Gamma_f^c , \Gamma_i^c \rangle$.  

The arguments for the saddle point action in $a$ are similar. In this case we can ignore region $C$. The corresponding path integral is shown in figure \ref{fig:trivial_evolution}. Once $C$ is ignored, the final state in $\bra{\Gamma_f^a}$ does not evolve in time and the saddle point contribution from $a$ is related to the inner product $\langle \Gamma_f^a , \Gamma_i^a \rangle$.

To summarise the above discussion, we can break the transition amplitude in equation \ref{eq:sec_renyi_trans_amp} into products of transition amplitudes in regions $a$ and $c$ both of which do not depend on $T$. Therefore, we have
\begin{malign}
    \Tr(\rho^2_{A \cup B \cup \bar A}) &=  \bra{\Gamma_f}  e^{ - H T} \ket{\Gamma_i} \\ &\approx  \bra{\Gamma_f^a} e^{-H_a T} \ket{\Gamma_i^a} \times  \bra{\Gamma_f^c} e^{-H_b T} \ket{\Gamma_i^c} \\ &=  \langle \Gamma_f^a,   \Gamma_i^a \rangle \times  \langle \Gamma_f^c,  \Gamma_i^c\rangle \\  &= \langle \Gamma_f , \Gamma_i \rangle \\ &= \langle \Gamma_f | e^{- 0 \times T} |\Gamma_i \rangle \, .
\end{malign}
Therefore, the second R\'enyi entropy of $A \cup B \cup \bar A$ for $v > v_B$ is the same as the second R\'enyi entropy of $A \cup B \cup \bar A$ under a trivial time evolution.  Under a trivial time evolution, $A$ remains maximally entangled with $\bar A$ and $B$ is maximally entangled with $\bar B$. Therefore, the second R\'enyi entropy of $A \cup B \cup \bar A$ is 
\begin{align}
    S_{A \cup B \cup \bar A}^{(2)} =  |B| \times \log d  = v T \log d \, .
\end{align}

Using the arguments of appendix \ref{app:vgvb}, it is straightforward to generalise this result for higher R\'enyi entropy. We conclude that for arbitrary $n$:
\begin{align}
    S^{(n)}_{A \cup B \cup \bar A} = v T \log d \,.
\end{align}

\section{Second R\'enyi entropy from operator growth}\label{sec:purity_prob}

We begin this appendix by defining the norm of an operator acting on the Hilbert space $\cH$ as
\begin{equation}
    ||O|| = \left(\Tr_\cH(O^\dagger O)\right)^{1/2}
\end{equation}

We are interested in the second R\'enyi entropy of the subsystem $A \cup B \cup \bar A$ in the state $\ket{\Psi(T)}$ defined in equation \eqref{eq:maxstate}. We can decompose $\ket{\Psi(T)}$ as follows:
\begin{malign} 
    &\ket{\Psi(T)} = \frac{1}{\sqrt{\dim \cH_A}}\sum_{i = 1}^{\dim \cH_A} \ket{i}_{\bar A} \otimes U(T) \left( \ket{i}_{A} \otimes \ket{\chi}\right),\\ &\ket{\chi} = \frac{1}{\sqrt{\dim \cH_{B \cup C}}}\sum_{j = 1}^{\dim \cH_{B \cup C}} \ket{j}_{B \cup C} \otimes \ket{j}_{\bar B \cup \bar C}\,.
\end{malign}
Now consider computing $\Tr (\rho^2(T)_{A \cup B \cup \bar A})$:
\begin{align}\label{eq:purity_sum}
    \Tr \left( \rho^2(T)_{A \cup B \cup \bar A} \right) =  \frac{1}{\dim \cH_A}\sum_{i,j = 1}^{\dim \cH_A}  \Tr_{A\cup B} \left(\Tr_C \,O^\dagger_{i,j}(T) \, \Tr_C \, O_{i,j}(T)   \right)\,,
 \end{align}
where \begin{malign}
O_{i,j}(T) &=  U(T) \le(\ket{i}_A\bra{j}_A \otimes \rho_{\text{MAX}} \ri)U^\dagger(T)  \, , \\
\rho_\text{MAX} &= \Tr_{\bar B \cup \bar C}(\ket{\chi} \bra{\chi}) = \frac{I_{B \cup C}}{\dim \cH_{B \cup C}} \,.
\end{malign}

Every term in the sum in equation~\eqref{eq:purity_sum} is a square of the norm of the operator $O_{i,j}(T)$ acting on $A \cup B$. This norm can be expanded in any basis of orthonormal operators acting on $A \cup B$. Let $\{ \Phi_J \}$ form one such basis of operators on $A \cup B$. Then, 
\begin{equation}
    \Tr_{A\cup B} \left(\Tr_C\,  O^\dagger_{i,j}(T) \,\Tr_C \,O_{i,j}(T)   \right) = \sum_{J} \Tr_{A \cup B} ( \Tr_C \,O_{i,j}^\dagger (T) \Phi_J)  \,\Tr_{A\cup B}(\Tr_C\, O_{i,j} (T) \Phi_J^\dagger)
\end{equation}
We can extend the action of $\Phi_J$ on the full chain $R = A \cup B \cup C$ by demanding that they act trivially on $C$:
\begin{equation}
    \Phi_J \rightarrow \Phi_J^e = \Phi_J \otimes I_C
\end{equation}
This allows us to extend the trace in the above equation to $R$. We can rewrite the equation as\begin{equation}
     \Tr_{A\cup B} \left( \Tr_C \, O^\dagger_{i,j}(T) \,\Tr_C \, O_{i,j}(T)   \right) = \sum_{J} \Tr_R ( O_{i,j}^\dagger (T) \Phi_J^e)  \Tr_R( O_{i,j} (T) \Phi_J^e { }^\dagger)
\end{equation}
Notice also that the set of operators $\{O_{i,j} (T)\}$ forms an orthogonal set of operators on $R$ which, at $T =0$, have non-trivial action only on $A$. On $R$, $O_{i,j}(T)$ have norm
\begin{malign}
    || O_{i,j}|| &= \sqrt{\Tr_R \left(O^\dagger _{i,j}(T) O_{i,j} (T)\right)} =  \frac {1}{\sqrt{\dim \cH_{B \cup C}}}
\end{malign}
We can rotate the set $\{O_{i,j}(T)\}$  to another orthogonal set of Hermitian operators, say, $\{\Theta_I(T)\}$ with the same properties.  This gives us the following formula for the purity of the state $\rho_{A \cup B \cup \bar A}$: 
\begin{equation}
    \Tr(\rho^2(T)_{A \cup B \cup \bar A})  = \frac{1}{\dim \cH_A} \sum_I \sum_J \Tr_R^2 (\Theta_I (T) \Phi_J^e)
\end{equation}
For the Brownian SYK chain, it is natural to generate the set $\{ \Phi_J^e \}$ and $\{\Theta_I(0)\}$ by Hermitian operators using products of Majorana fermions:
\begin{malign}
\{\Phi_J^e \} &= 
\left\{\frac{ (P_{C})^{|J|} \Psi_J}{\sqrt{ \dim \cH_{A \cup B}}} \text{   s.t.   } \Psi_J = i^{|J|/2} \prod_{\psi_\alpha \in J} \psi_{\alpha } \quad \text{    for    } J \subseteq  A \cup B\right\}\,, \\ 
\{\Theta_I (0)\} &= 
\left\{\frac{(P_{B \cup C})^{|I|} \Psi_I }{\sqrt{ \dim \cH \dim \cH_{B \cup C}} } \text{   s.t.   } \Psi_I = i^{|I|/2} \prod_{\psi_\beta \in I} \psi_{\beta }  \quad \text{   for   } I \subseteq A \right\}\,, \\ 
\{\Theta_I(T) \} &= \{U(T) \Theta_I(0) U(T)^\dagger\}
\end{malign}
where $J$ (respectively $I$) runs over all possible sets of Majorana fermions in $A \cup B$ and $A$ respectively. $P_C$ and $P_{B \cup C}$ are parity operators defined as 
\begin{malign}
P_C &= i^{|C|/2} \prod_{\psi_\alpha \in C} \psi_\alpha, \quad 
P_{B \cup C} &= i^{|B \cup C|/2} \prod_{\psi_\alpha \in B \cup C} \psi_\alpha
\end{malign}
The parity operators are essential to ensure that $\{ \Phi_J^e\}$ and $\{\Theta_I(0)\}$ act trivially on $C$ and $B \cup C$ respectively.\footnote{To see why, consider a representation of Majorana fermions on a two qubit Hilbert space $\cH_1 \otimes \cH_2$. A possible representation of the fermions is $\psi_1  = X \otimes I, \psi_2 = Y \otimes I, \psi_3 = Z \otimes X, \psi_4 = Z \otimes Y.$  Operators acting only on $\cH_2$ are generated by $I\otimes X, I \otimes Y$. In terms of the fermions, they are given by $I \otimes X = -i \psi_1 \psi_2 \psi_3$ and $I \otimes Y = - i \psi_1 \psi_2 \psi_4$. $-i \psi_1 \psi_2$ is the parity operator for fermions acting on $\cH_1$.} 

In terms of the products of fermions, the purity is
\begin{malign}\label{purity_prob}
 &\dim^2 \cH_A \dim \cH_{ B}   \times \Tr (\rho^2(T)_{A \cup B \cup \bar A}) \\  &= \sum_{I \in A} \sum_{J \in A \cup B} \frac{ \Tr^2 \left(U(T) (P_{B \cup C})^{|I|} \Psi_I  U^\dagger(T) (P_C)^{|J|}\Psi_J \right)}{(\dim \cH)^2 } \\ &= \sum_{I \in A} \sum_{J \in A \cup B} \frac{ \Tr^2 \left(U(T) (P_R P_{A})^{|I|} \Psi_I  U^\dagger(T) (P_R P_{A \cup B})^{|J|}\Psi_J \right)}{(\dim \cH)^2 } \\&= \sum_{I \in A} \sum_{J \in A \cup B} \frac{ \Tr^2 \left(U(T) (P_{A})^{|I|} \Psi_I  U^\dagger(T) ( P_{A \cup B})^{|J|}\Psi_J \right)}{(\dim \cH)^2 } \\&=  \sum_{I \in A} \sum_{J \in A \cup B} \frac{ \Tr^2 \left(U(T) \Psi_I  U^\dagger(T)  \Psi_J \right)}{(\dim \cH)^2 } = \sum_{I \in A} \sum_{J \in A \cup B} \frac{ \Tr^2 \left(\Psi_I(T)   \Psi_J \right)}{(\dim \cH)^2 } 
\end{malign}
In the second step, we used the parity  operator $P_R$ defined as the product of all fermions on chain $R$.

To conclude, we can write the purity as:
\begin{malign}
    \Tr (\rho^2(T)_{A \cup B \cup \bar A})  &= \frac{1}{\dim^2 \cH_A \dim \cH_B} \sum_{I \in A} \sum_{J \in A \cup B} P_{I \rightarrow J}\,, \\ P_{I \rightarrow J} &=  \frac{ \Tr^2 \left(\Psi_I(T)   \Psi_J \right)}{(\dim \cH)^2 } 
\end{malign}
 
For every $I \in A$, the sum \begin{equation} \sum_{J \in A \cup B} P_{I \rightarrow J} \end{equation} is the probability that $\Psi_I(T)$ is supported in the region $A \cup B$. A typical operator acting on $\cH_A$, when expanded in a linear combination of an orthonormal basis of operators acting on $A$, would have an equal support on all basis elements. Therefore, 
\begin{equation} 
\dim \cH_B \times \Tr(\rho^2(T)_{A \cup B \cup \bar A})  = \frac{1}{\dim^2 \cH_A} \sum_{I \in A} \sum_{J \in A \cup B} P_{I \rightarrow J} 
\end{equation}
is the probability that a typical operator in $A$, when evolved by time $T$, has support in $A \cup B$. In~\cite{Mezei_Stanford_2017} essentially the same result, relating the second R\'enyi entropy to operator growth, was derived for spins.

\section*{} % Bibliography 
\bibliographystyle{JHEP}
\bibliography{refs}

\end{document}